\documentclass{aa}
\usepackage[varg]{txfonts}
\usepackage{natbib}
\usepackage{graphicx}
\usepackage{hyperref}
\usepackage{makecell}
\usepackage{makecell}
\usepackage{cellspace} %
\usepackage{tablefootnote}
\usepackage{longtable}
\usepackage{xtab}
\usepackage{float}
\usepackage{xcolor}

\begin{document}

\title{Bottom-up dust nucleation theory in oxygen-rich evolved stars}

\subtitle{I. Aluminum oxide clusters\\}

\author{David Gobrecht\inst{1}
  \and John M. C. Plane\inst{2}
     \and Stefan T. Bromley\inst{3,4}
       \and Leen Decin\inst{1}
        \and Sergio Cristallo\inst{5,6} 
          \and Sanjay Sekaran \inst{1}} 
\offprints{D. Gobrecht, \email{dave@gobrecht.ch}}

\institute{Institute of Astronomy, KU Leuven,
  Celestijnenlaan 200D, B-3001 Leuven, Belgium
\and School of Chemistry, University of Leeds, Leeds LS2 9JT, United Kingdom
\and Departament de Ci\`encia de Materials i Qu\'imica F\'isica \& Institut de Qu\'imica Te\'orica i Computacional (IQTCUB), Universitat de Barcelona, E-08028 Barcelona, Spain
\and Instituci\'o Catalana de Recerca i Estudis Avan\'cats (ICREA), E-08010 Barcelona, Spain
\and INAF – Osservatorio Astronomico d’Abruzzo, Via mentore maggini s.n.c., I-64100 Teramo, Italy
\and INFN – Sezione di Perugia, via A. Pascoli, I-06123 Perugia, Italy}

\date{Received -- / Accepted --}

\abstract{Aluminum oxide (alumina, Al$_{2}$O$_{3}$) is a promising candidate as a primary dust 
condensate in the atmospheres of oxygen-rich evolved stars. Therefore, alumina \textit{seed} particles  
might trigger the onset of stellar dust formation and of stellar mass loss in the wind. 
However, the formation of  alumina dust grains is not well understood.}
{To shed light on the initial steps of cosmic dust formation (i.e. nucleation) in oxygen-rich environments by a quantum-chemical bottom-up approach.}
{Starting with an elemental gas-phase composition, we construct a detailed chemical-kinetic network describing the formation and destruction 
of aluminium-bearing molecules and dust-forming (Al$_{2}$O$_{3}$)$_n$ clusters up to the size of dimers 
($n$=2) coagulating to tetramers ($n=$4).   
Intermediary species include the prevalent gas-phase molecules AlO and AlOH,
and  Al$_x$O$_y$ clusters with $x=$1$-$5, $y=$1$-$6. 
The resulting extensive network is applied to two model stars, representing a semi-regular variable and a Mira-type, 
and to different circumstellar gas trajectories including a non-pulsating outflow and a pulsating model.
The growth of larger-sized (Al$_{2}$O$_{3}$)$_n$ clusters with $n=$4$-$10 is described by the temperature-dependent Gibbs free energies of the most favourable structures (i.e. the global minima clusters) as derived from global optimisation techniques and calculated by density functional theory.
We provide energies, bond characteristics, electrostatic properties and vibrational spectra of the clusters as a function of size $n$ 
and compare these to corundum corresponding to the crystalline bulk limit ($n\rightarrow \infty$).
} 
{The circumstellar aluminium gas-phase chemistry in oxygen-rich giants is primarily controlled by AlOH and AlO, which are tightly coupled by the reactions AlO+H$_2$, AlO+H$_2$O, and their reverse. Models of semi-regular variables show comparatively higher AlO abundances, and a later onset and a lower efficiency of alumina cluster formation when compared to Mira-like models.
The Mira-like models exhibit an efficient cluster production accounting for more than 90\% of the available aluminium content, which is in agreement with the most recent ALMA observations. Chemical equilibrium calculations fail to predict
the alumina cluster formation, as well as the abundance trends of AlO and AlOH in the AGB dust formation zone.  
Furthermore, we report the discovery of hitherto unreported global minima candidates and low-energy isomers for cluster sizes $n=$7, 9, and 10. 
A homogeneous nucleation scenario, where Al$_2$O$_3$ monomers are successively added, is energetically viable. 
However, the formation of the Al$_2$O$_3$ monomer itself represents an energetic bottleneck. 
Therefore, we provide a bottom-up interpolation of the cluster characteristics towards the bulk limit by excluding the monomer, approximately following a $n^{-1/3}$ dependence.} 
%
{}{}
\keywords{Physical data and processes: Astrochemistry, Molecular data, Molecular processes -- Solid state: refractory -- Stars: abundances, AGB and post-AGB, atmospheres, mass-loss, winds, outflows -- dust }

\maketitle 

\section{Introduction}
Asymptotic Giant Branch (AGB) stars are a major contributor to the global dust budget in 
galaxies \citep{2018A&ARv..26....1H}. 
Owing to their refractory nature, alumina (stoichiometric formula Al$_2$O$_3$) 
is a promising candidate to represent the first dust condensate in oxygen-rich AGB stars. 
Related alumina clusters are thought to initiate dust formation in these environments and 
are often referred to as \textit{seed particles} \citep{2013pccd.book.....G}. 
However, the sizes and compositions of these aluminium oxide clusters are not well characterised.
In this study, we investigate a range of Al:O stoichiometries in order to review these predictions
and to construct realistic models of these initial dust seeds.
The emergence of a specific condensate is predicted by its condensation temperature 
\citep{2005pcim.book.....T} and depends on the thermal stability of the solid, as well as the gas 
density and its composition. 
Usually, the evaluation of the stability of the likely condensates
is based on macroscopic bulk 
properties such as the vapour pressure, which is a measure of the volatility of 
a substance. Hence, the most refractory condensate is expected to have the lowest vapour 
pressure. Corundum ($\alpha$-alumina), corresponding to the most stable crystalline bulk form of alumina,  
fulfills this condition \citep{GailWetzel13}.    
The growth and size distribution of dust grains is commonly described by Classical 
Nucleation Theory (CNT). However, the applicability of CNT in an expanding circumstellar envelope has been 
questioned \citep{1985ApJ...288..187D,doi:10.1111/j.1365-2966.2011.20255.x,C6CP03629E,
2017ApJ...840..117G}.
In particular, the concept of vapor pressures and the universal assumption of thermodynamic equilibrium 
are in contradiction with the synthesis and growth of dust grains in highly dynamical AGB atmospheres.  
Moreover, in CNT, the properties of small solids are derived from the (crystalline) bulk material.
However, the properties of nano-sized clusters often differ significantly from bulk 
analogues. 
The constraints associated with extreme small size leads to clusters with 
non-crystalline structures, whose 
characteristics (e.g. energy, geometry, bond lengths and angles, atomic coordination) differ 
substantially from the bulk material \citep{BromleyZwijnen}. 
In particular, the energetic stability of such nano-clusters are typically higher than that of clusters with structures directly obtained from "top down" cuts from the parent bulk crystalline material, which represent 
meta-stable, or even unstable, configurations \citep{C6NR05788H}. 
In addition, the concept of surface free energy (or tension), which is fundamental in CNT, is not applicable to small clusters, where it is difficult to differentiate between surface and bulk. 
Surface energies can only be applied to clusters with fairly large sizes (e.g. facetted bulk cut clusters).
We understand nucleation as the formation and growth of stable seed nuclei (i.e., clusters) from prevalent gas-phase molecules, whose abundance varies in time (i.e. is often not in equilibrium). 
Therefore, a cluster is intermediate in size between a molecule and a bulk solid.\\
Pure oxygen in the gas phase has no other stable forms than atomic O, molecular O$_2$ and 
O$_3$. Though solid O$_2$ ice exists in the interstellar medium, its condensation temperature
is far too low to instigate circumstellar dust nucleation.  
A homo-atomic monomeric nucleation, as occurs in the case of carbon (see e.g. \citep{1984A&A...133..320G}), is thus not 
applicable. Therefore, the nucleation likely proceeds via several chemical elements (i.e. via a hetero-atomic scenario). 
Inorganic metal oxides are promising nucleation candidates, as they are particularly 
thermally and structurally stable.
In fact, the major part of oxygen-rich stardust is in the form of 
silicates \citep{2010LNP...815.....H}, which are composed of oxygen, silicon and at least 
one other metal (usually Mg or Fe). However, they do not represent the first dust species 
emerging in the atmospheres of oxygen-rich AGB stars.
The thermal stability of solid 
enstatite (MgSiO$_3$) and forsterite (Mg$_2$SiO$_4$), corresponding to Mg-rich members 
of the pyroxene and olivine silicate family, is lower than that of alumina, 
gehlenite (Ca$_2$Al[AlSiO$_7$]) and spinel (MgAl$_2$O$_4$) \citep{doi:10.1021/je300199a,GailWetzel13}.
Despite their refractory nature, the latter dust species are limited by the availability of 
the elements Ca and Al, both being approximately one order of magnitude less abundant than Si.
Therefore, refractory Al-bearing condensates could represent seed nuclei in oxygen-rich circumstellar 
envelopes. 
Less refractory, but more abundant materials like Mg- and Fe-silicates can condense on the seeds 
at later stages of the wind acceleration, further radially outwards in the circumstellar envelope. 
Iron-rich silicates are unlikely to be condensation seeds, as their large opacity 
to stellar radiation would lead 
to the subsequent heating up and evaporation of the dust grains \citep{Woitke06}. 
Consequently, the inclusion of iron in silicates allows them to act as a thermostat and tends to occur at later stages of the wind acceleration, further radially outwards in the stellar wind of AGB stars.
Furthermore, nano-sized (Mg-rich) silicates are thought to become important in the interstellar medium \citep{doi:10.1021/acsearthspacechem.9b00139}.\\                         
Some oxygen-rich AGB stars show a spectral emission feature around 13 $\mu$m \citep{1990AJ.....99.1173L},  
which is commonly attributed to Al-O vibrational stretching and bending modes \citep{1997ApJ...476..199B}. 
The carrier of this dust feature has been hypothetised to be spinel 
\citep{1999A&A...352..609P,2001A&A...373.1125F} or alumina \citep{2003ApJ...594..483S}. 
The strength of the 13 $\mu$m feature correlates with some CO$_{2}$ emission lines in 
the range of 3.3$-$16.3 $\mu$m \citep{1998A&A...330L..17J}. 
Moreover, in many stars the 13 $\mu$m feature is accompanied by emission features around 11, 20, 28, and 32 $\mu$m, 
respectively \citep{2003ApJ...594..483S}. As potential carriers for these additional emissions,   
different polymorphs (i.e. crystal structures) of alumina have been suggested \citep{2019AAS...23341104S}.
A recently conducted microgravity experiment in a sounding-rocket has shown that solid Al$_2$O$_3$ 
exhibits broad emission in the 11$-$12 $\mu$m wavelength range \citep{2018NatCo...9.3820I}.
Other dust features often seen in oxygen-rich AGB stars are located at around 10 $\mu$m and 18 
$\mu$m and they are attributed to Si$-$O streching and Si$-$O$-$Si bending modes, respectively 
\citep{1970Natur.227..822H}.
Observational studies have shown that the features of silicate and alumina can appear together, 
but also seperately \citep{Karov2013,refId0D,Takigawaeaao2149}. 
Stellar sources, showing the 13 $\mu$m feature only, include S Ori and RCnC, which exhibit low mass-loss rates of the order of 10$^{-8}$ M$_{\odot}$/yr to 10$^{-7}$ M$_{\odot}$/yr.  
There are a number of M-type AGB stars that show both, the Si$-$O and the Al$-$O vibration
modes, including GX Mon, W Hya and R Dor.
Stars showing only the silicate feature might also bear alumina, but the emission at 
13 $\mu$m could be blended with a mantle of silicate material on the grains.
This could, for example, be the case in the high mass-loss rate AGB star IK Tau.

The dust shell, that is associated with the 13 $\mu$m feature, is loacted at 1.4$-$3 stellar radii (R$_\star$), whereas the 
silicate shell associated with the 10 $\mu$m and 18 $\mu$m features is located further 
out at distances of about 5 R$_\star$ \citep{Karov2013,Ohnaka16,Takigawaeaao2149}. These findings observationally 
confirm the higher thermal stability of alumina in comparison with Mg-rich silicates. 
\cite{1997ApJ...476..199B} derived infrared optical constants for amorphous types of 
alumina from laboratory experiments and found an emission peaking at 11.5$-$11.8 $\mu$m.
Albeit the amorphous laboratory-synthesized grains reflect the non-crystalline character of the 
clusters investigated in this study, we note a substantial difference in size. 
The clusters we consideqr are 
(sub)nanometer sized, whereas amorphous alumina particles produced by the sol-gel technique are 
micron-sized (factor of 1000 larger).   
\cite{demyk04} investigated the vibrational properties of (Al$_2$O$_3$)$_n$ by cluster beam experiments and found that the band positions depend on the cluster size $n$. Small clusters ($n\le$8) exhibit vibrational bands around 11 $\mu$m and the larger sized clusters around 15 $\mu$m, pointing towards similarities with the spectra of crystalline $\gamma$-alumina, but not with $\alpha$-alumina.
\\
The evolutionary progression of a star in the AGB phase is reflected by its increasing 
mass-loss rate that will eventually end in a relatively short superwind phase with a high 
mass-loss rate \citep{doi:10.1111/j.1745-3933.2008.00535.x}.
The AGB mass-loss rate is also correlated with the regularity and in particular 
the period of the stellar pulsations, showing smaller rates for semi-regular AGB stars and 
larger rates for Mira-type stars \citep{McDonald_2016}.
A strikingly large fraction of the stars showing the 13 $\mu$m emission feature are semi-regularly variable
AGB stars, suggesting that these stars have not yet reached the tip of the AGB and therefore evolutionarily precede
 Mira-type AGB stars exhibiting regular long-period pulsations.
\citep{1996ApJ...463..310S}.\\ 
The metal Aluminium (Al) is the 11th most abundant element in the solar system and has 
an abundance of $\sim$ 3 $\times$ 10$^{-6}$ with respect to the total gas
\citep{2009ARA&A..47..481A}. Hence, the overall amount of aluminium-bearing molecules, 
alumina clusters and dust is limited by the availability of aluminium. 
In the past decades, several Al-bearing molecules have been found in circumstellar 
environments including AlF \citep{1994ApJ...433..729Z}, AlCl 
\citep{1987A&A...183L..10C}, AlO \citep{2009ApJ...694L..59T} and AlOH \citep{2010ApJ...712L..93T}.
AlF has the highest bond energy (681 kJ mol$^{-1}$) of all Al-containing diatomic 
molecules, followed by AlCl (515 kJ mol$^{-1}$) and AlO (499 kJ mol$^{-1}$). 
These bond energies have been computed in the present study and are in agreement with
the compilation of bond energies provided by \citet{Luo2007,2013pccd.book.....G}. 
AlCl (and tentatively AlF) were detected in the envelope of the carbon-rich AGB star 
IRC 102+16 \citep{1987A&A...183L..10C}. AlCl is also found in two oxygen-rich 
AGB stars with different mass-loss rates \citep{refId0D}. The aluminium-bearing molecules 
AlO and AlOH were first detected in the envelopes of the red supergiant VY Canis 
Majoris \cite{2009ApJ...694L..59T}.
In subsequent studies, AlO (and AlOH) were identified by their rotational transitions in 
the circumstellar envelopes of several of low-mass oxygen-rich AGB stars with different mass-loss rates 
\citep{2016AA...592A..42K,2017AA...598A..53D,refId0D}.
Moreover, a visible AlO transition from an electronically excited state has been observed in absorption 
and emission in the spectra of the prototypical star Mira (o Ceti) \citep{2016AA...592A..42K}. 
Related transition dipole moments and radiative lifetimes of the excited states have been the subject of recent experimental 
studies, see e.g. \citet{LAUNILA201110,Bai_2020}. 
The photon absorption cross section at 4823 \AA, corresponding to a strong electronic transition (B$^{2}\Sigma^{+}\rightarrow$X$^{2}\Sigma^{+}$) in AlO, was recently
experimentally determined by \cite{GOMEZMARTIN201756}.\\

In a previous study, four different species (TiO$_2$, MgO, SiO and Al$_2$O$_3$) and their role as nucleation candidates in oxygen-rich 
circumstellar envelopes were examined \citep{2019MNRAS.tmp.2040B}. The authors assumed a quasi-stationary circumstellar envelope and 
applied a chemical-kinetic network to a model grid with constant pressures and temperatures. 
Furthermore, homogeneous and homo-molecular cluster growth was adopted for each of the nucleation candidates.
\citet{2016A&A...585A...6G} studied the kinetic nucleation and subsequent coagulation of two dust components, alumina and forsterite. 
The authors assumed a two-step process, where alumina nucleates first homogeneously, and subsequently, forsterite condenses heterogenously on the surface of the alumina seeds.
In this study, we focus on Al$_2$O$_3$ as a nucleation candidate (Al$_2$O$_3$) by developping an extensive chemical-kinetic network that makes use of benchmark quantum calculations at a high level of theory (CBS-QB3, \citet{doi:10.1063/1.481224}), 
includes the reaction rate estimates derived from recent experiments and statistical rate theory,
and goes beyond the formation of the smallest formula units (monomers). 
We then apply the chemical network to two model stars, a semi-regular variable AGB star and a Mira-type AGB star. 
The circumstellar gas trajectories include a non-pulsating outflow model, described by a $\beta$-velocity law, and a pulsating model, 
described by Lagrangian flows corresponding to the pulsationally-induced excursions of a circumstellar post-shock gas.\\
This paper is organized as follows. In Section \ref{meth}, we describe the methods used to derive 
the lowest-energy candidate structures and their refinement with quantum-chemical density 
functional theory (DFT) methods. In Section \ref{results}, we present the results of the cluster 
calculations including energy, structure, and vibrational spectra as well as the kinetic networks
applied to circumstellar gas trajectories. 
We summarise our findings in Section \ref{concl}.

\section{Methods}
\label{meth}
\subsection{Global optimisation searches}
\label{22}
The computational cost for a geometry optimisation of an aluminium oxide cluster increases with its size 
(or with the number of atoms or electrons), typically following a power-law. 
However, our ability to explore the potential energy landscapes is more limited 
by the number of possible isomers, which exponentially increases with cluster size \citep{PhysRevA.28.2408,Arslan_2005}.
To reduce the number of possible structural configurations 
to explore (and hence also the computational effort), global optimisation searches for 
low-energy aluminium-oxide clusters are performed.
We employ the Monte-Carlo Basin-Hopping (MC-BH) global optimisation technique
\citep{1998cond.mat..3344W} with interionic pair potentials of an Al-O system to find 
candidate low-energy clusters. For our purposes, we used an in-house, modified version of the GMIN 
programme \citep{PhysRevLett.95.185505}.\
The general form of the interionic Buckingham pair potential
(including the Coulomb potential) is

\begin{equation}
\centering
U(r_{ij}) =  \frac{q_{i}q_{j}}{r_{ij}} + A\exp\left(-\frac{r_{ij}}{B}\right) - \frac{C}{r_{ij}^6} ,
\label{buck}
\end{equation}

\noindent
where $r_{ij}$ is the relative distance between two atoms, $q_i$ and $q_j$ the charges of atom 
$i$ 
and $j$, respectively, and $A$, $B$ and $C$ the Buckingham parameters.
The first term in Eq. \ref{buck} describes the interionic electrostatic interactions,
the second term the 
short-range, steric repulsion term due to the Pauli exclusion principle, and the last term 
describes the attractive van der Waals interaction.\
The potential describes the repulsion and attraction of charged particles, in this case, 
of aluminium and oxygen ions within an Al$-$O containing cluster. 
To reduce the probability of missing stable configurations in our searches, 
we used a large number of structurally diverse initial geometries.
Moreover, we performed test calculations by swapping the Al and O atoms in the most stable 
configurations accounting for atomic segregation (i.e. covalent bonds between 
identical atoms). 
\begin{table}
	\caption{The parameter ranges used in this study to compute the interionic Buckingham 
	pair potential are listed as (1) charges of aluminium, q(Al), (2) charges of oxygen q(O), given in atomic units, (3) A in eV, (4) B in $\rm{\AA}$, (5) C in eV $\rm{\AA}^{-6}$\label{tab1}.}	
\begin{tabular}{ l l | l l l }
	q(Al) & q(O) & A(Al-O) & B(Al-O) & C(Al-O) \\
	      &      & A(O-O)  & B(O-O)  & C(O-O) \\
	\hline
	\rule{0pt}{4ex}    
	+3 & -2 &  4534.2  & 0.2649 &  0.0  \\
	   &    &  25.410  & 0.6937 & 32.32 \\
\end{tabular}
\end{table}
We applied the parameter set listed in Table \ref{tab1}, commonly used for structure optimisation of Al$-$O systems \citep{A901227C}. 
The searches cover diverse structural families including compact geometries, void cages and open-cage-like clusters by choosing various seed structures (i.e. initial geometries).  
In systems like aluminium oxides, electronic polarisation should also play a role. 
Therefore, we re-optimised our candidate isomers using a potential that describes polarisation via a 
core-shell model 
according to the parameters of \citet{JM9940400831} using the General Utility Lattice Program (GULP) developed by \citet{A606455H}. We did not find any additional structural isomers, but achieved a more realistic energetic ordering of different cluster isomers with respect to 
more accurate quantum calculations (see next subsection).
Although the use of interionic potentials is an approximation, it enabled us to 
perform tractable yet thorough searches. 
With our approach we aimed to minimise the probability of missing stable alumina cluster configurations.
For a comparison of the performance of different alumina interionic potentials (i.e. force fields), we refer to \citet{PhysRevB.101.045427}.\

\subsection{Quantum chemical calculations}
For the smallest molecular systems (i.e. up to 12 atoms, slightly more than the size of an alumina dimer), 
we performed quantum-chemical compound method calculations. Compound 
methods 
combine a high level of theory and a small basis set with methods that employ lower 
levels 
of theory with larger basis sets. We use the benchmark Complete Basis Set (CBS-QB3) method that 
extrapolates several single-point energies to a more accurate CBS-QB3 energy in the basis set limit
\citep{doi:10.1063/1.481224}.
We note that the CBS-QB3 method is prohibitive for large clusters ($>$ 15 atoms).
CBS-QB3 calculations were performed for the following species: Al, O, H, Cl, F, H$_2$, H$_2$O, OH, 
AlO, AlOH, SiO, AlCl, AlF, Al$_2$O, AlO$_2$, OAlOH, Al(OH)$_2$, Al$_2$O$_2$, Al$_2$O$_3$, Al$_2$O$_4$, 
Al$_3$O$_2$, Al$_3$O$_3$, Al$_3$O$_4$, Al$_3$O$_5$, Al$_4$O$_3$, Al$_4$O$_4$,  Al$_4$O$_5$, 
Al$_4$O$_6$, Al$_4$O$_7$, Al$_5$O$_4$, Al$_5$O$_5$, Al$_5$O$_6$, Al$_5$O$_7$, AlSiO$_3$, and HAlSiO$_3$.  
Once multiple sets of candidate structures with different initial geometries, temperatures 
and parameters were found, we refined the $\sim$ 50-100 most favourable candidate 
structures for each size in subsequent optimisations at a DFT level of theory using
two different hybrid density functionals, B3LYP \citep{1993JChPh..98.1372B} and PBE0 
\citep{1996JChPh.105.9982P}, in combination with the 6-311+G(d) basis set.
We performed 
these calculations using the computational chemistry software package 
Gaussian09 \citep{g09}. We used f-type orbitals (7F) for the basis functions and an 
ultrafine grid corresponding to the standard input. 
The DFT calculations were performed at 0 K and a pressure of 0 atm. In the 
Born-Oppenheimer 
approximation used here, the Potential Energy Surface (PES) does not depend on 
temperature.
Hence, the optimised cluster geometry is also temperature-independent. However, the 
vibrational population and the computation of the thermodynamic quantities (i.e. enthalpy, 
entropy, and Gibbs free energy) depend on temperature. 
Moreover, the entropy and the Gibbs free energy are also pressure-dependent.
We include a vibrational analysis to calculate the vibrational zero-point energy as 
well as appropriate partition functions for any 
other conditions. The partition functions are composed of electronic, translational, 
rotational and vibrational contributions and are used to compute the enthalpy, the entropy 
and eventually the Gibbs free energy.
Moreover, a vibrational analysis helps to identify and exclude possible transition 
states characterised by an imaginary frequency.
The predicted vibrational spectra of the clusters can then be compared with 
astronomical observations and laboratory experiments.\\

\subsection{Transition State Theory and RRKM}
\label{rrkm}
Rate coefficients for reactions with intermediate local minima on their potential energy surfaces 
were calculated with Rice-Ramsperger-Kassel-Markus (RRKM) theory, using the Master Equation Solver 
for Multi-Energy well Reactions (MESMER) program \citep{doi:10.1021/jp3051033}. 
The geometries of the Al-containing molecules (reactants, products and intermediates) were first optimized at the B3LYP/6-311+g(2d,p) level of theory within the Gaussian 16 suite of programs \citep{g09}, 
and the resulting rotational constants and vibrational frequencies used for the MESMER calculations. 
The CBS-QB3 method was used to obtain more accurate relative energies of these stationary points. 
Each intermediate species formed during the reaction was assumed to dissociate back to the reactants or forward to the products, or be stabilised by collision with H$_{2}$ as a third body (for the astrochemical environment). 
The internal energy of each intermediate was divided into a contiguous set of bins (typical width of 110 cm$^{-1}$) containing a bundle of rovibrational states. The density of these bundels was calculated with the theoretical vibrational frequencies and rotational constants, without making a correction for anharmonicity and using a classical density of states treatment for the rotational modes. Each bin was then assigned a set of microcanonical rate coefficients for dissociation to reactants or products (as appropriate). These rate coefficients were determined using an inverse Laplace transformation to link them directly to the capture rate coefficient $-$calculated using long-range transition state theory \citep{doi:10.1063/1.1899603}. The probability of collisional transfer between bins was estimated using the exponential-down model, where the average energy for downward transitions is designated $<\Delta E>_{down}$, and the probabilities for upward transitions were determined by detailed balance $<\Delta E>_{down}$ down was treated as temperature-independent, 
with a value of 200 cm$^{-1}$ for H$_2$ \citep{Gilbert1990TheoryOU}. 
The Master Equation, which describes the evolution with time of 
the adduct bin populations, was then expressed in matrix form and solved to yield the 
rate coefficient for bimolecular reactions and recombination at a specified pressure and temperature.

\section{Results}
\label{results}
\subsection{Precursors of alumina dust}
\label{precursors}
The molecular gas-phase precursors of stoichiometric (Al$_2$O$_3$)$_n$ clusters
are likely AlO and AlOH, which are the most abundant and prevalent aluminium-bearing molecules in an oxygen-rich circumstellar gas. 
Furthermore, we include other related aluminum-oxygen-hydrogen species containing 
Al$_2$O, AlO$_2$, Al$_2$O$_2$, OAlOH, and Al(OH)$_2$, as well as aluminum halides like AlF and AlCl.
As a first step, we asses the accuracy of the employed electronic structure methods (CBS-QB3, B3LYP, PBE0) and the basis sets 
(6-311+G(d) and cc-pVTZ). For this purpose, we compare the calculated enthalpies of formation at 0~K ($\Delta H_{f}^0$) with experimental data of the NIST-JANAF database\footnote{https://janaf.nist.gov/} (see Table \ref{Janaf}). Note, that in contrast to the binding energies, $\Delta H_{f}^0$(0) is scaled with respect to the atomic heats of formation. Overall, the CBS-QB3 method results in the most accurate energies for the considered aluminium-bearing molecules. \\

\subsubsection{Molecular dust precursors}
In the following, we use the term \textit{binding} energy corresponding to the CBS-QB3 energy with respect to the constituent atoms and which is normalised with respect to the number of atoms in the respective molecule or cluster. Note that the binding energies are not scaled to the atomic heats of formation, unlike the JANAF enthalpies of formation. \textit{Relative} energies correspond to the energy difference between two isomers of a given composition and size, typically between the lowest-energy structure and a higher-lying isomer. 

\subsubsection*{AlO}
\begin{figure}[H]
  \includegraphics[scale=0.19]{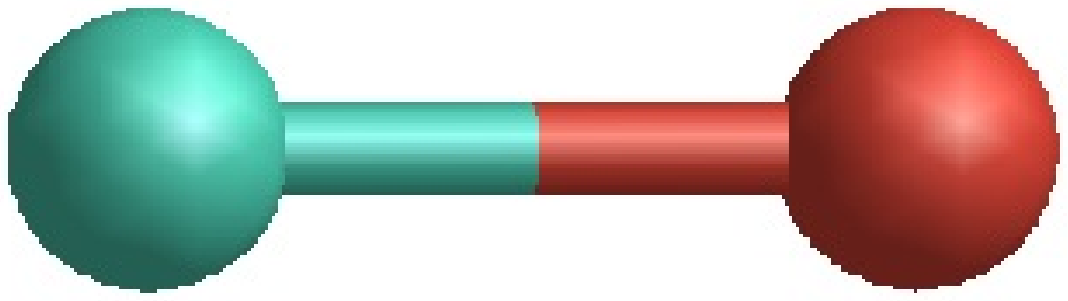}
\newline
   \includegraphics[scale=0.3]{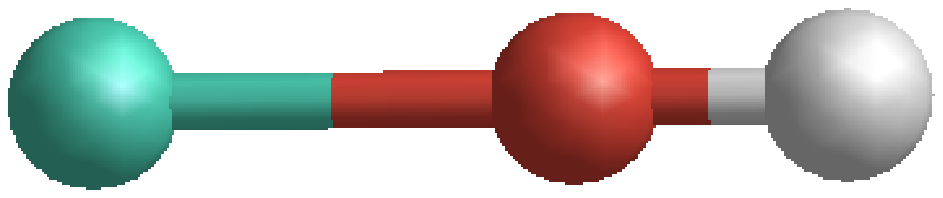}
\newline
   \includegraphics[scale=0.40]{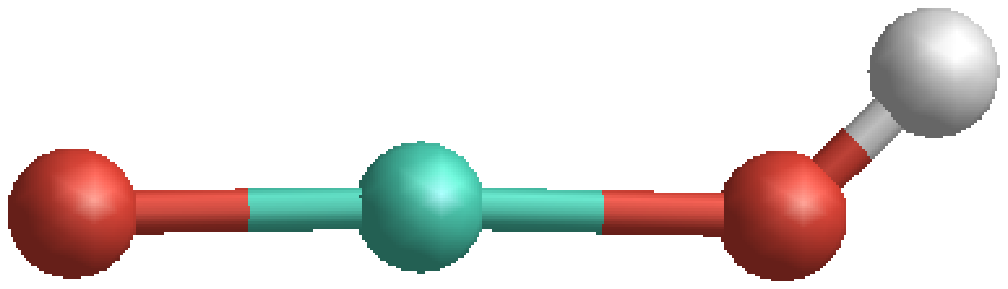}
\caption{Structures of the molecules AlO, AlOH and OAlOH. Al atoms are displayed in turquoise, O atoms in red, and H atoms in white.\label{molprec}} 
\end{figure}

The ground state of AlO (displayed in Figure \ref{molprec} has an unpaired electron and has thus a spin multiplicity of 2 (doublet state). Its equilibrium bond length is 1.630 \AA{}, which is in good agreement with the experimental
value of 1.618 \AA{} \citep{Huber1979}. The AlO bond dissociation energy is 498 kJ mol$^{-1}$ (i.e., a binding energy of 249 kJ mol$^{-1}$ per atom), which in very good agreement to the experimental dissociation energy of 501.9$\pm$10.6 \citep{Luo2007}. 
\subsubsection*{AlOH}
The singlet state AlOH molecule (also displayed in Figure \ref{molprec}) can be present in linear or in bent form. At 0 K the linear isomer is more stable than the bent form by only 2.8 kJ mol$^{-1}$.
Although recent findings of \citet{0004-637X-863-2-139} indicate a slightly bent form as the 
AlOH ground state, we find the linear form to be more stable for all temperatures 
considered in this study.\
The AlOH binding energy is 327 kJ mol$^{-1}$. 
The AlOH dissociation leads either to the products Al+OH with a dissociation energy of 554 kJ mol$^{-1}$, or to AlO+H with a dissociation energy of 482 kJ mol$^{-1}$. The two AlOH dissociation channels are highly endothermic and therefore, the molecule is relatively stable once it has formed.     

\subsubsection*{OAlOH}
The energetically most favourable geometry of OAlOH (once again illustrated in Figure \ref{molprec}) shows
the terminal H atom bent at an angle of $\sim$ 45$^{\circ}$ with respect to the O-Al-O inter-atomic 
axis. The binding energy of OAlOH is 352 kJ mol$^{-1}$.
OAlOH can be formed by the reaction AlO+H$_2$O. This reaction was studied in detail by \citet{Mangan2021} showing that OAlOH is a minor product of the reaction.

The stability and structure of ground state OAlOH has been investigated by 
\citet{Cobos2002}.
We consider three possible dissociation channels of OAlOH. 
With an energy barrier of 429 kJ mol$^{-1}$, 
O + AlOH is the least endothermic reaction, 
followed by AlO+OH with 485 kJ mol$^{-1}$ and 
OAlO+H with 520 kJ mol$^{-1}$.\\

\begin{table*}
\caption{Enthalpies of formation, $\Delta H_{f}^0$, at $T=$0 K in kJ mol$^{-1}$ of Al-bearing molecules for density functionals/basis sets used in this study, compared with the JANAF thermochemical database\label{Janaf}.}
\setcellgapes{2pt}\makegapedcells
\begin{tabular}{c | c c c c c c c c c c}
              & AlO      & AlO$_2$ & Al$_{2}$O & Al$_2$O$_2$ & AlOH & OAlOH  & AlH & AlCl & AlF & Al$_{2}$\\
\hline
PBE0/6-311+G  & +113.  &  +8.  & -86.  & -240. & -117. & -253. & +269. & -38. & -224. & +590.\\ 
B3LYP/6-311+G & +98.   & -21.  & -98.  & -249. & -134. & -277. & +253. & -31. & -238. & +599.\\
PBE0/cc-pVTZ  & +101.  & -20.  & -99.  & -281. & -143. & -292. & +266. & -44. & -237. & +586.\\
B3LYP/cc-pVTZ & +85.   & -50.  & -113. & -292. & -159. & -316. & +250. & -38. & -251. & +595.\\
CBS-QB3       & +76.   & -70.  & -116. & -395. & -190. & -372. & +242. & -68. & -277. & +597.\\
\hline
JANAF         & +67.04   & -85.01  & -144.48 & -391.31 & -175.63 & -454.51 & +259.51 & -51.66 & -265.62 & +486.28\\

\end{tabular}
\end{table*}

The enthalpies of formation (i.e. binding energies), derived from our DFT calculations, seem to be systematically lower than those derived empirically in JANAF (\ref{Janaf}). 
However, we note that the JANAF values are often based on incomplete or extrapolated experimental data 
dating back to 1970's. Therefore, the tabulated JANAF energies may not be accurate. 
Nevertheless, it is a standard reference database, and we include those values for comparison purposes.
We find better agreement with JANAF by using a larger numerical basis set (cc-pVTZ) compared with a Gaussian type basis set (6-311+G(d)). In addition, we find that the B3LYP functional to provide values in better agreement with the JANAF values than PBE0.
The enthalpies derived using the composite CBS-QB3 method shows the closest agreement with the JANAF values, particularly for the case of Al$_2$O$_2$.
However, CBS-QB3 tend to be prohibitive for large systems with more than $\sim$ 15 atoms. 
Therefore, we use the B3LYP/cc-pVTZ method as a compromise between accuracy and computational feasibility.

\subsubsection{Al$_x$O$_y$ (where $x=$2$-$5 and $y=$1$-$6) clusters}
\label{non-stoichm} 
The choice of the stoichiometric range ($x=$2$-$5, $y=$1$-$6) of Al$_x$O$_y$ clusters (displayed in Table \ref{ALXoYs}) is based on stoichiometric reasoning (i.e. realistic Al:O ratios that are in the vicinity of 2:3) and on the abundances of aluminium and oxygen (and hydrogen).
Primarily, the energetic and kinetic stability of the Al-O-H containing species determines their abundances. 
However, on shorter timescales, the species abundances can be set by the availabilty of their component 
(i.e., the elements). 
For example, despite its lower bond energy ($\sim$300 kJ/mol), AlH is temporarily more abundant than the strongly bound AlO molecule ($\sim$500 kJ/mol), due to the omnipresence of hydrogen.
Therefore, we consider Al$_x$O$_y$ clusters with no more than one Al atom in excess i.e., $x \le y+1$.\  
The molecule Al$_2$O is a linear molecule
with a CBS-QB3 binding energy of 355 kJ mol$^{-1}$, which is relatively large. 
As a symmetric linear molecule, Al$_2$O cannot be observed by pure rotational spectroscopy, owing to the lack of a permanent dipole moment. A non-zero quadrupole moment is present, but the related transitions are `forbidden'. 
Other Al$_2$O isomers with permanent dipole moments (including triplet states) lie $\sim$ 300 kJ mol$^{-1}$ above the electronic ground state and are thus unlikely to be observed by rotational lines \citep{2020ApJ...904..110D}. 
However, we note that asymmetric and bending stretches in Al$_2$O could be observed by (ro-)vibrational spectroscopy.  

\begin{table*}[!ht]
\caption{Global minimum (GM) candidate structures of Al$_x$O$_y$, $x=$1$-$5 $y=$1$-$6 clusters including binding energies per atom (in kJ mol$^{-1}$). The color-coding of the atoms follows that used in Figure \ref{molprec} and holds for all Figures.\label{ALXoYs}}
\setcellgapes{2pt}\makegapedcells
\begin{tabular}{p{1cm} | p{2.7cm} | p{2.7cm} | p{2.7cm} | p{2.7cm} | p{2.7cm}}
& AlO$_{y}$ ($x=$1) & Al$_{2}$O$_{y}$ ($x=$2) & Al$_3$O$_{y}$ ($x=$3) & Al$_4$O$_{y}$ ($x=$4) & Al$_5$O$_{y}$ ($x=$5)\\
\hline
Al$_x$O ($y=$1) & \includegraphics[width=0.10\textwidth]{AlO.eps} \newline 249. &
\includegraphics[width=0.15\textwidth]{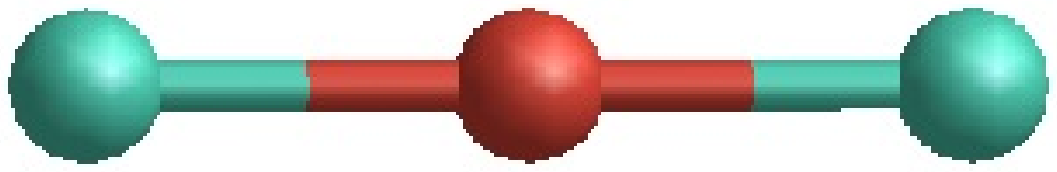} \newline 355. & & & \\
\hline
Al$_x$O$_2$ ($y=$2) &
 \includegraphics[width=0.15\textwidth]{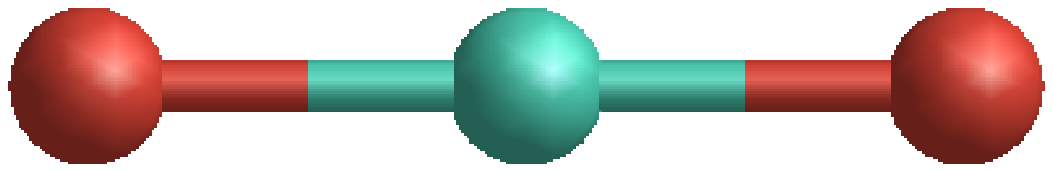} \newline 296. &
 \includegraphics[width=0.10\textwidth]{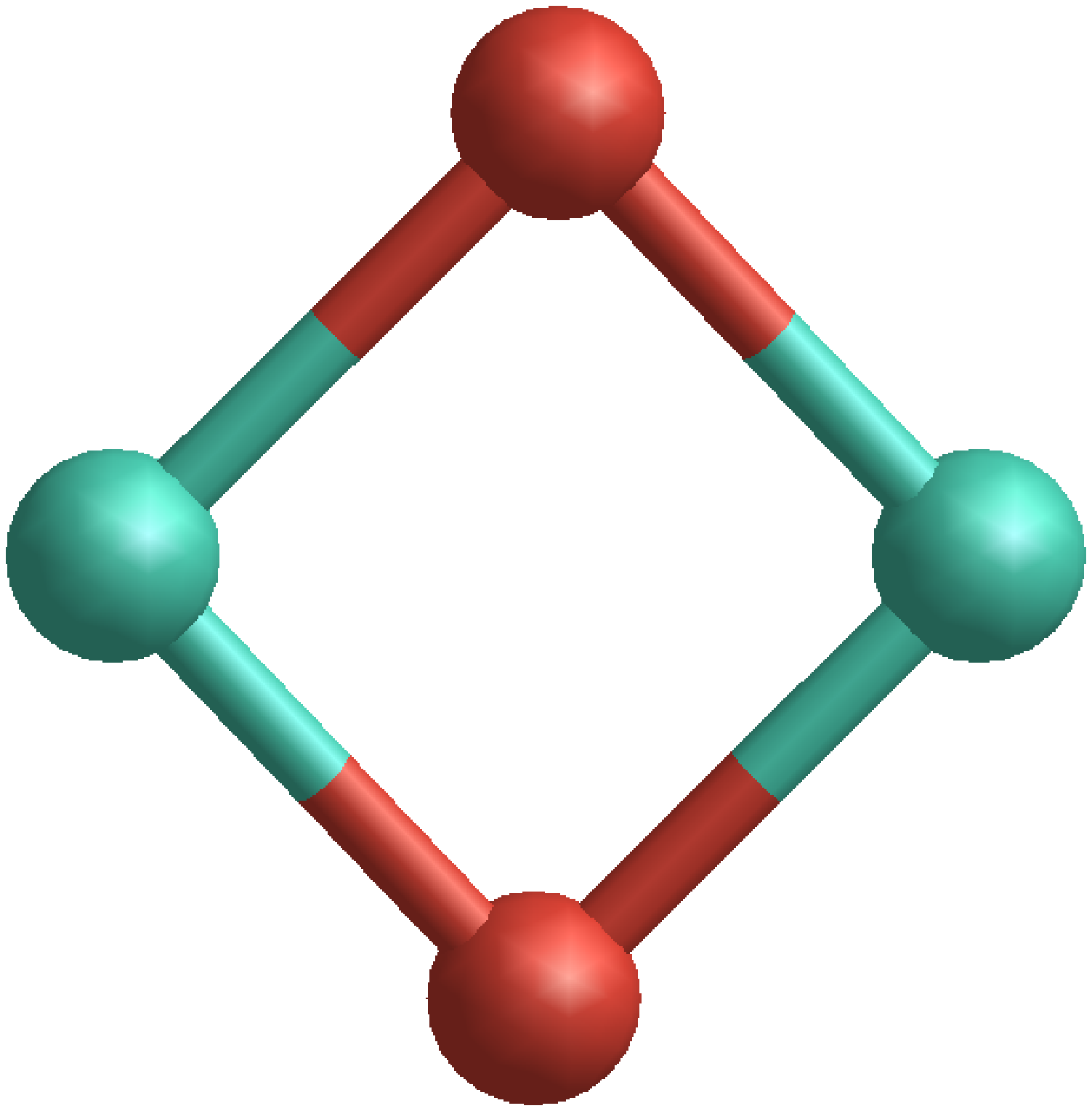} \newline 386.  &
 \includegraphics[width=0.15\textwidth]{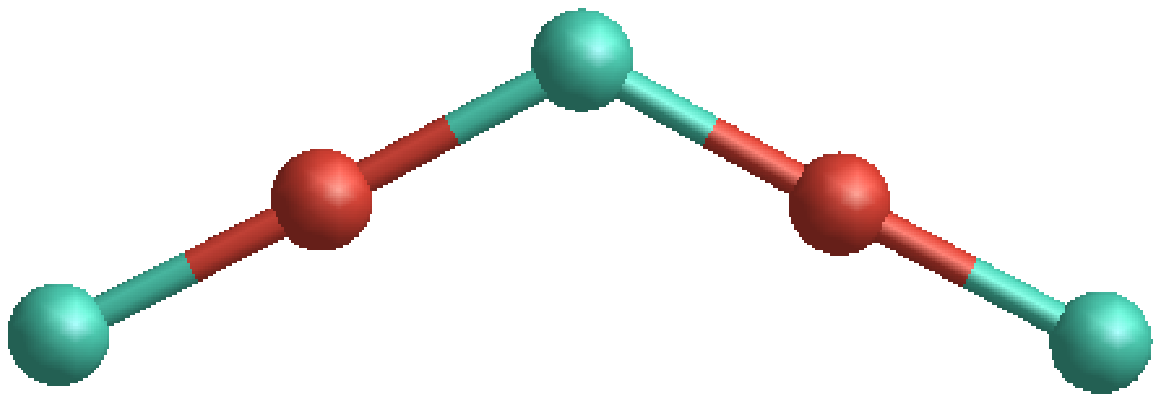} \newline 384. & 
& \\
\hline
Al$_x$O$_3$ ($y=$3) & 
 \includegraphics[width=0.15\textwidth]{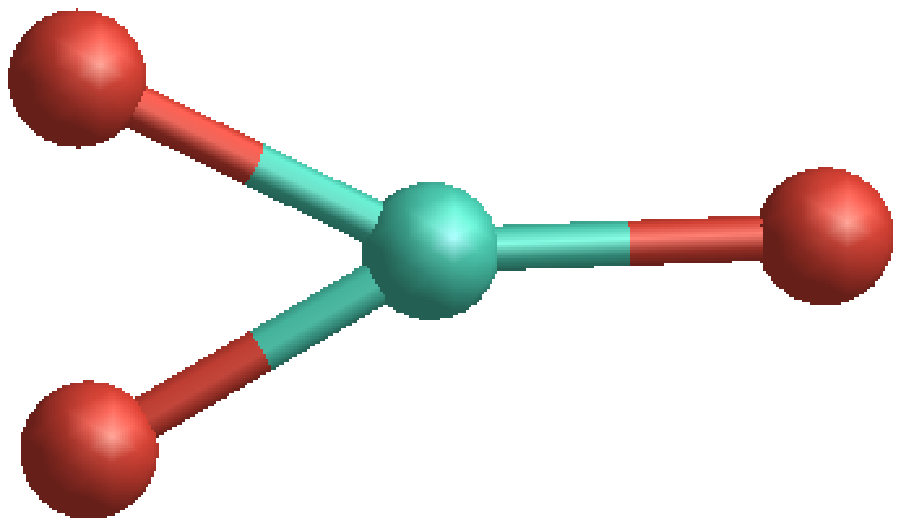}  294. &
 \includegraphics[width=0.15\textwidth]{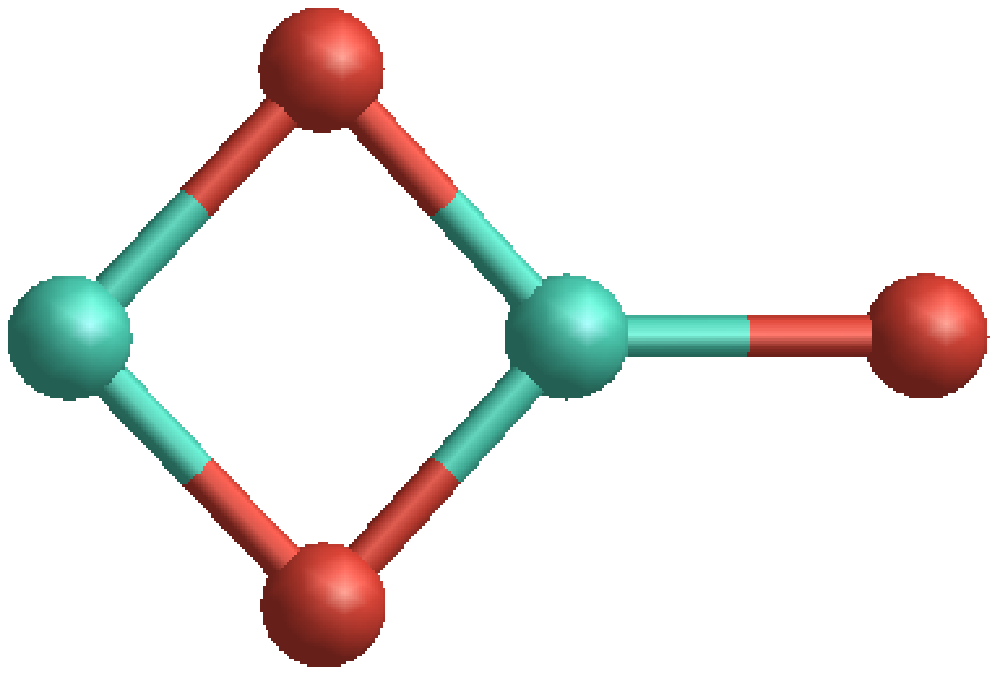} 388. &
 \includegraphics[width=0.15\textwidth]{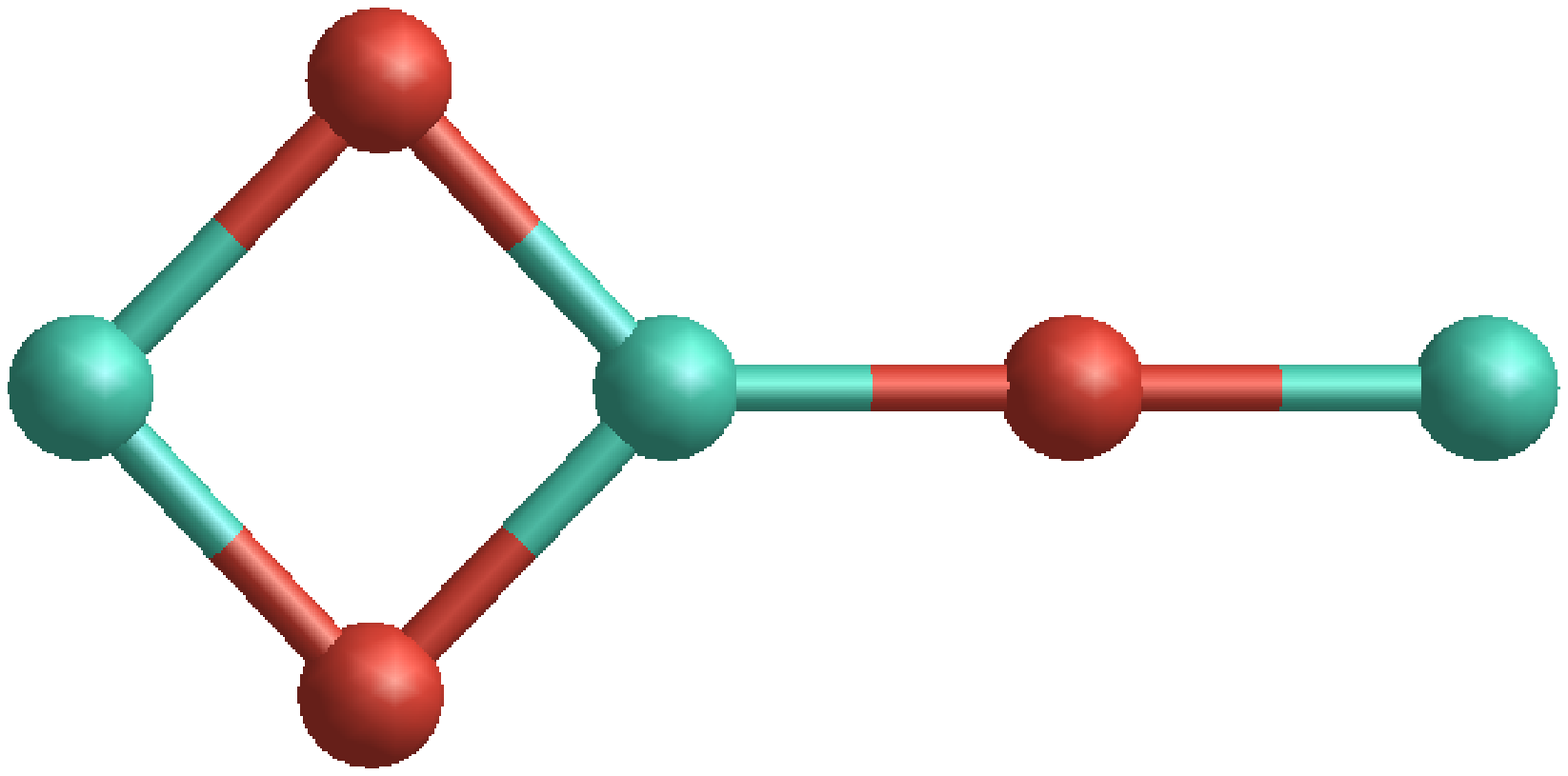} 383. &
 \includegraphics[width=0.13\textwidth]{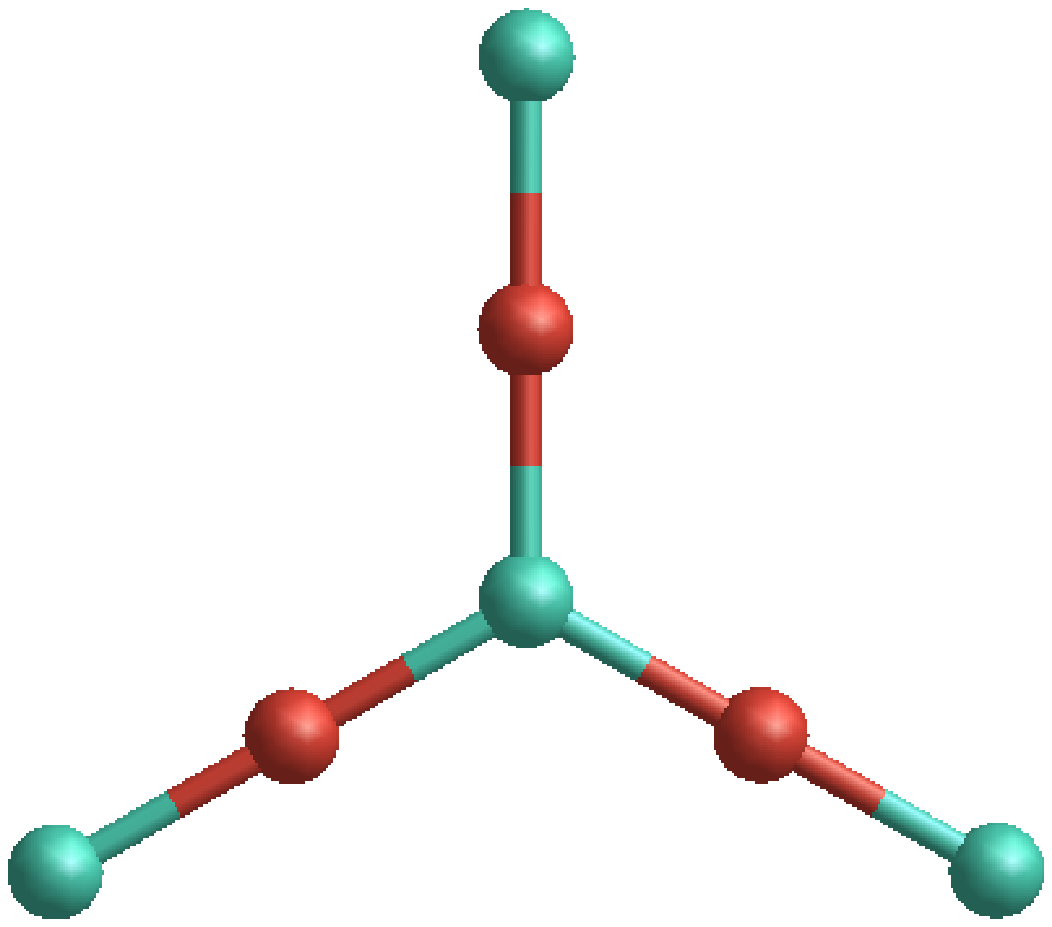} 425. & \\
\hline
Al$_x$O$_4$ ($y=$4) & &
 \includegraphics[width=0.15\textwidth]{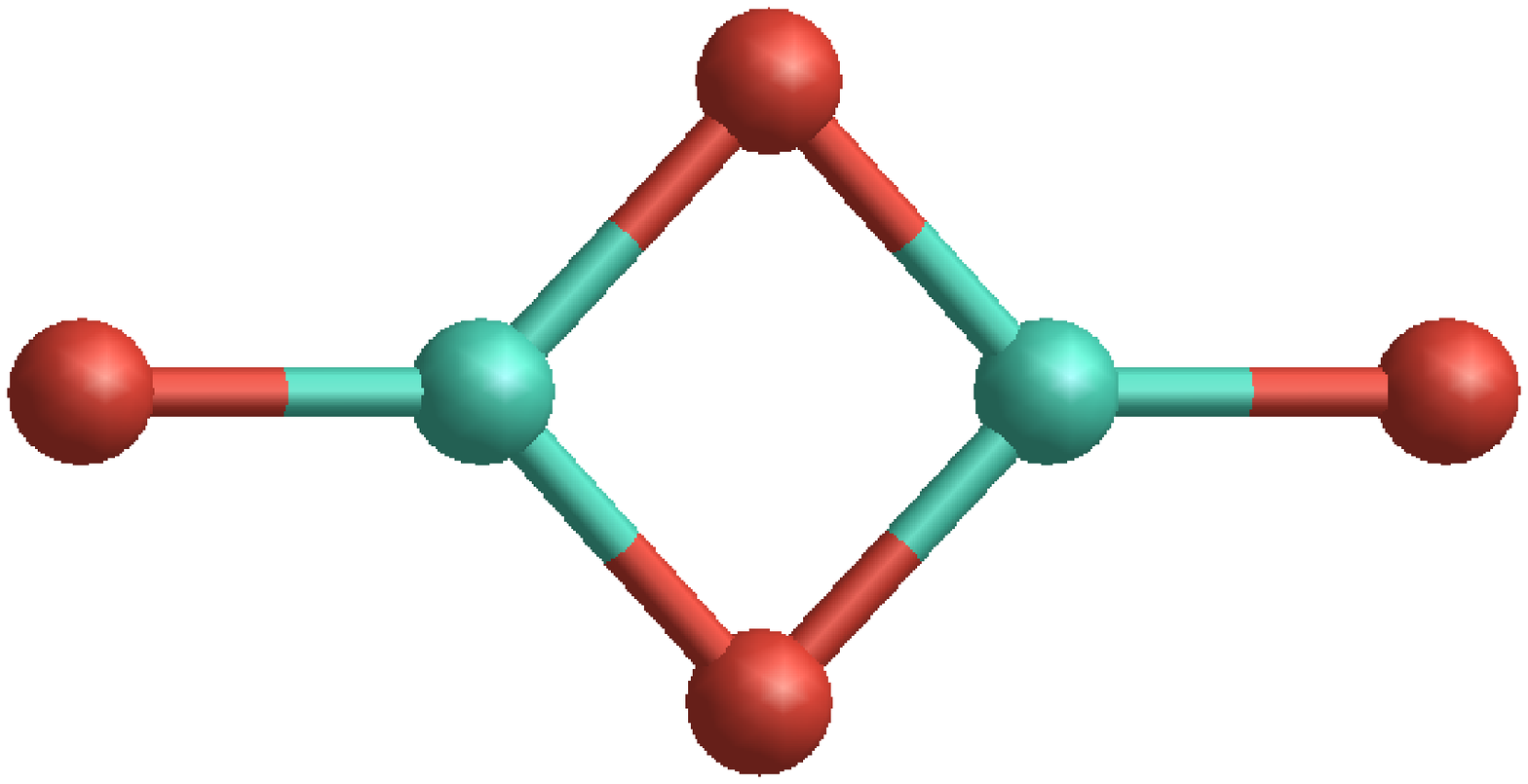} 414. &
 \includegraphics[width=0.15\textwidth]{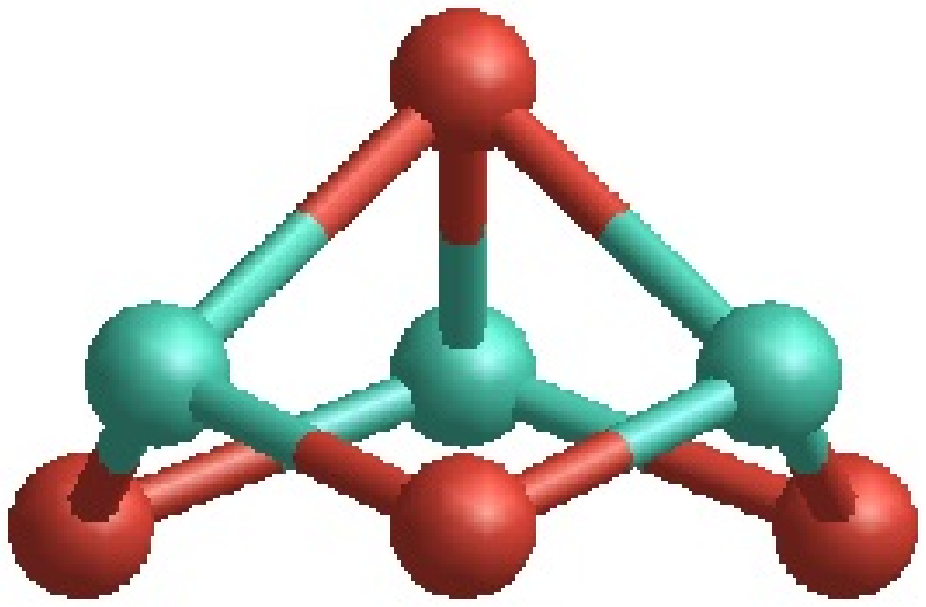} 437. &
 \includegraphics[width=0.15\textwidth]{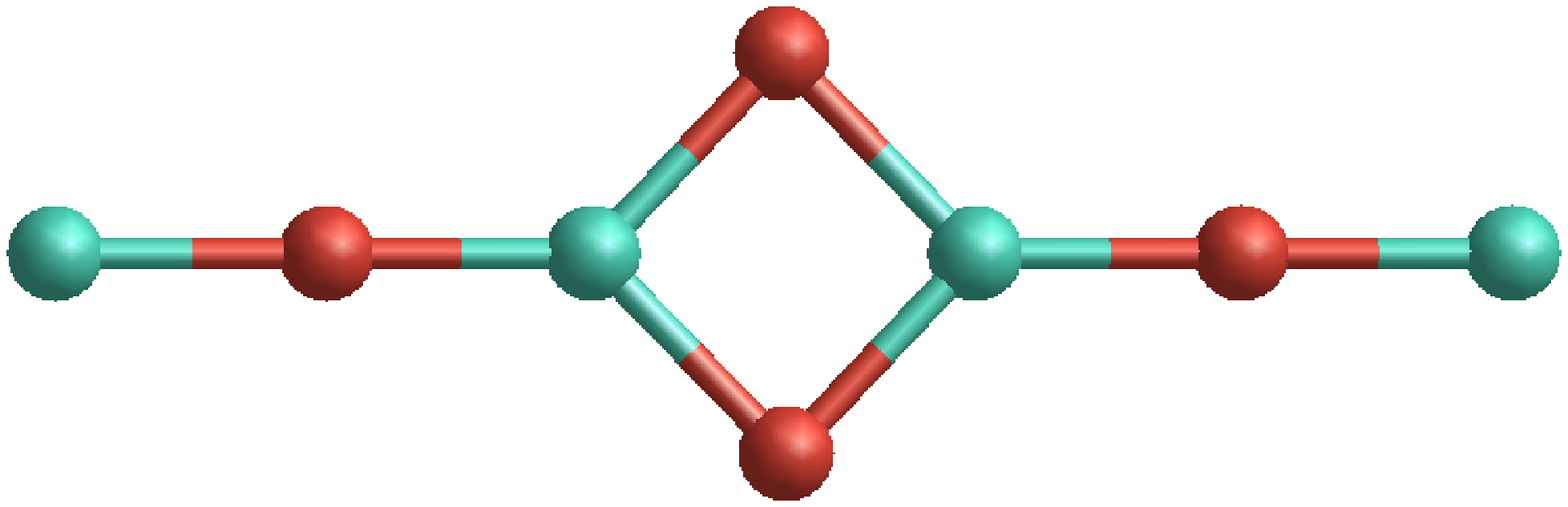} 452. & 
 \includegraphics[width=0.15\textwidth]{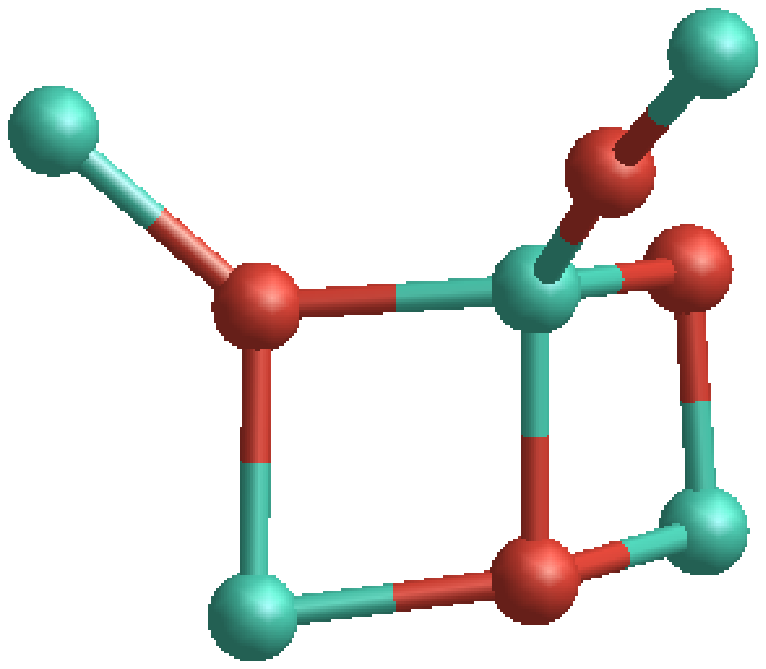} 426. \\
\hline
Al$_x$O$_5$ ($y=$5) & & &
 \includegraphics[width=0.15\textwidth]{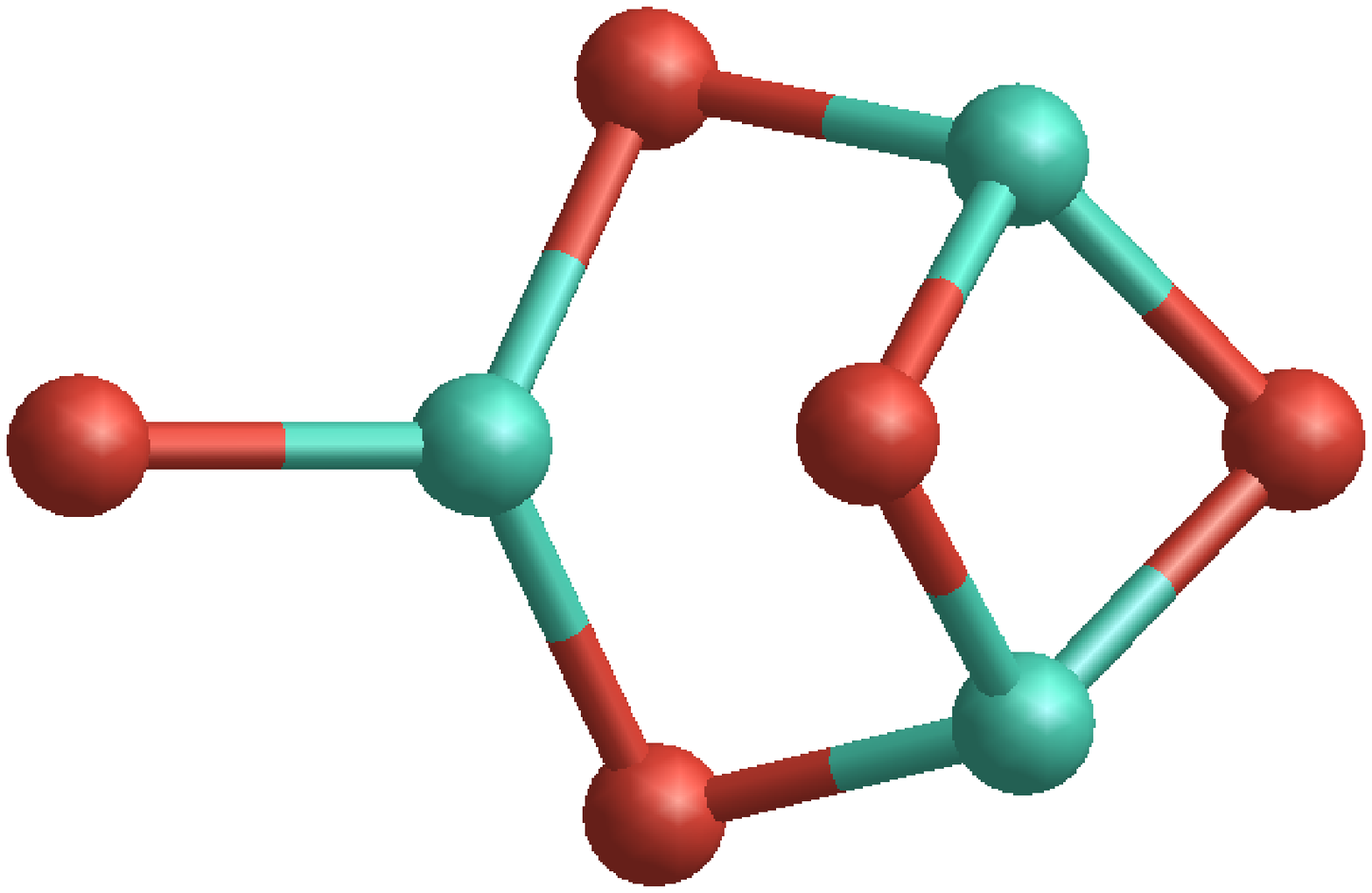} 437. &
 \includegraphics[width=0.12\textwidth]{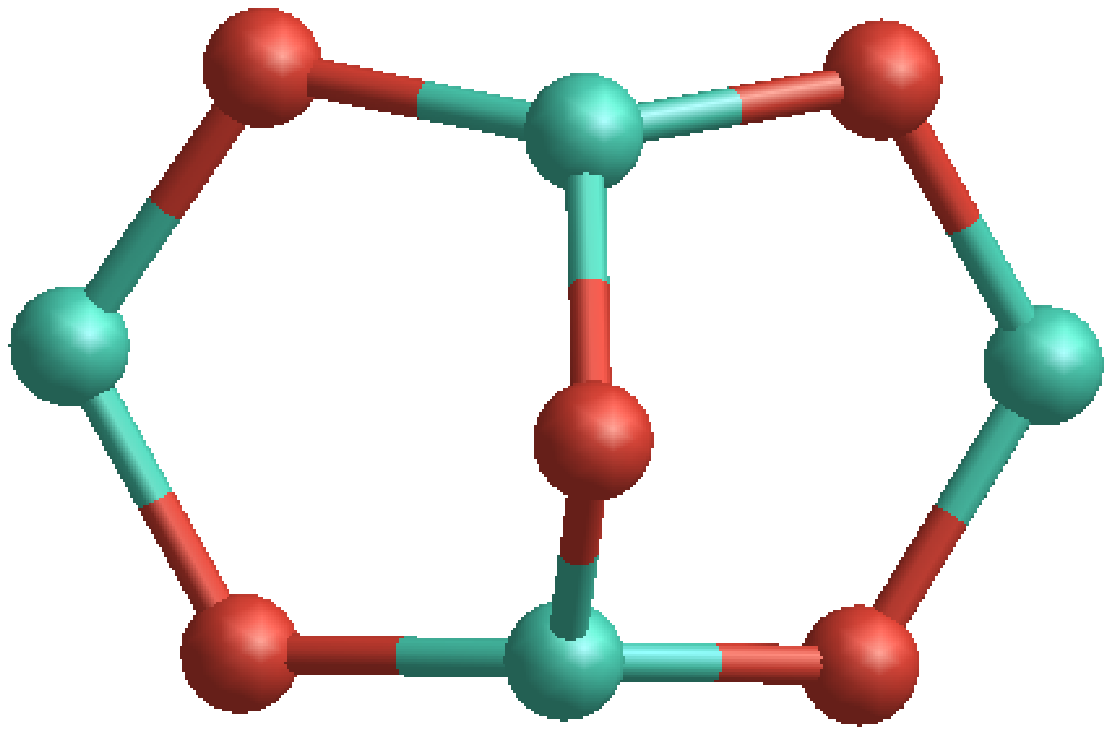} 465. &
 \includegraphics[width=0.15\textwidth]{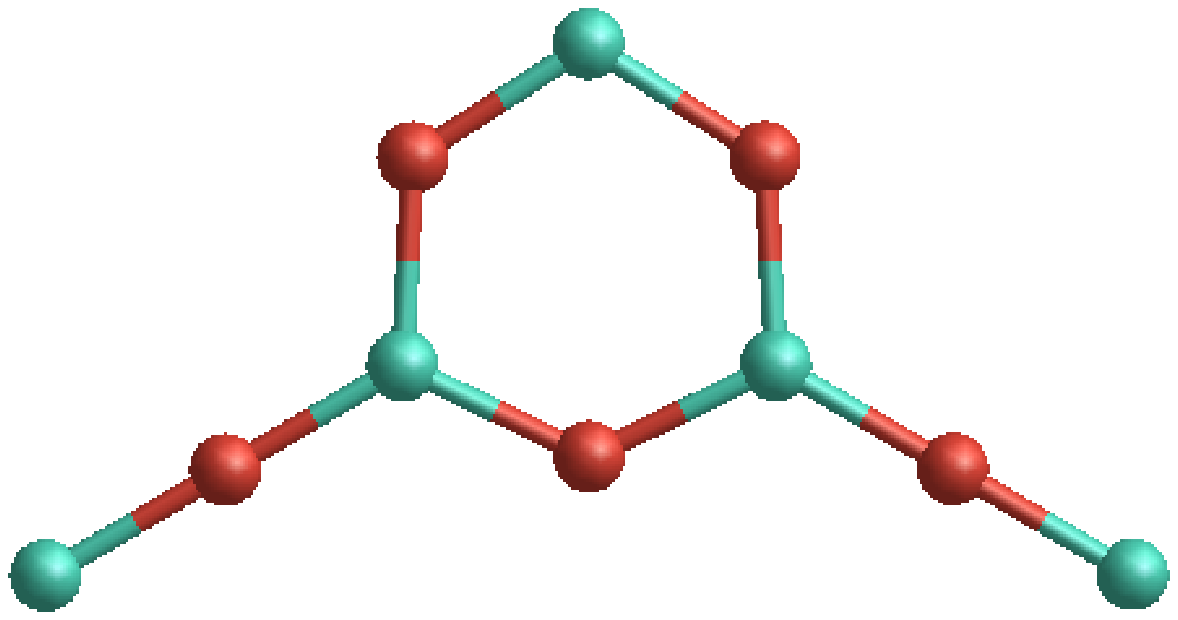} 460. \\
\hline
Al$_x$O$_6$ ($y=$6) & & & 
 \includegraphics[width=0.12\textwidth]{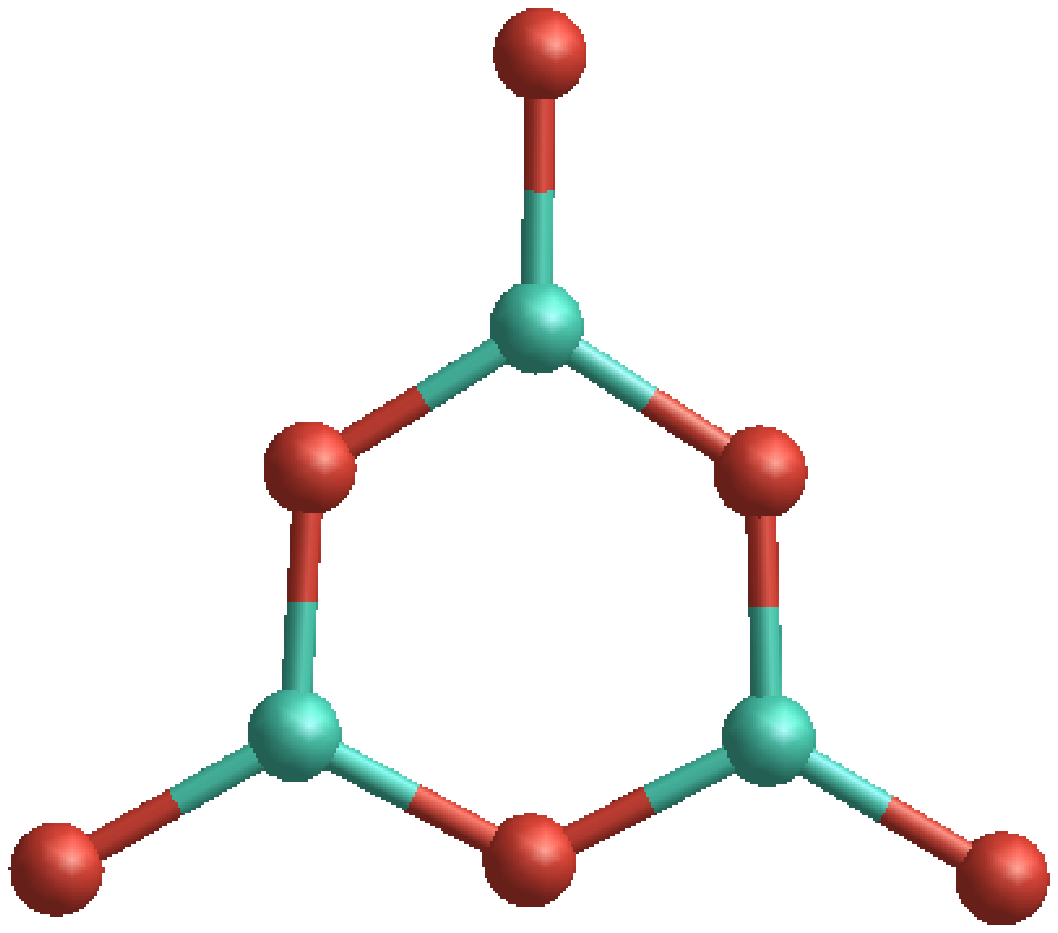} 429. & 
 \includegraphics[width=0.12\textwidth]{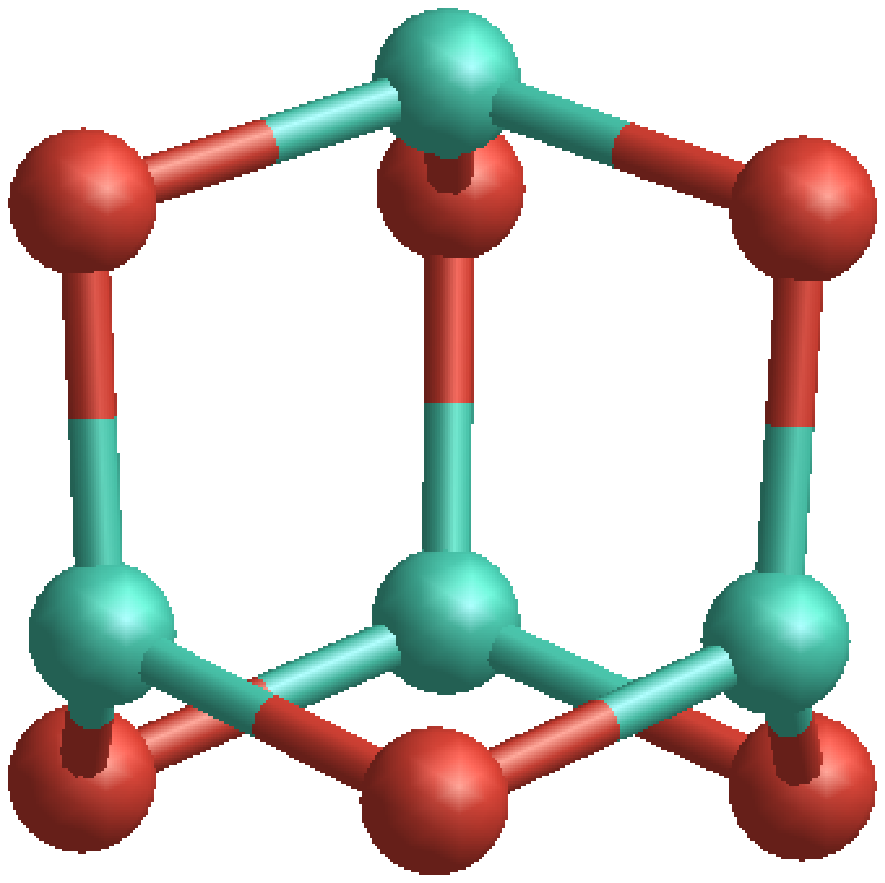} 484. &
  \includegraphics[width=0.12\textwidth]{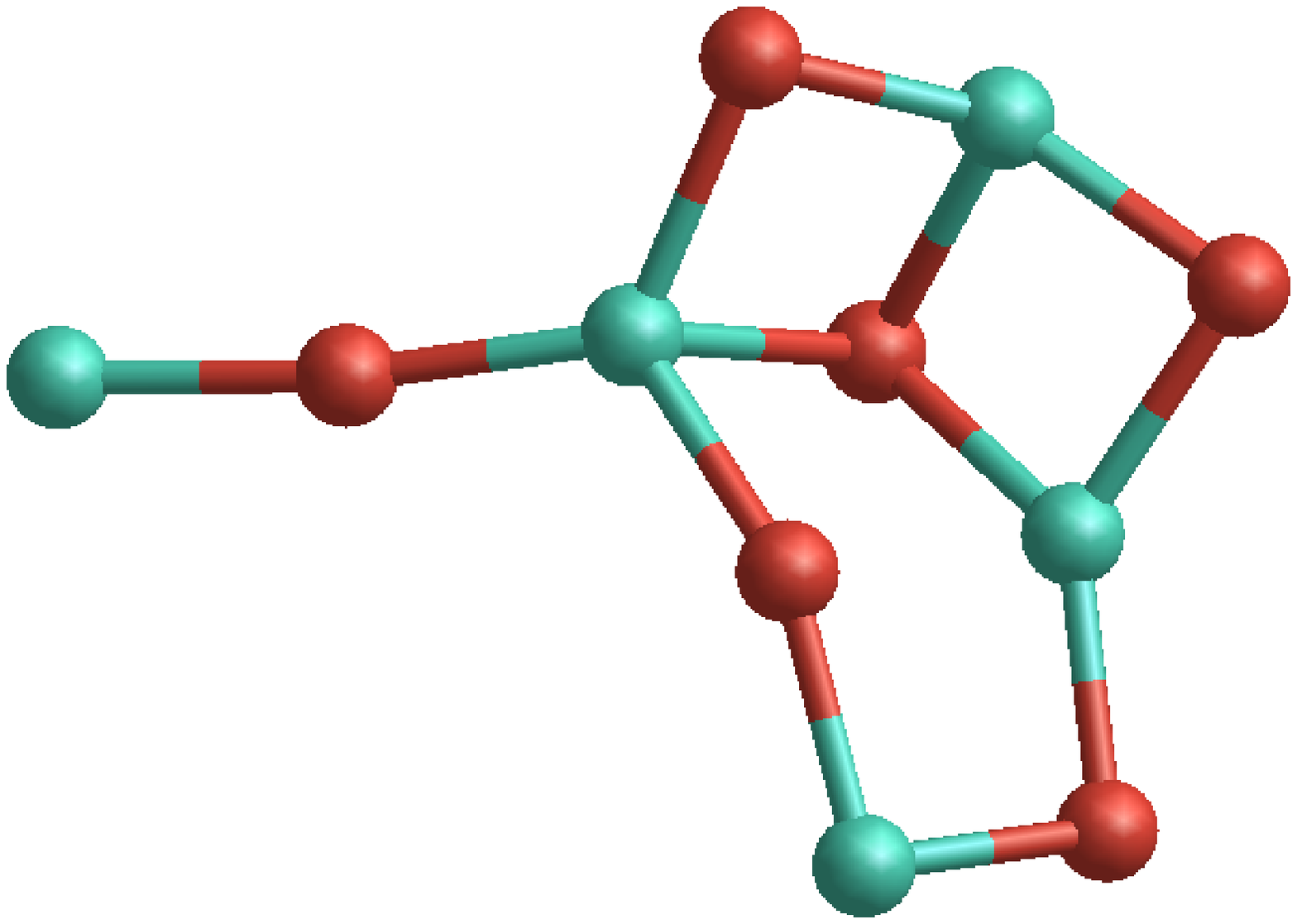} 471. \\
\end{tabular}
\end{table*}

Also, the global minimum (GM) structure of AlO$_2$ is symmetric and linear 
and so is not observable by its pure rotational spectrum.
Although AlO$_2$ isomers with relative energies below 200 kJ mol$^{-1}$ exist, 
their dipole moments are modest making them difficult to detect.
The binding energy of AlO$_2$ is 296 kJ mol$^{-1}$.\
The lowest-energy Al$_2$O$_2$ configuration has a diamond-shaped, rhombic form and has
a binding energy of 386 kJ mol$^{-1}$ 
Owing to its symmetry this molecule has no permanent dipole moment. 
Hence, Al$_2$O$_2$ is not observable by rotational transitions. 
Asymmetric and bending vibration modes of Al$_2$O$_2$ can be observed,
but the related intensities of these vibrations modes are comparatively rather low.  
The linear Al$_2$O$_2$ isomer has a relative energy of 32 kJ mol$^{-1}$ above the GM.
Moreover, we also find a triplet diamond-shaped Al$_2$O$_2$ molecule with a relative energy of 
232 kJ mol$^{-1}$ above the GM.
We conclude that circumstellar Al$_2$O$_2$ is predominantly in the form of the GM structure.\
Al$_2$O$_3$ corresponds to the monomer of stoichiometric alumina and will be discussed 
separately in Section \ref{mono}.
The Al$_{2}$O$_{4}$ GM candidate shows a flat D$_{2h}$ symmetric structure consisting of a 
diamond-shaped Al$_2$O$_2$ ring with an extra oxygen atom attached to each of the Al atoms. 
The Al$_{2}$O$_{4}$ binding energy is 414 kJ mol$^{-1}$.\\

The lowest-energy  Al$_{3}$O$_{2}$ configuration has a binding energy of 384 kJ mol$^{-1}$. 
It is a bent linear chain with a bond angle of 122.5$^{\circ}$ on the central Al atom.
Similarly, the GM Al$_3$O$_3$ structure 
is flat and kite-shaped. It can be seen as a composite of Al$_2$O$_2$ and AlO and has
a binding energy of 383 kJ mol$^{-1}$.
The Al$_3$O$_4$ GM has a C$_{3v}$ symmetric pyramidal form. 
It differs from the other Al$_x$O$_y$ molecules in the sense that it is the smallest 
three dimensional molecule in this study and has a 3-coordinated O atom (top of the 
pyramid), as has been noticed previously \citep{2005EPJD...32..329P}. 
The binding energy of this structure is 437 kJ mol$^{-1}$.\\

The GM Al$_3$O$_5$ structure 
has a C$_s$ symmetry and was recently reported by \citet{doi:10.1021/acs.jpca.9b01729}.
The binding energy per atom is 438 kJ mol$^{-1}$. The geometry as well as the bond energy 
is very close to that of Al$_3$O$_4$. The dangling oxygen atom is expected to be the most 
reactive site of the molecule.\
We also find another structurally similar isomer that is just 8.1 kJ mol$^{-1}$ higher in energy.
It has a three-dimensional geometry with a C$_{2v}$ symmetry and was firstly found by \citet{Martinez2001}. 
The most favourable Al$_3$O$_6$ is a flat structure consisting of a Al$_3$O$_3$ ring having a terminal oxygen atom on each of the three Al atoms \citep{doi:10.1021/jp038040n}.
The lowest-energy Al$_4$O$_3$ isomer is a flat arrangement around a central Al atom with 
a perfect three-fold symmetry (point group D$_{3h}$). All other 
investigated geometries did not converge, or, are found to be transition states relaxing 
to the three-fold GM. The binding energy is 426 kJ mol$^{-1}$.
The most stable Al$_{4}$O$_{4}$ configuration is a chain with a rhombus in the center. 
The flat C$_{2v}$-symmetric geometry can be regarded as an Al$_2$O$_2$ diamond with a 
terminal AlO group on each side. Structurally, Al$_{4}$O$_{4}$ 
can be regarded as a compound of AlO+Al$_2$O$_2$+AlO.   
The binding energy at the CBS-QB3 level of theory is 452 kJ mol$^{-1}$. With 
respect to the AlO molecule, the Al$_4$O$_4$ dissociation energy is 406 kJ mol$^{-1}$.
The most favourable Al$_4$O$_5$ cluster is a three-dimensional C$_{2v}$ symmetric 
configuration consisting of two 6-member rings sharing two AlO bonds 
as previously reported by \citet{doi:10.1080/00268976.2010.542777}.
In this configuration all O atoms are 2-coordinated and all Al atoms are 3-coordinated 
Its binding energy per atom is quite high (465 kJ mol
$^{-1}$), but still below the alumina dimer (484 kJ mol$^{-1}$ ).
The isomer with the second lowest energy is still 89 kJ mol$^{-1}$ above the GM candidate and has a C$_s$ symmetry.
Other Al$_4$O$_5$ GM candidates reported in the literature show even larger relative energies and are far
above our GM candidate. 
For example, the GM candidate reported by \citep{2015JPCA..119.8944L} is still 97 kJ mol$^{-1}$ above our finding.
The structure of the most stable Al$_5$O$_4$ isomer 
has been predicted by \citet{doi:10.1021/ct800232b}. We find a CBS-QB3 binding energy of 426 kJ mol
$^{-1}$ per atom. Its negative ion (Al$_5$O$_{4}^{-}$) is found to be a highly symmetric planar structure  
with strong electron affinity. A very similar quasi-planar structure has also been predicted as a GM candidate 
for neutral Al$_5$O$_4$ \citep{doi:10.1021/acs.jpca.9b01729}.
However, its CBS-QB3 energy is 32 kJ mol $^{-1}$ above our non-planar GM candidate. 
The GM isomer candidate of Al$_{5}$O$_{5}$ is a flat C$_{2v}$-symmetric structure consisting of a six-member 
ring with two \textit{cis}-oriented terminal Al-O groups. 
Its CBS-QB3 binding energy is 460 kJ mol$^{-1}$ per atom and 422 kJ mol$^{-1}$ per AlO unit. 
Another GM candidate, reported by \citep{doi:10.1021/acs.jpca.9b01729}, consisting of a Al$_4$O$_4$ cube with an O-Al chain on one of the Al edges, has a CBS-QB3 energy which is 29 kJ mol$^{-1}$ above our GM candidate.
The lowest-energy Al$_5$O$_6$ isomer 
was also previously found by \citet{doi:10.1021/acs.jpca.9b01729}.
It has a binding energy of 472 kJ mol$^{-1}$ and shows no symmetry. 
A quasi-planar `heart'-shaped isomer with a C$_s$ symmetry lies 14 kJ mol$^{-1}$ above the 
GM candidate.

\subsubsection{Alumina (Al$_2$O$_3$)$_n$, $n=$1$-$10, clusters }
\subsubsection*{Alumina monomer (Al$_2$O$_3$)}
\label{mono}
\begin{figure}[H]
	\includegraphics[scale=0.3]{Al2O3A.eps}1A
	\includegraphics[scale=0.3]{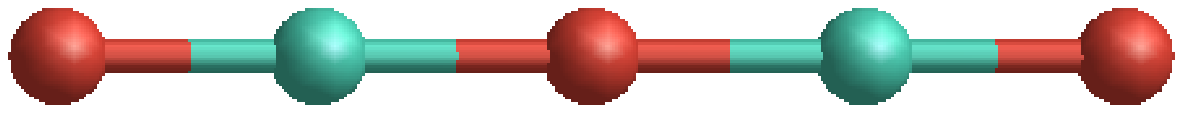}1B \\
         \hspace*{3.4cm}GM\hspace*{3.8cm}11.
	\caption{Left: GM structure of the alumina monomer Al$_2$O$_3$ (1A) 
		Right: Second lowest-energy isomer of the alumina monomer (1B). \label{monomer}}
\end{figure}

The most favourable alumina monomer structure (1A) has a kite-shaped form and is in a 
triplet state (see Figure \ref{monomer}). Its geometry is flat and obeys a C$_{2v}$ point 
symmetry. 
The binding energy (per atom) on the CBS-QB3 level of theory is 388 kJ mol$^{-1}$, 
which is higher than for DFT calculations using the B3LYP functional (359 kJ mol$^{-1}$) and the 
PBE0 functional (360 kJ mol$^{-1}$).
The linear singlet alumina monomer (1B) has a relative energy of 15  kJ mol$^{-1}$, 11 kJ mol$^{-1}$, and 44 kJ mol$^{-1}$  above 1A at the CBS-QB3, B3LYP and PBE0 levels of theory, respectively.
All other isomers have significantly higher energies than 1A and 1B, and are metastable with respect to 1A and 
1B.

\subsubsection*{Alumina dimer (Al$_2$O$_3$)$_2$}
\begin{figure}[H]
	\includegraphics[scale=0.5]{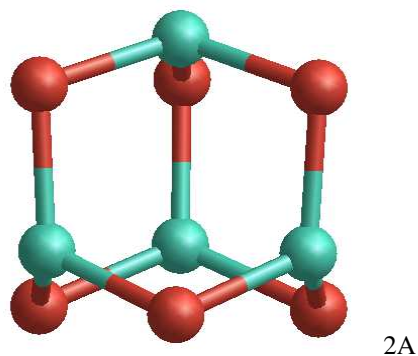}2A
	\caption{GM structure of the alumina dimer (Al$_2$O$_3$)$_2$ (2A) \label{2A}}
\end{figure}

The GM alumina dimer cluster (2A) is displayed in Figure \ref{2A}.
The geometry of the GM cluster is a cage composed of four 6-membered rings and shows a 
tetrahedral symmetry (point group T$_d$). All Al atoms are 3-coordinated and all O atoms 
are 2-coordinated corresponding to the valence of the atoms. 
Moreover, owing to its symmetry, 2A is characterised by one single bond distance of 
1.744 \AA{} for all 12 bonds of the clusters. 
Owing to its energetic stability, its symmetry and its approximate sphericity, 2A 
is a natural and logical link between a molecular regime controlled by chemical-kinetics and cluster coagulation 
(see also \citet{2016A&A...585A...6G}). Unfortunately, as a consequence of the tetrahedral symmetry, 2A has no permanent dipole moment and obervations in the IR are challenging.  
The energetically second-lowest dimer isomer have CBS-QB3, B3LYP and PBE0 energies that are 42 kJ mol$^{-1}$, 45 kJ mol$^{-1}$, and 32 kJ mol$^{-1}$ above 2A, respectively. Owing to this 
considerable energy difference, only 2A is considered to contribute to the dimer abundance. 
The CBS-QB3 binding energy of 2A is 484 kJ mol$^{-1}$ corresponding to the largest binding energy in Table \ref{ALXoYs}.
As for the monomer, the CBS-QB3 binding energy is higher than the predictions of B3LYP (442 kJ mol$^{-1}$) and PBE0 (446 kJ mol$^{-1}$).

\subsubsection*{Alumina trimer (Al$_2$O$_3$)$_3$}
\begin{figure}[H]
	\includegraphics[scale=0.3]{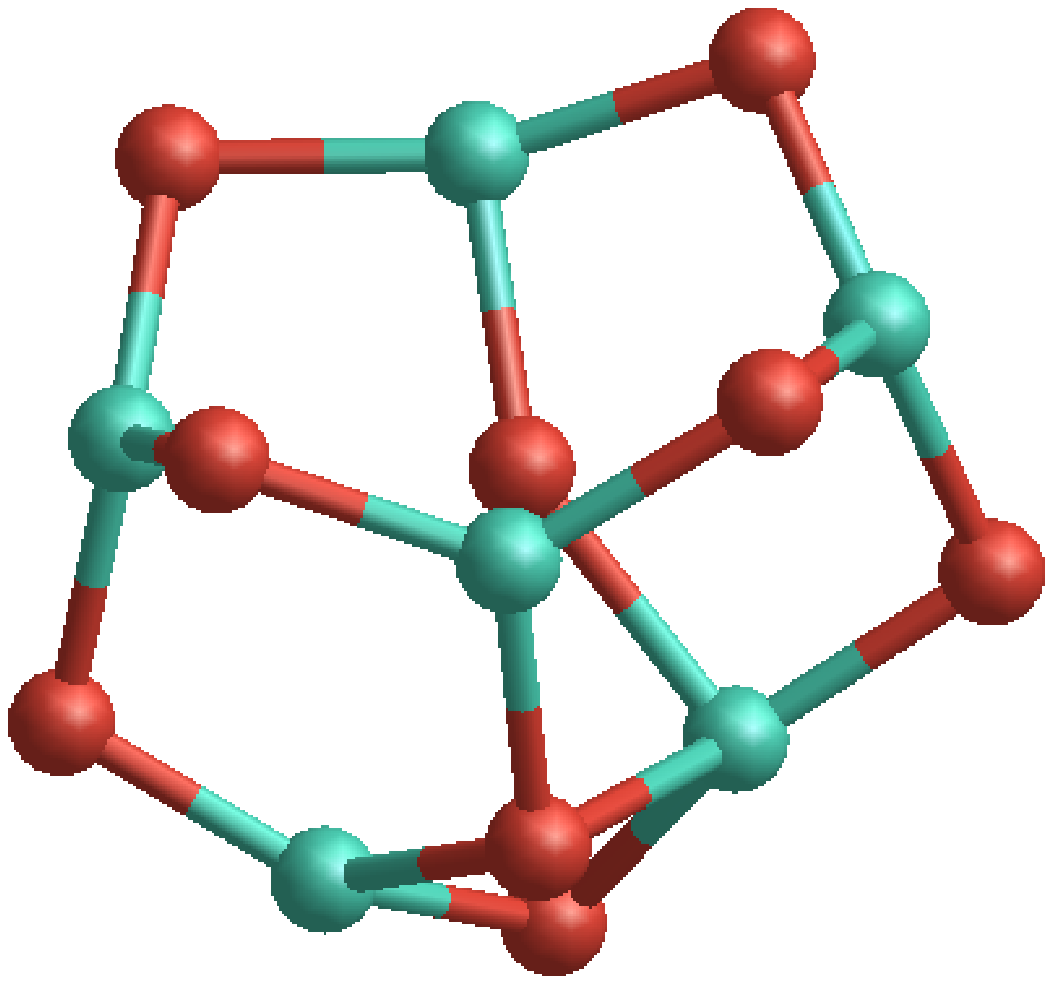}3A
	\includegraphics[scale=0.3]{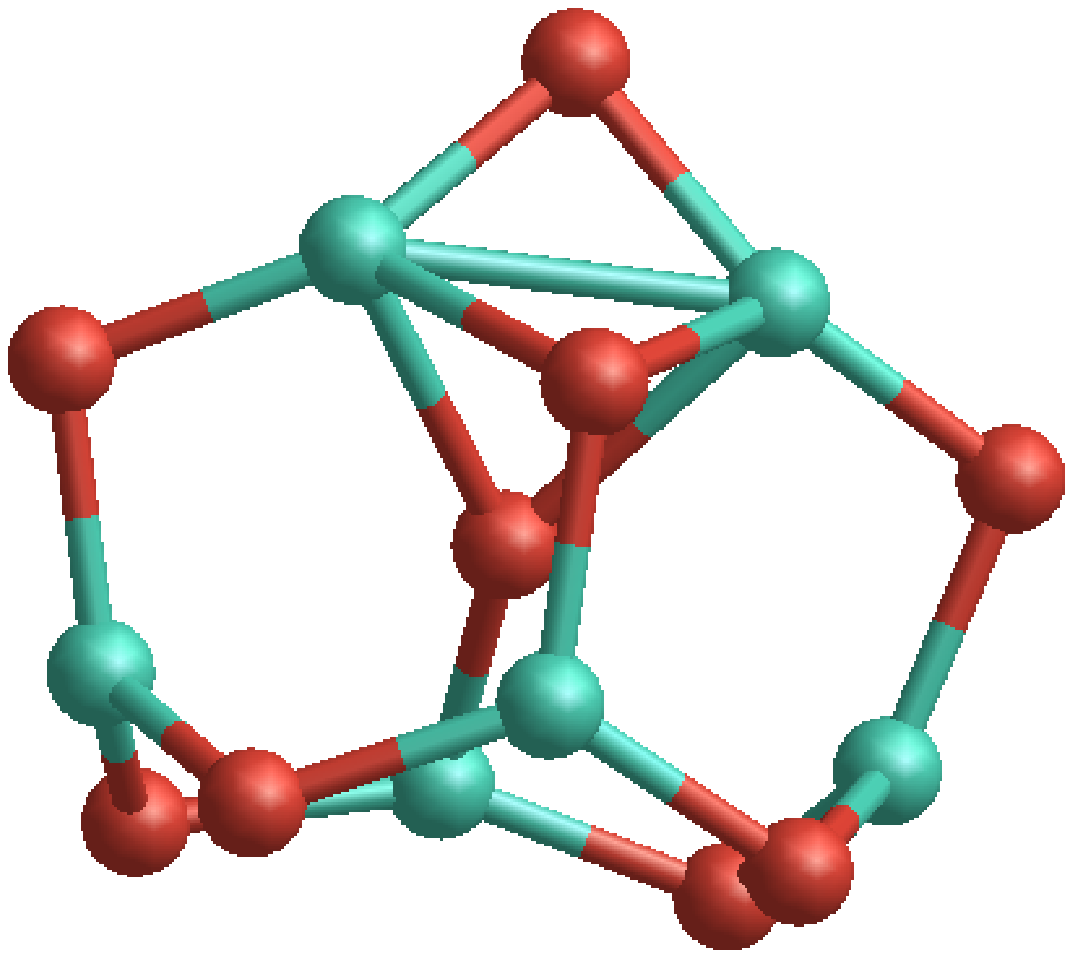}3B
         \hspace*{3.4cm}GM\hspace*{4.8cm}2.
	\caption{GM candidate structure of the alumina trimer (Al$_2$O$_3$)$_{3}$. \textit{Left:} 3A ; \textit{Right:} 3B\label{trimer}.}
\end{figure}

The most favourable alumina trimer (3A) is a `tea-cosy'-shaped structure (see Figure \ref{trimer}). 
The binding energy at the CBS-QB3, B3LYP and PBE0 levels of theory are 515 kJ mol$^{-1}$, 470 kJ mol$^{-1}$ and 
476 kJ mol$^{-1}$), respectively.
As \citet{LI2012125} have shown, there are four energetically low-lying structural isomers 
that are close in energy (all within an energy range of 5.5 kJ mol$^{-1}$ at the B3LYP/6-311+G(d) level of theory). 
Consequently, all four isomers are expected to contribute to the overall abundance
of alumina trimers. We confirm the narrow spacing in energy of the two most stable 
configurations (3A and 3B) by more accurate CBS-QB3 calculations. 
However, we find that 3C, as predicted in \citet{LI2012125}, lies 21 kJ mol$^{-1}$ (for B3LYP) and 44 kJ mol$^{-1}$ (for PBE0) above 3A; 
for 3D we found an imaginary frequency indicating a transition state and not a real minimum.      
As pointed out, the second most stable trimer (3B) is energetically degenerate as its CBS-QB3, B3LYP and PBE0 energies 
lie only 0.7 kJ mol$^{-1}$, 2.1 kJ mol$^{-1}$ and 0.2 kJ mol$^{-1}$ above 3A respectively. 

\subsubsection*{Alumina tetramer (Al$_{2}$O$_{3}$)$_4$}
\begin{figure}[H]
	\includegraphics[scale=0.3]{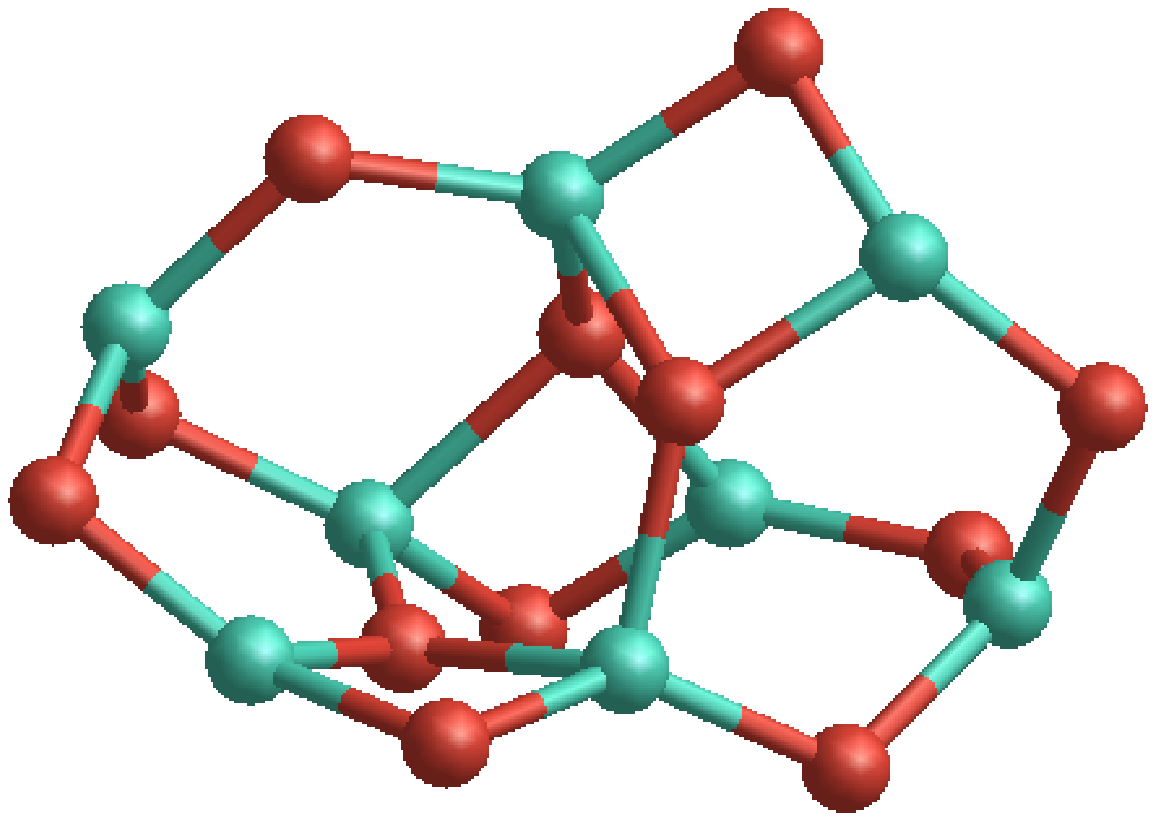}4A
	\includegraphics[scale=0.2]{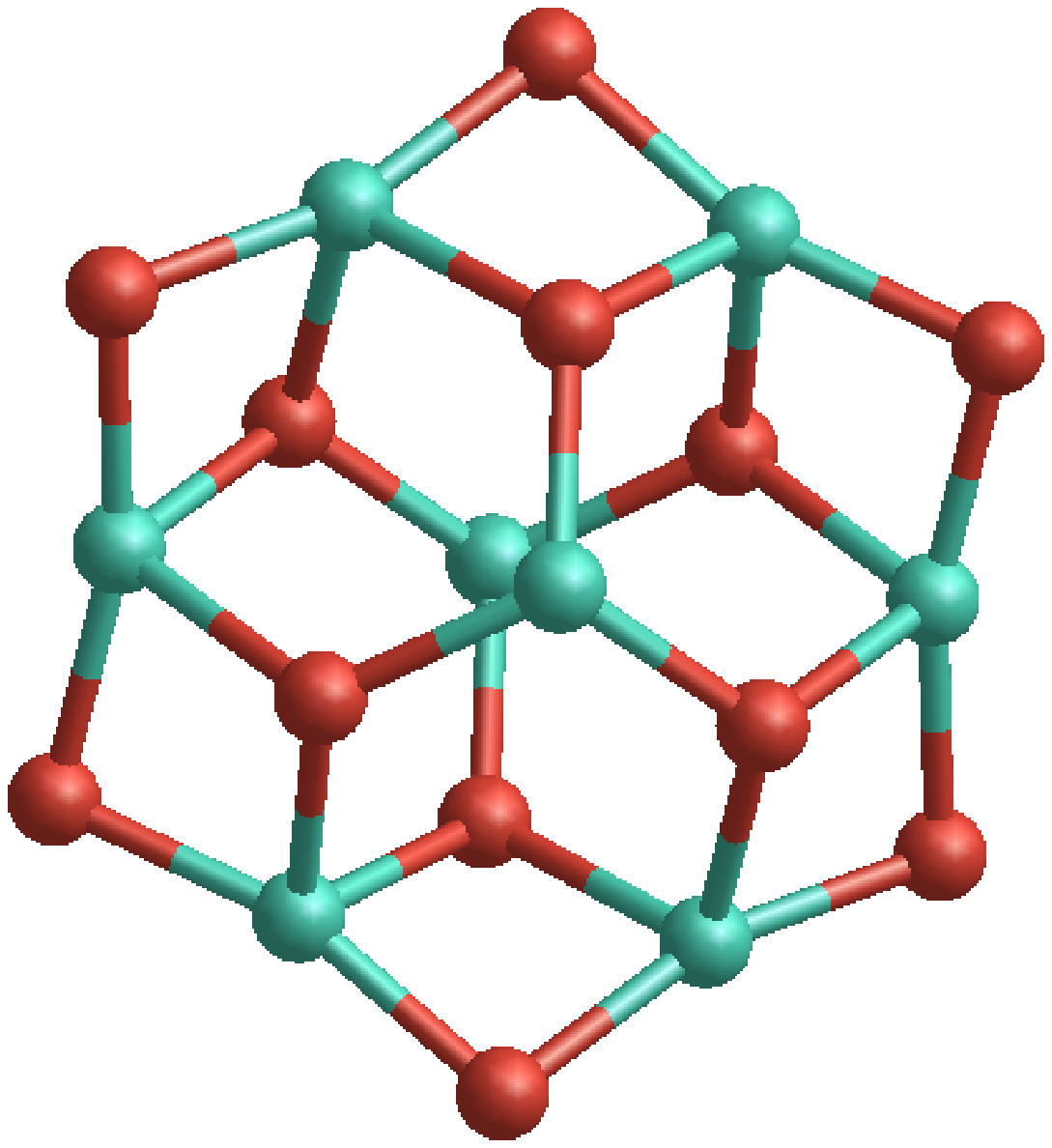}4B \\
         \hspace*{4.1cm}GM\hspace*{2.5cm}10.
	\caption{Global minimum candidates of the alumina tetramer (Al$_2$O$_3$)$_{4}$. \textit{Left:} 4A; \textit{Right:} 4B \label{tetra}}
\end{figure}

The lowest energy configuration of the alumina tetramer (4A) found with the B3LYP functional has 
no particular symmetry (point group C$_1$). 
Its B3LYP and PBE0 binding energy are 484 kJ mol$^{-1}$ and 488  kJ mol$^{-1}$, respectively.
With the PBE0 functional we find a different lowest-energy isomer (4B) showing a highly symmetric D$_{3d}$ structure (see Figure \ref{tetra}).  
The binding energy of 4B is found to be 483 kJ mol$^{-1}$ (B3LYP) and 497 kJ mol$^{-1}$ (PBE0).
Experiments have shown that 4A is the GM structure and indicate that the B3LYP method accurately describes Al$_{2}$O$_{3}$ structures \citep{doi:10.1002/anie.200604823}.

\subsubsection*{Alumina pentamer (Al$_{2}$O$_{3}$)$_5$}
\begin{figure}[H]
	\includegraphics[scale=0.3]{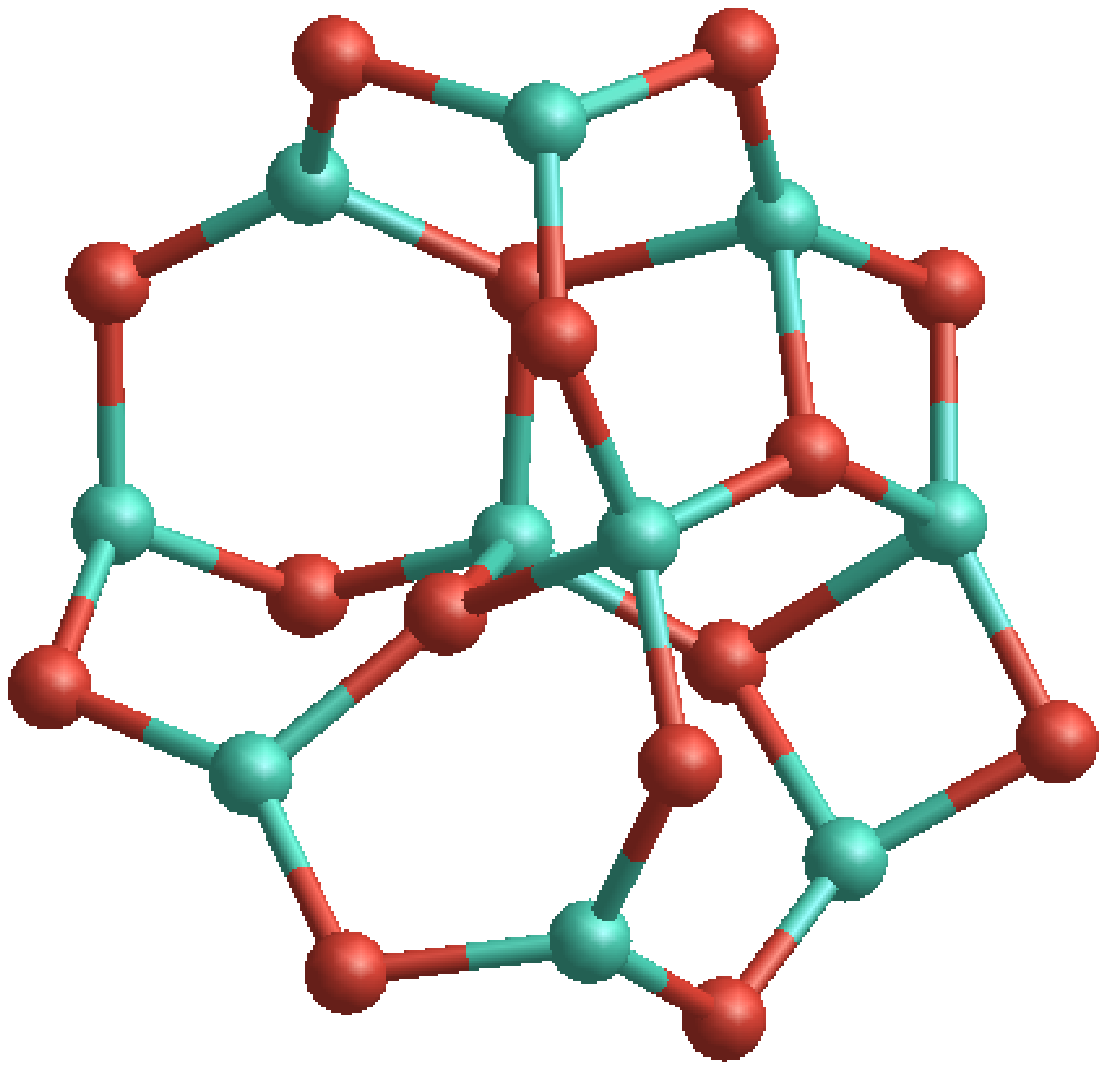}5A 
	\includegraphics[scale=0.3]{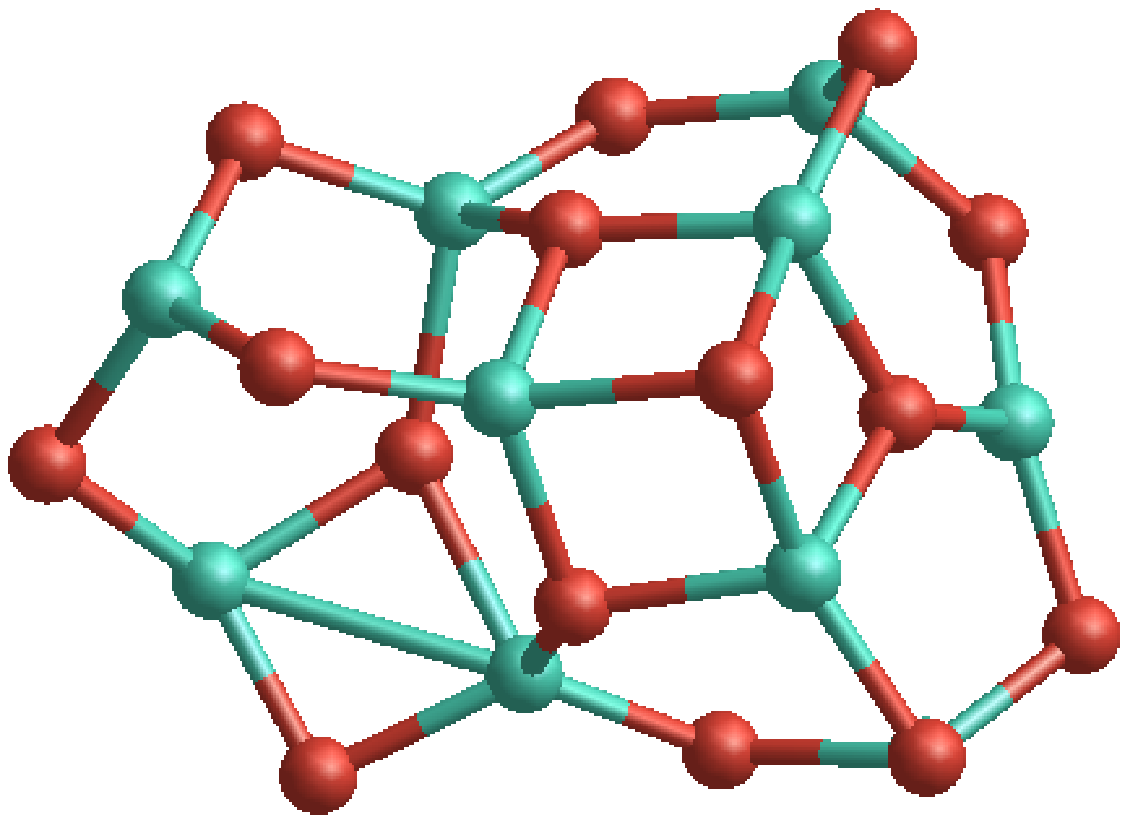}5B \\
         \hspace*{4.2cm}GM\hspace*{3.6cm}6.
	\caption{Global minimum candidates of the alumina pentamer (Al$_2$O$_3$)$_{5}$. \textit{Left:} 5A; \textit{Right:} 5B \label{penta}.}
\end{figure}

The lowest-energy alumina pentamer shows no symmetry (C$_1$) and is displayed in Figure \ref{penta}. 
The B3LYP and PBE0 binding energies of 5A are 492 kJ mol$^{-1}$ and 497 kJ mol$^{-1}$, respectively. 
As for the tetramer ($n=$4) we find a different GM candidate (5B) at the PBE0/6-311+G level of theory. 
5B has no particular symmetry and lies 5.5 kJ mol$^{-1}$ above 5A using B3LYP, but is 3.4 kJ mol$^{-1}$
lower in energy using PBE0.
As the energy differences between 5A and 5B are small in calculations using both functionals, B3LYP and PBE0, 
we conclude that these degenerate structures could both be considered as GM structures and they contribute equally to the pentamer 
abundance.

\subsubsection*{Alumina hexamer (Al$_{2}$O$_{3}$)$_6$}
\begin{figure}[H]
	\includegraphics[scale=0.3]{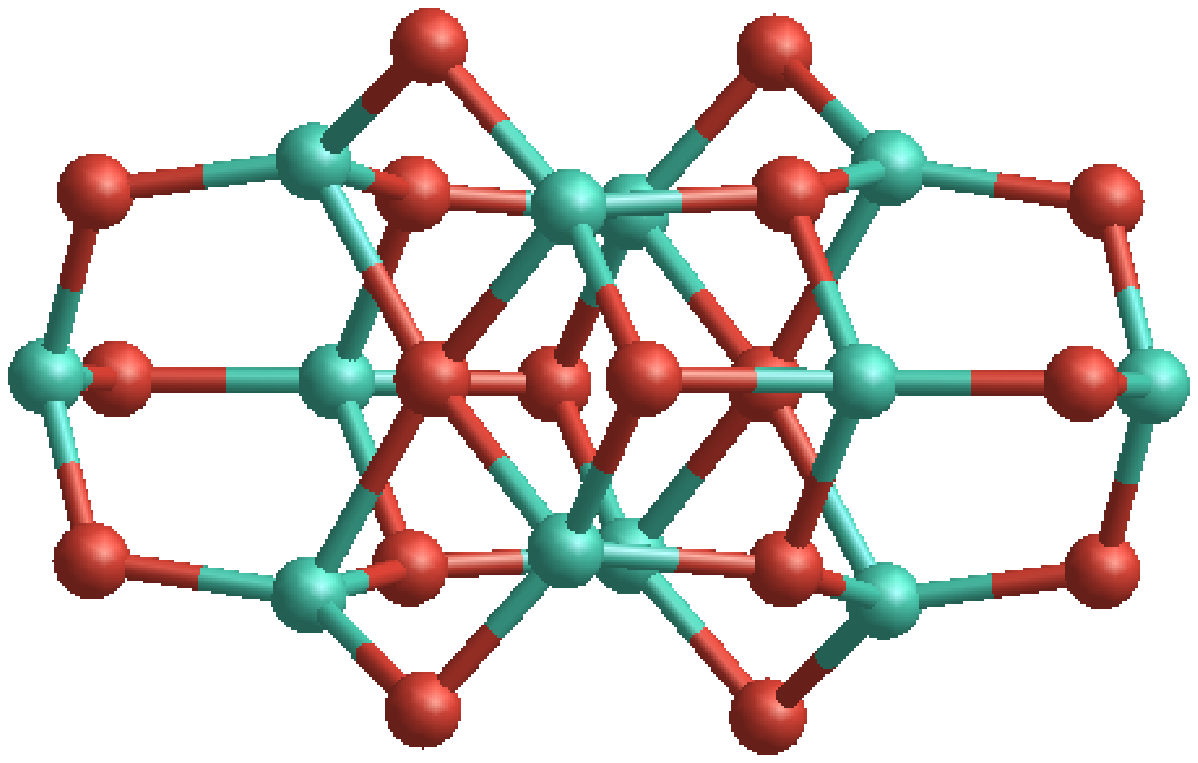}6A
	\includegraphics[scale=0.3]{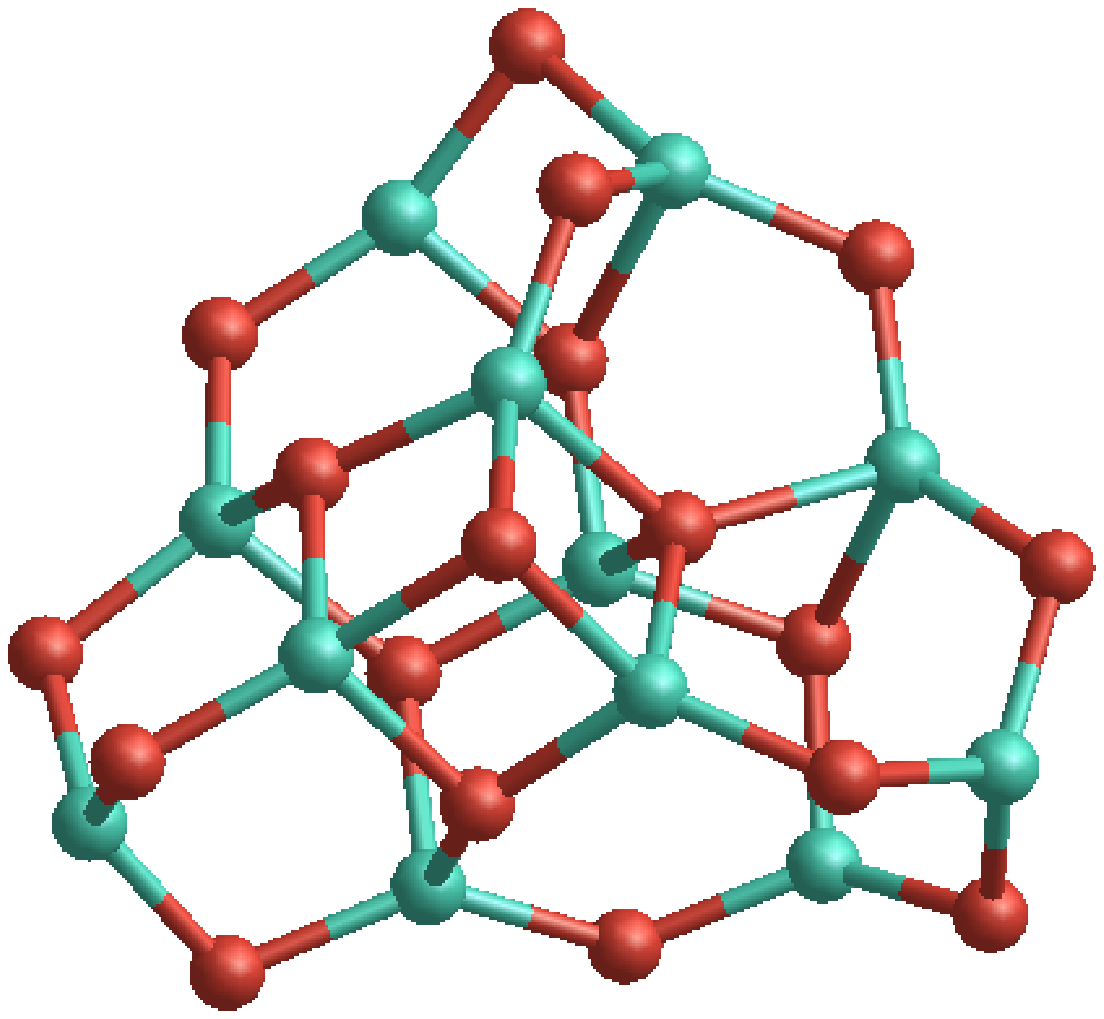}6B\\
         \hspace*{4.2cm}GM\hspace*{3.9cm}1.
	\caption{Global minimum candidates of the alumina hexamer (Al$_2$O$_3$)$_{6}$. \textit{Left:} 6A; \textit{Right:} 6B.
		\label{hexamer}}
\end{figure}

The most favourable alumina hexamer is shown in Figure \ref{hexamer}.
It exhibits a C$_{2h}$ symmetric structure. 
The binding energy per atom is 499 kJ mol$^{-1}$ (B3LYP) and 505
kJ mol$^{-1}$ (PBE0). Due to its symmetry 6A has no net dipole moment.
It should be noted that another hexamer (6B) structure is essentially degenerate with an 
energy of only 0.5 kJ mol$^{-1}$ above 6A (B3LYP). With the PBE0 functional, the 
energy difference between 6A and 6B is larger (14.5 kJ mol$^{-1}$) though also not 
significantly different. This degeneracy has already been noted by \citet{LI2012125}.  
We applied a larger basis set (cc-pVTZ) to 6A and 6B to test the reliability of our 
results with the 6-311+G(d,p) basis set. We find that 6A is lower than 6B by 5.9 kJ mol$^{-1}$ 
(B3LYP/cc-pVTZ) and 21.0 kJ mol$^{-1}$ (PBE0/cc-pVTZ). These results indicate that the calculations with
the 6-311+G(d,p) basis set are well-founded and that 6A is indeed the lowest-energy isomer 
for $n=6$ though structure 6B is relatively close in energy.
Structure 6B has no symmetry (C$_{1}$).
Its binding energy at the B3LYP and PBE0 level of theory are 499 kJ mol$^{-1}$  
and 505 kJ mol$^{-1}$, respectively.


\subsubsection*{Alumina heptamer (Al$_{2}$O$_{3}$)$_7$}
\label{338}
\begin{figure}[H]
	\includegraphics[scale=0.6]{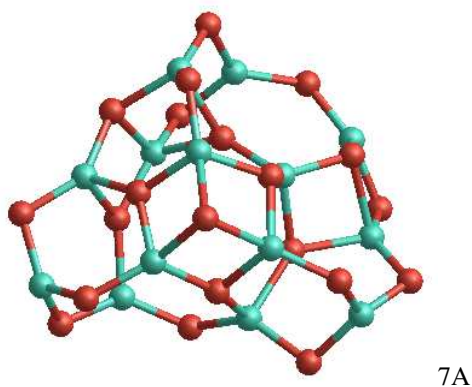}7A
	\caption{Global minimum candidate (7A) of the alumina heptamer (Al$_2$O$_3$)$_{7}$.
		\label{hepta}}
\end{figure}

\begin{figure}[H]
	\includegraphics[scale=0.7]{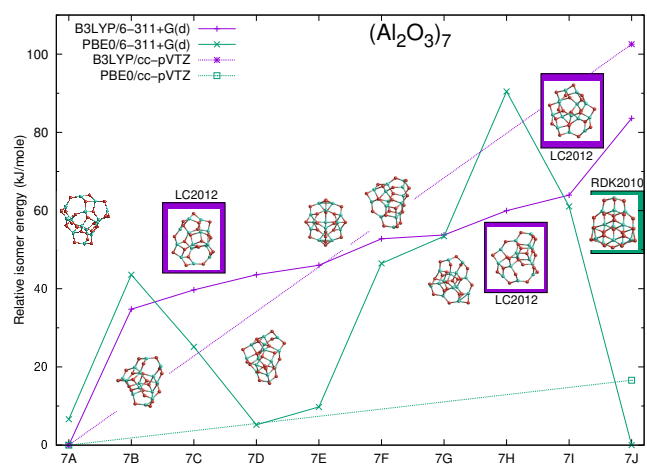}
	\caption{Relative energies of the lowest-energy alumina heptamer isomers (Al$_2$O$_3$)$_{7}$. Structures reported by \citet{LI2012125} are indicated in violet as LC2012 and the GM candiate reported by \citet{doi:10.1021/jp2050614} in green as RDK2010\label{n7compa}.}
\end{figure}

We report the discovery of six energetically low-lying isomers (7A, 7B, 7D, 7E, 7F, 7G) including a GM candidate (7A), 
which are shown in Figures \ref{hepta} and \ref{n7compa}. To our knowledge, these six isomers were hitherto not reported. 
7A does not show any symmetry and have B3LYP and PBE0 binding energies of 504 kJ mol$^{-1}$ and 509 kJ mol$^{-1}$ per atom, respectively. 

Using the PBE0 functional, we find a different lowest-energy isomer (7J) that was previously reported by \citet{doi:10.1021/jp2050614}.
As for $n=6$, we perform benchmark calculations of 7A and 7J with a larger, numerical basis set (cc-pVTZ) revealing that 7A is lower in energy (B3LYP/cc-pVTZ: 102.6 kJ mol$^{-1}$, PBE0/cc-pVTZ: 16.6 kJ mol$^{-1}$) than 7J (see Figure \ref{n7compa}). Therefore, we assume 7A to be the GM candidate. 

Using the cc-pVTZ basis set, we note that the relative energies of 7A and 7J differ by more than 80 kJ mol$^{-1}$ , comparing B3LYP with PBE0.
Moreover, the energetic ordering of the structural isomers is very different for PBE0/6-311+G(d) (7J,7A) and B3LYP/6-311+G(d) (7A,7B,7C,7D,7E,7F,7G,7H,7I,7J). 
This is a rather surprising result, as the hybrid density functionals B3LYP and PBE0 differ primarily by the amount of Hartree-Fock (HF) exchange (PBE0: 25 \%, B3LYP: 20 \%), which is rather small. 
We conclude that the alumina $n=$7 isomer energies are very sensitive to the choice of the functional (amount of HF exchange). A thorough study on the sensitivity is beyond the scope of this paper,
but the set of isomers 7A$-$7J could be useful as a test case for the performance of (hybrid) density functionals. 
\subsubsection*{Alumina octamer (Al$_{2}$O$_{3}$)$_8$}

\begin{figure}[H]
	\includegraphics[scale=0.3]{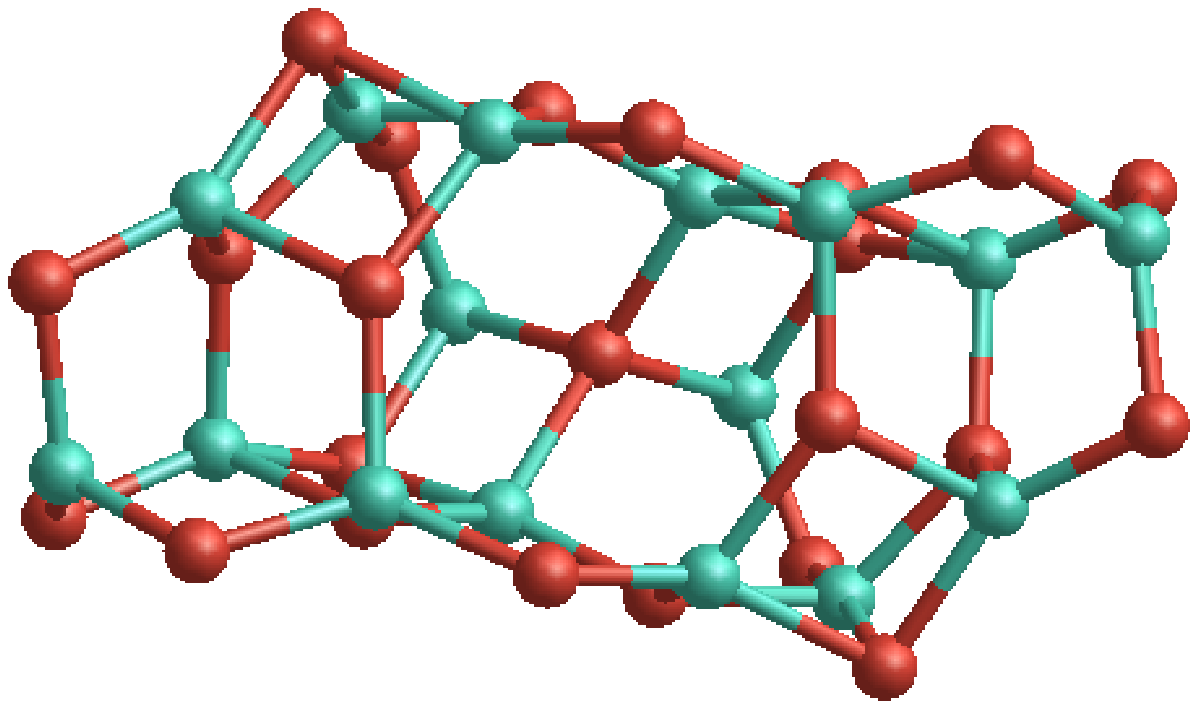}8A
	\includegraphics[scale=0.3]{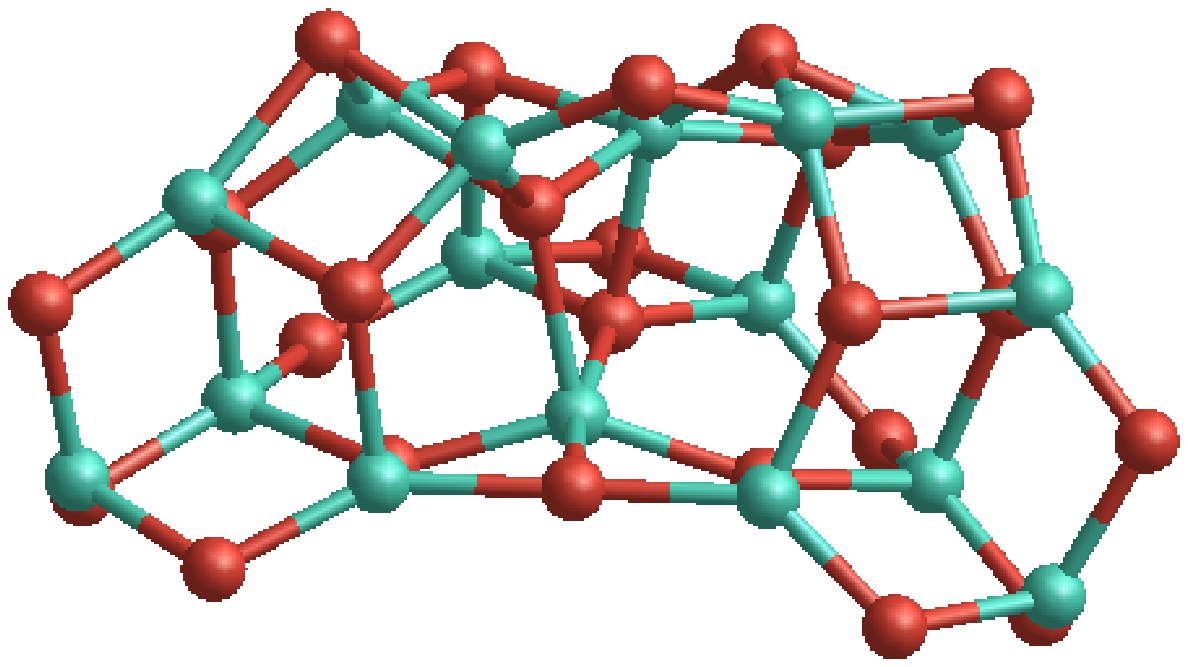}8B \\
         \hspace*{3.8cm}GM\hspace*{3.8cm}14.
	\caption{Global minimum candidate structures of the alumina octomer (Al$_2$O$_3$)$_{8}$\label{octo}.}
\end{figure}

The lowest-lying alumina octomer clusters are extensively discussed and reported in \citet{GOBRECHT2018138}. 
Here we summarise our main findings. 
As for $n=4$, 5, and 7 we also find for $n=$8 different GM structures depending on whether the PBE0 or the B3LYP functional is applied in combination with the 6-311+G(d) bais set. With the B3LYP functional, the lowest-energy isomer (8A) is a C$_2$ symmetric 
structure (see Figure \ref{octo}). Its B3LYP and PBE0 binding energies (per atom) are 507 kJ mol$^{-1}$ and 513
kJ mol$^{-1}$, respectively. Isomer 8B has no special symmetry and B3LYP and PBE0 energies of 507 kJ mol$^{-1}$ and 514
kJ mol$^{-1}$, respectively. Both structures, 8A and 8B, show geometries with large aspect ratios.

\subsubsection*{Alumina nonamer (Al$_{2}$O$_{3}$)$_9$}

\begin{figure}[H]
	\includegraphics[scale=0.7]{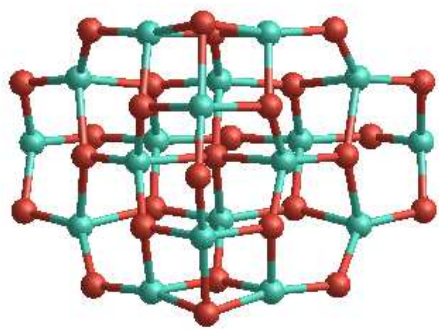}9A
	\caption{GM candidate (9A) of the alumina nonamer (Al$_2$O$_3$)$_{9}$\label{nonamer}.}
\end{figure}

\begin{figure}[H]
	\includegraphics[scale=0.2]{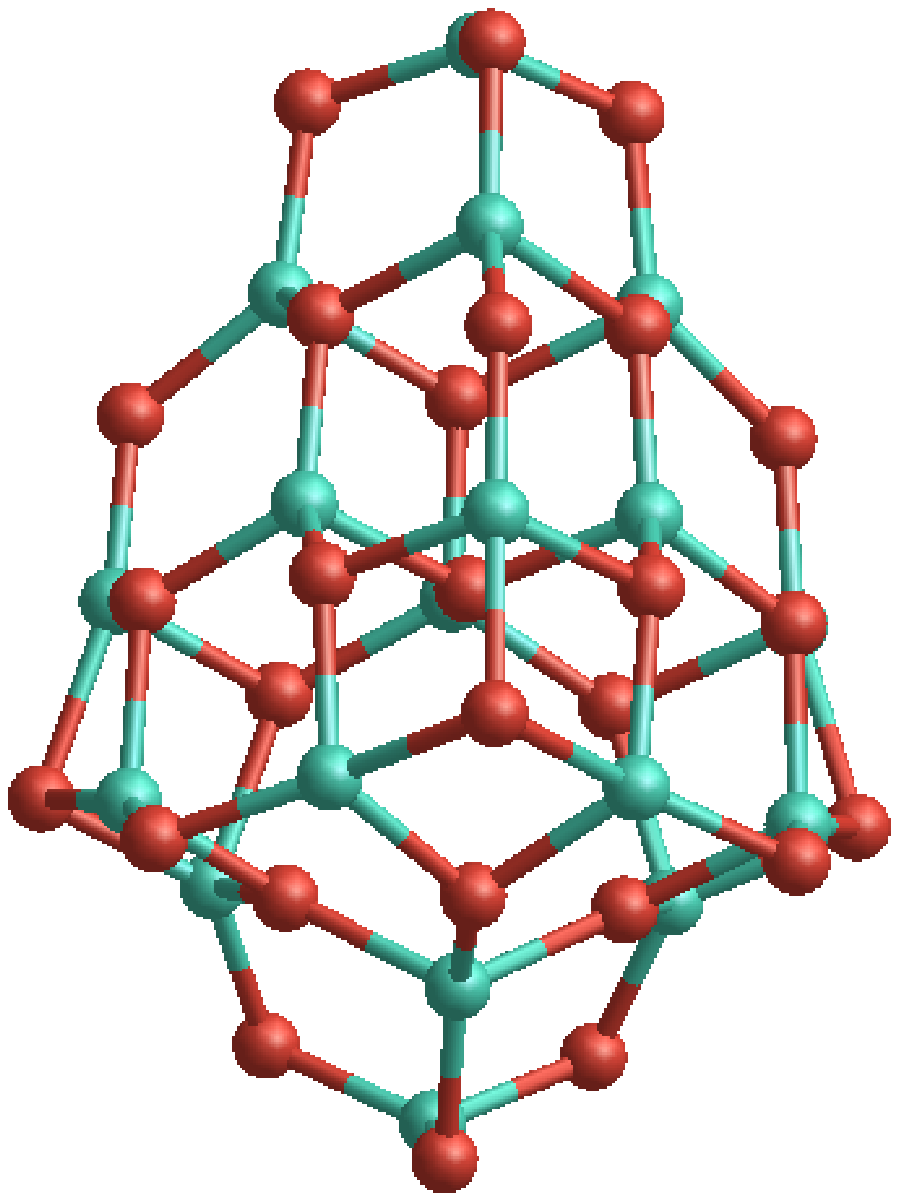}9B
	\includegraphics[scale=0.2]{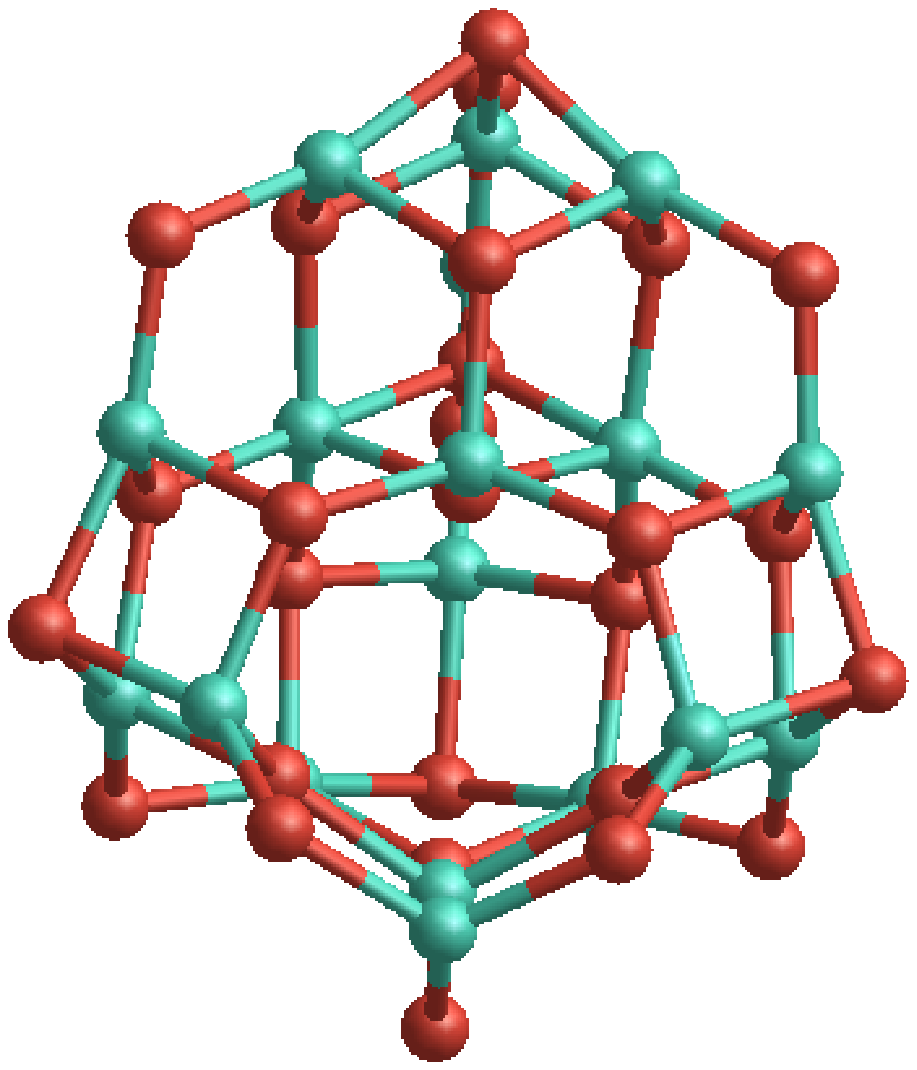}9C
	\includegraphics[scale=0.2]{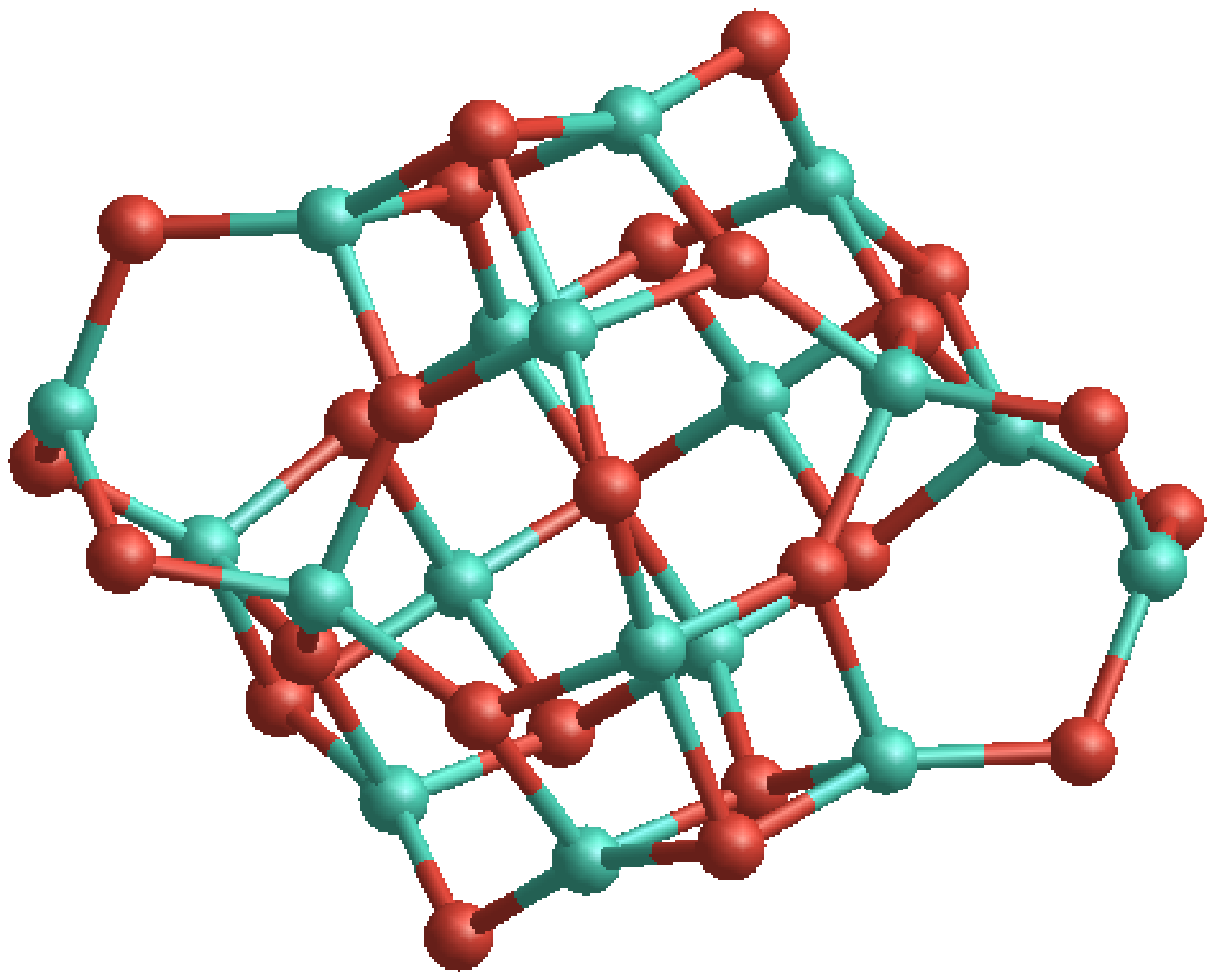}9D
	\includegraphics[scale=0.2]{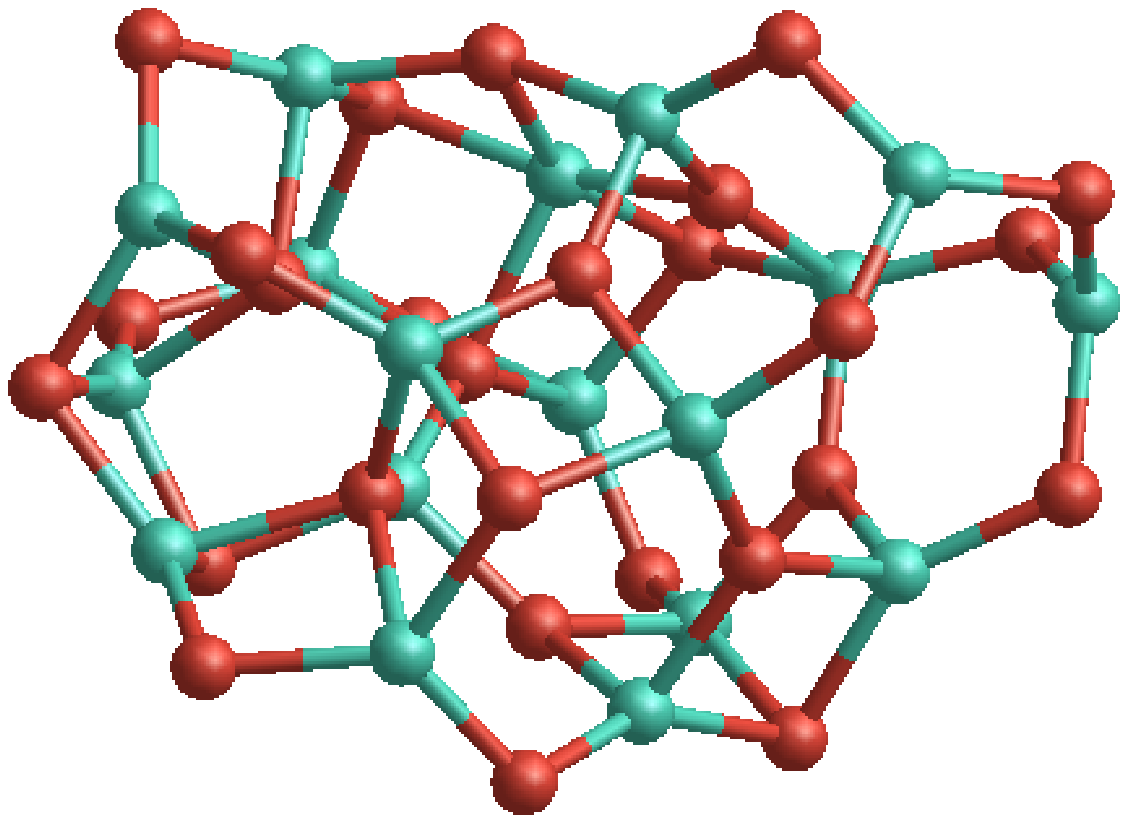}9E
	\includegraphics[scale=0.2]{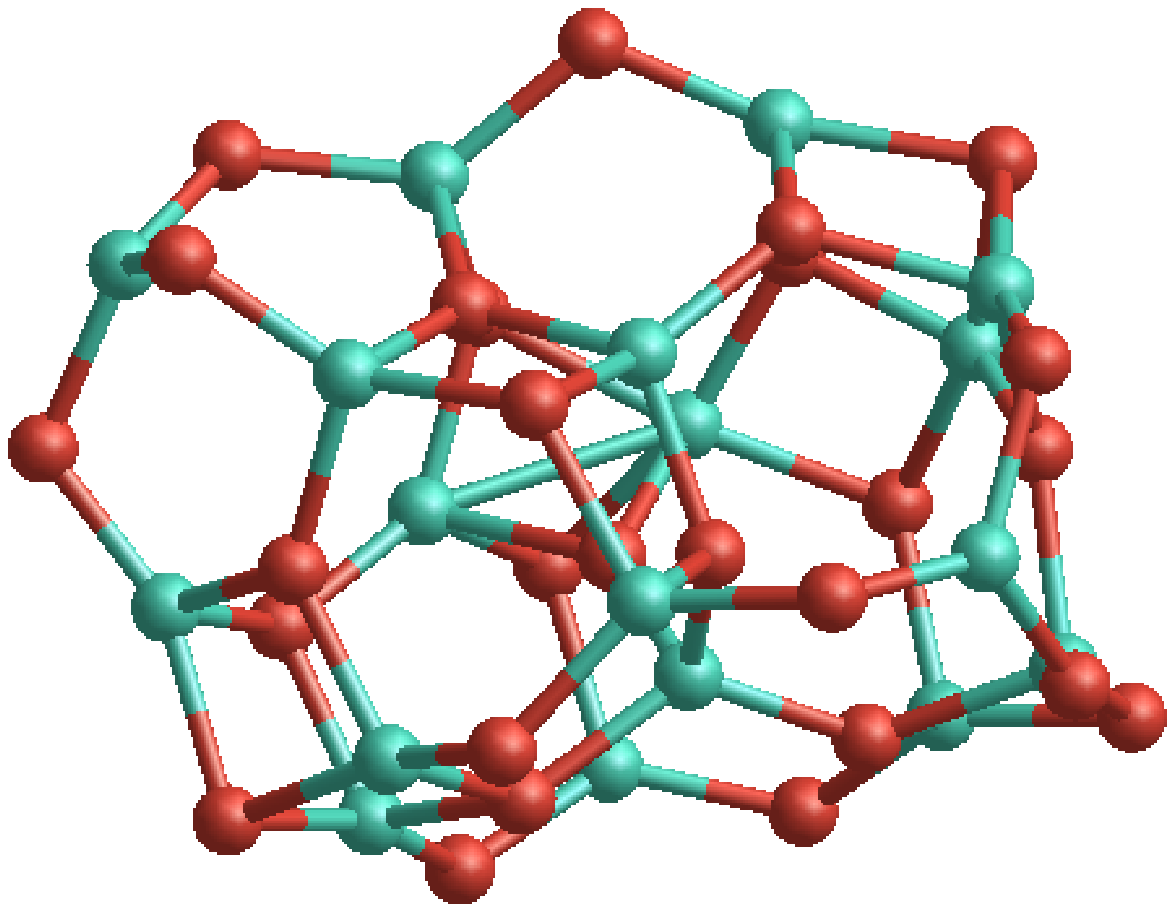}9F
	\includegraphics[scale=0.2]{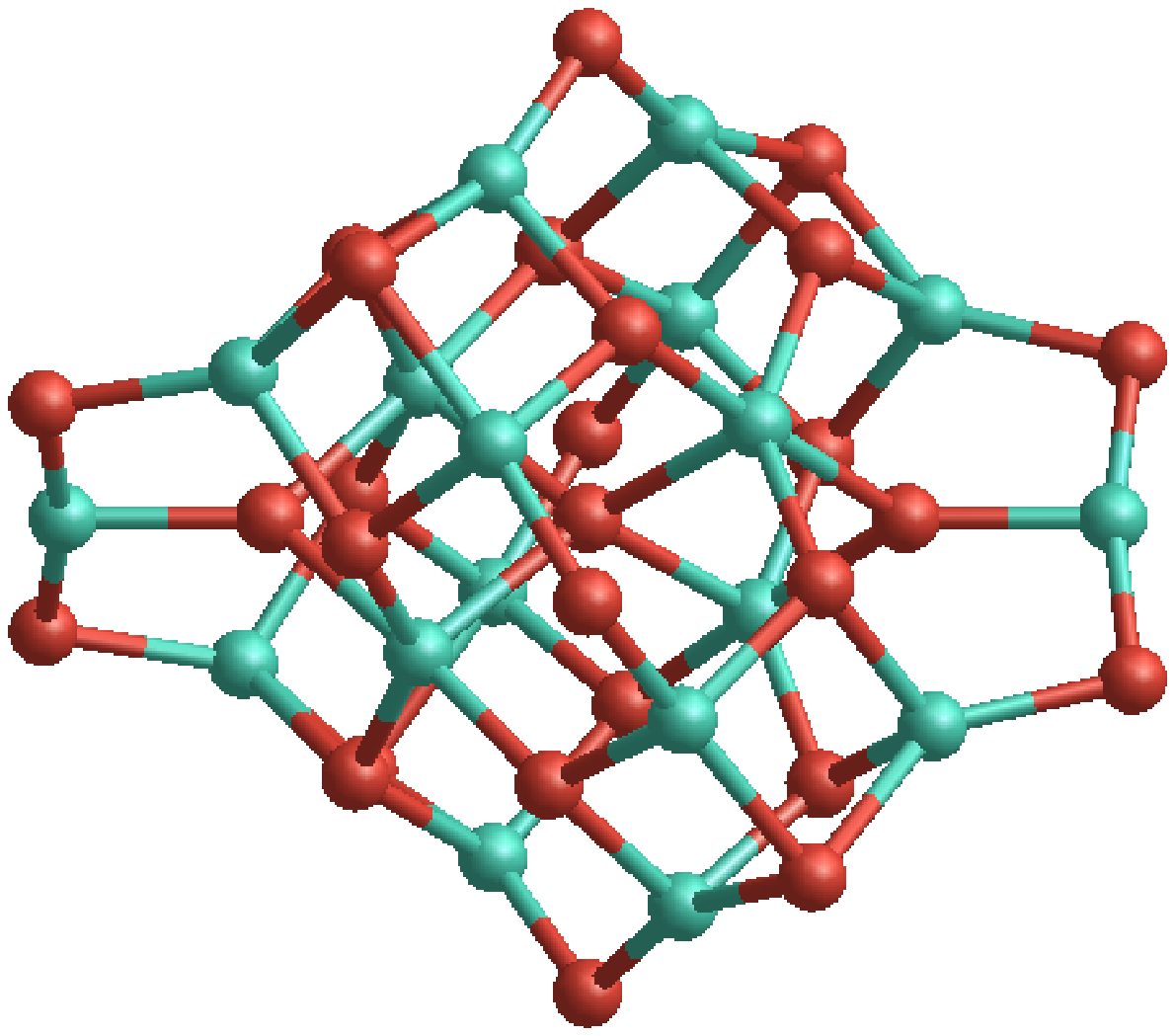}9G
	\caption{Low-energy isomers of the alumina nonamer (Al$_2$O$_3$)$_{9}$\label{nonamers}.}
\end{figure}

The most favourable isomer for $n=9$ is a C$_s$ symmetric structure depicted in Figure 
\ref{nonamer}. Its overall shape resembles a tetrahedron with a 4-coordinated oxygen atom in the center. 
This cluster structure strongly resembles a truncated block of $\alpha$-alumina also showing 4-
coordinated oxygen atoms. 
With two 3-coordinated exceptions, the Al atoms are 4-coordinated and are located
at the surface of the cluster. By the method of mirror images one could artificially increase the size 
of 9A resulting in 6-coordinated Al atoms and 4-coordinated Al atoms with similar bond angles 
as in $\alpha$-alumina. 
For 9A, we find binding energies of 514 kJ mol$^{-1}$ (B3LYP) and  520 kJ mol$^{-1}$ (PBE0), 
respectively. 
Furthermore, we find five other hitherto unreported, energetically-metastable structures (9B, 9C, 9D, 9E, 9F) with relative energies $>$ 45 kJ mol$^{-1}$ above 9A (see Figure \ref{nonamers}).  
It is unexpected that four out of the six newly discovered low-energy isomers show a high degree of symmetry. Structure 9G corresponds to the GM candidate reported by \citet{doi:10.1021/jp2050614}.

\subsubsection*{Alumina decamer (Al$_{2}$O$_{3}$)$_{10}$}
\begin{figure}[H]
	\includegraphics[scale=0.7]{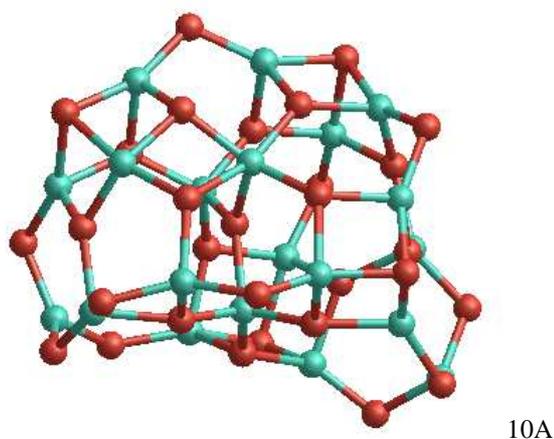}10A
	\caption{Global minimum candidate (10A) of the alumina decamer (Al$_2$O$_3$)$_{10}$\label{decamer}.}
\end{figure}

We present a new GM candidate isomer for $n=10$ without a particular symmetry (space group C$_1$, see Figure \ref{decamer}).
On the B3LYP and PBE0 level of theory, the binding energy of 10A is 515 kJ mol$^{-1}$ and 521 kJ mol$^{-1}$, respectively. 

\begin{figure}[H]
	\includegraphics[scale=0.2]{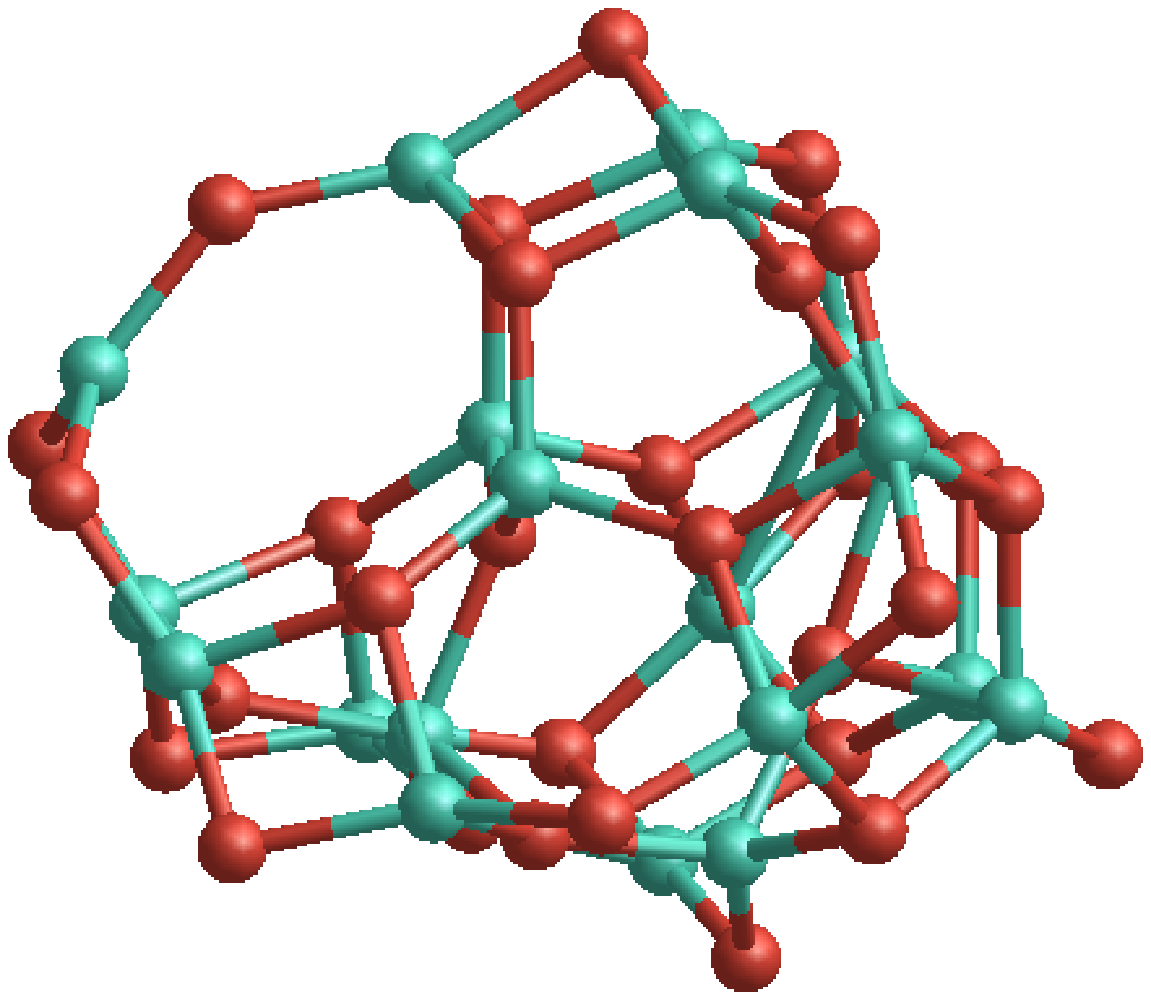}10B
	\includegraphics[scale=0.2]{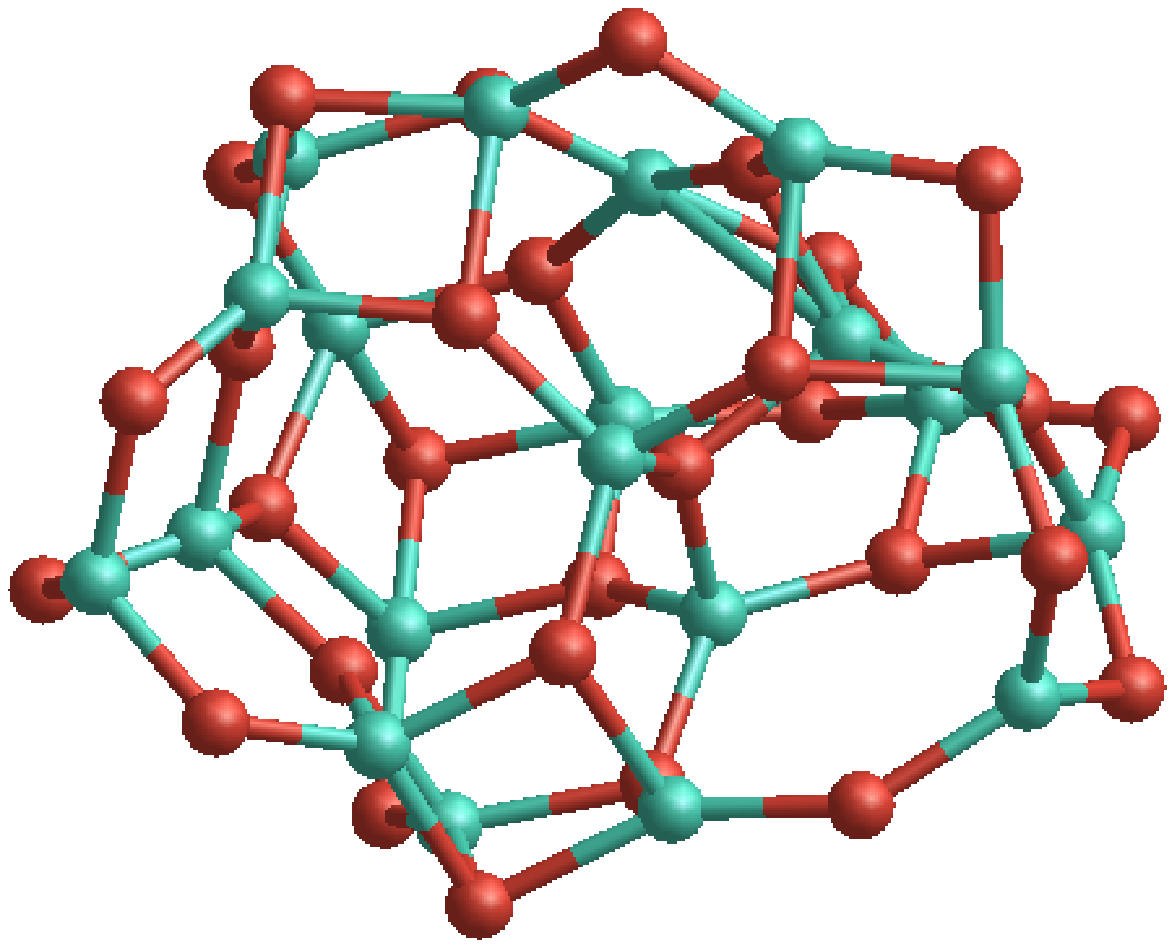}10C
	\includegraphics[scale=0.15]{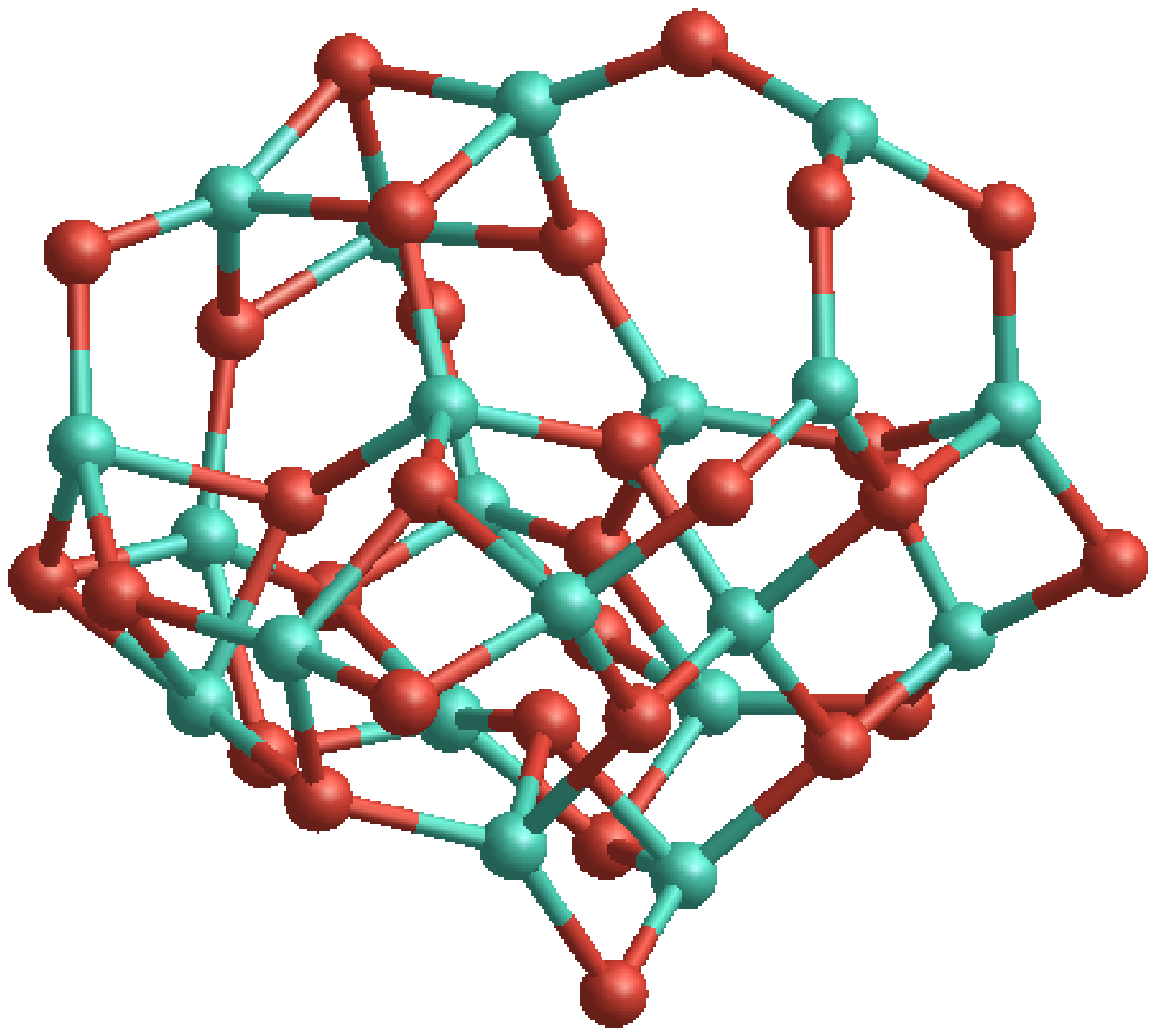}10D
	\includegraphics[scale=0.2]{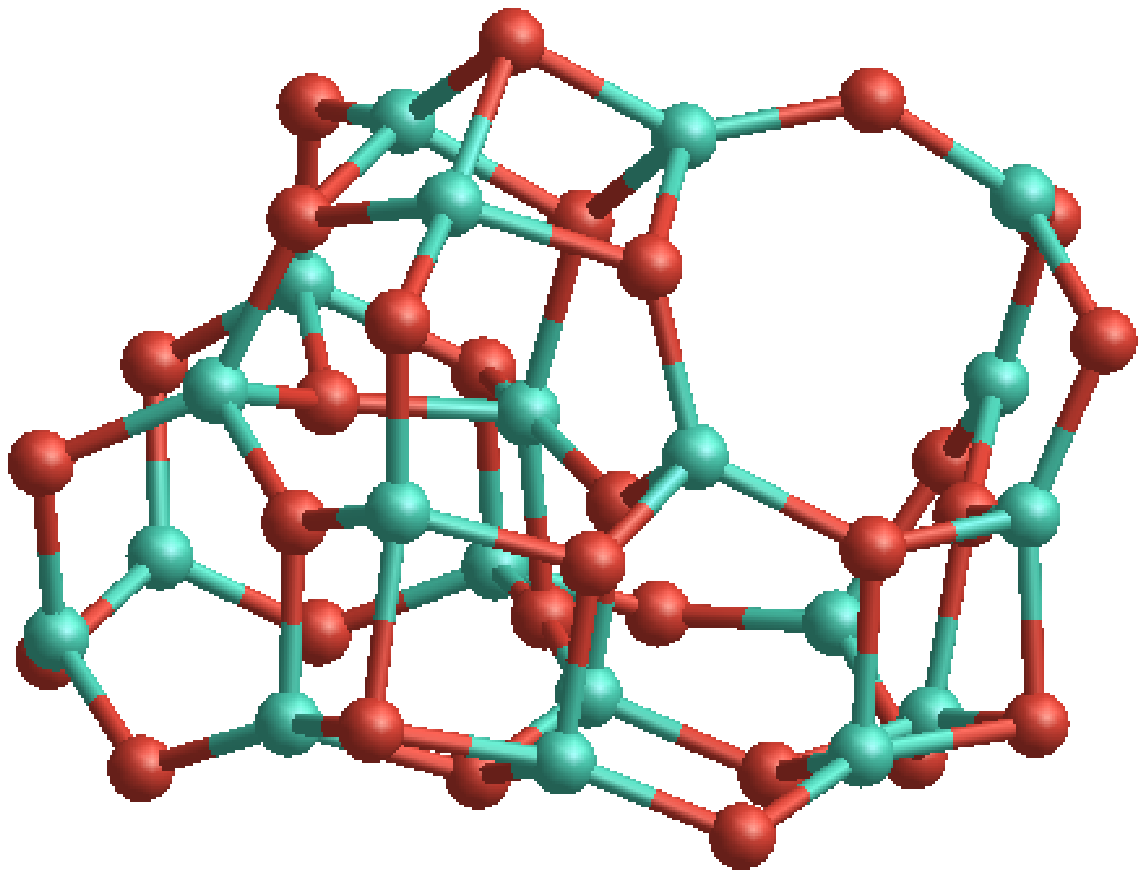}10E
	\includegraphics[scale=0.2]{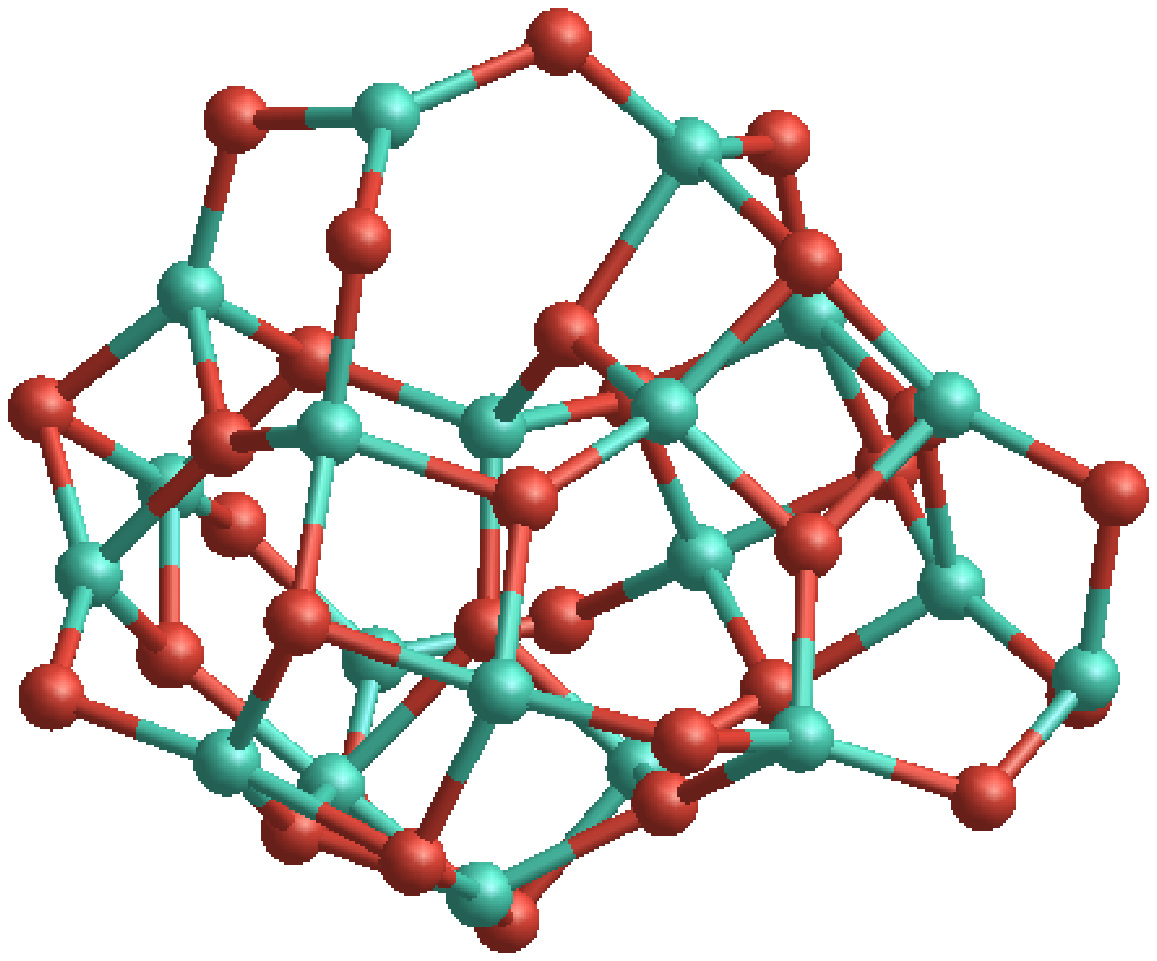}10F
	\includegraphics[scale=0.2]{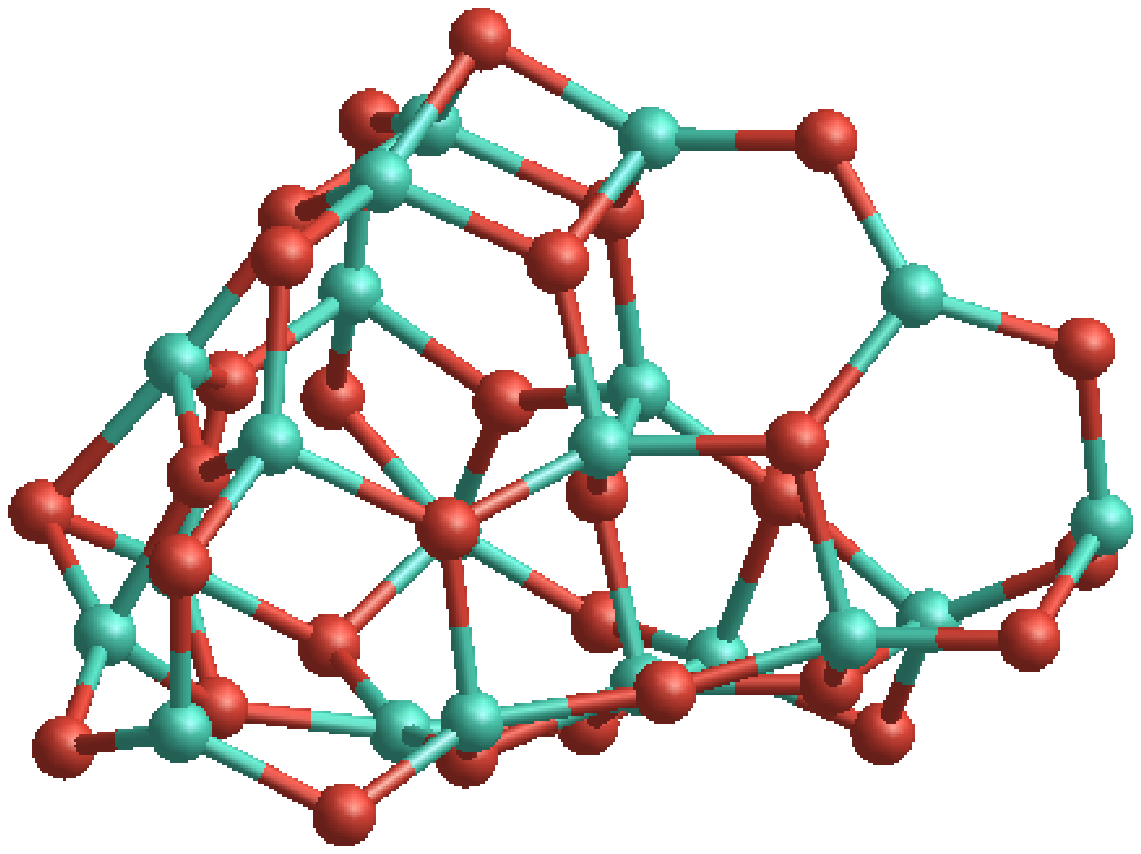}10G
	\caption{Low-energy isomers (\textit{Top}:10B, 10C, 10D, \textit{Bottom}: 10E, 10F, 10G) of the alumina decamer ((Al$_2$O$_3$)$_{10})$\label{decamerS}.}
\end{figure}

Our search also resulted in 22 further structures that lie below the GM candidate of \citet{doi:10.1021/jp2050614} having a relative energy of 18 kJ mol$^{-1}$ with respect to 10A.
We show the next higher-lying decamer isomers (10B, 10C, 10D, 10E, 10F, 10G) in Figure \ref{decamerS} to give 
the reader an impression of the structural complexity, the narrow energy spacing, and the extent of our searches. 
The energetic ordering and the relative energies of the investigated alumina decamers are largely independent of the used functional (B3LYP und PBE0). 

\subsection{Thermodynamic viability of the monomer formation}
\label{viability}
We investigate several chemical-kinetic formation routes towards the monomers (Al$_2$O$_3$) and dimers (Al$_4$O$_6$)
alumina clusters. The dimers can coagulate to tetramers (Al$_8$O$_{12}$) marking the end point of our chemical-kinetic description. The subsequent cluster nucleation and growth is treated homogeneously as function of cluster size $n$ (see Section \ref{Nucleation}).
A recent study investigated the kinetic formation of alumina tetramers via elementary reactions by means of TST and RRKM calculations \citep{SABA2021535}. 
However, in contrast to our current study, the authors do not consider the nucleation at astrophysical sites like circumstellar envelopes.
Consequently, \citet{SABA2021535} do not include hydrogenated aluminium oxides, descibe the oxidation reactions by O$_2$ and O only, and uses the high pressure limit.\\

\begin{figure}[H]
  \includegraphics[scale=0.4]{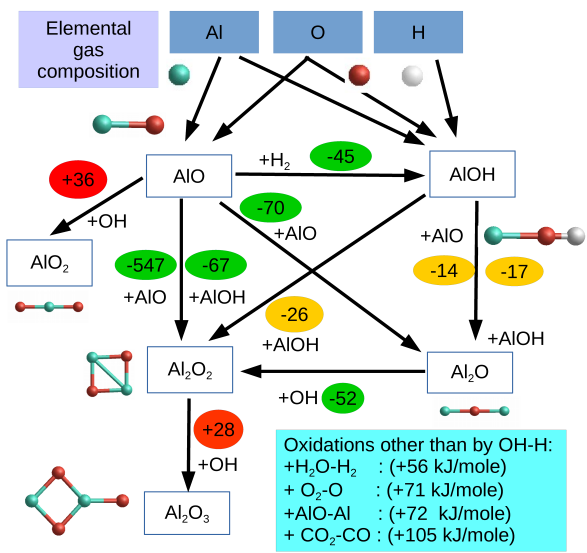}
\caption{Reaction scheme: Formation routes towards the alumina monomer (Al$_{2}$O$_{3}$). Reaction enthalpies are given in kJ mol$^{-1}$ and colour-coded according to their values. Red corresponds to supressed exothermic reactions ($\Delta$H$>$0 kJ mol$^{-1}$) , yellow to slightly exothermic reactions ($\Delta$H$>-$30 kJ mol$^{-1}$), and green to very exothermic reactions ($\Delta$H$<-$30 kJ mol$^{-1}$)\label{Reactscheme}.}
\end{figure}

As a first step we asses the viability of a reaction according to its reaction enthalpy and its temperature-dependent Gibbs free reaction energy as derived from the CBS-QB3 calculations. 
In contrast to the previous subsections, the \textit{reaction} energy or enthlapy corresponds to the sum of the total energies,which are not normalised to the atoms.
The structural viability is subsequently investigated by reaction trajectory calculations as descibed in Section \ref{rrkm}, accounting for geometrical rearrangements (i.e. breaking and formation of chemical bonds).
In the end, it is the kinetics and related energy barriers controlling the chemistry. 
However, the energetic viability (i. e. the exogonicity) is a necessary but insufficient 
prerequisite for a chemical reaction to occur. These principal reaction viabilities are addressed in the following.
The reaction scheme linking these molecular precursors to the Al$_2$O$_3$ monomer is displayed in Figure \ref{Reactscheme}. All considered  Al$_x$O$_y$H$_{z}$ molecules and clusters up to the size of a dimer are linked to each other by various chemical reactions that can proceed in both directions, forward and backward (see Figures \ref{Reactscheme} and \ref{dimerdirect}).
AlO can react with itself to form Al$_2$O+O, AlO$_2$+Al, or, Al$_2$O$_2$ via the termolecular channel ($-$548 kJ mol$^{-1}$). 
The first product channel towards Al$_2$O+O is 
exothermic by 70 kJ mol$^{-1}$ at 0 K and becomes less favourable (endergonic) around 1800 K. The products AlO$_2$+Al are suppressed for all temperatures by at least 100 kJ mol$^{-1}$.\
AlOH might also react with itself in order to form Al$_2$O$_2$ and H$_2$ which has a heat of reaction of $-$21 kJ mol$^{-1}$.
However, at temperatures above 300 K the latter process is suppressed as it becomes increasingly endergonic. Moreover, this latter 
reaction involves the breaking of two O$-$H bonds and the recombination of two H atoms to H$_2$. 
Instead, it is more likely that two AlOH molecules form Al$_2$O + H$_2$O ($\Delta$ H$_r$(0K) = $-$26 kJ mol$^{-1}$). The latter 
process becomes endergonic above 700 K.
AlO and AlOH can also react with each other to form either Al$_2$O$_2$+H or Al$_2$O+OH. 
The channel AlO$_2$+AlH is not considered here as it has a large reaction endothermicity ($\sim$ 300 kJ mol$^{-1}$) and it involves 
multiple bond breaking (Al$-$O and O$-$H). Also, the possible products OAlOH and Al are suppressed by energy barriers ($>$ 70 kJ mol$^{-1}$) and by structural hindrance.
Therefore, the primary products of AlO and AlOH are Al$_2$O and Al$_2$O$_2$.\\ 

The oxidations in Figure \ref{Reactscheme} are described by reactions with the prevalent species H$_2$O and OH, but 
also with CO$_2$, and O$_2$. 
In terms of electronic energies (at $T=0$ K), oxidations by OH are most favourable.
If the oxidation reaction proceeds instead via H$_2$O, the reaction enthalpy is +56.5 kJ mol$^{-1}$ higher.
Oxidations by O$_{2}$ (+71.0 kJ mol$^{-1}$) and CO$_2$ (+105.5 kJ mol$^{-1}$) are less energetically favourable than 
oxidations by OH. 

To form the monomer (i. e. Al$_2$O$_3$), Al$_2$O or Al$_2$O$_2$ need to be oxidised. 
The double oxidation of Al$_2$O by water has been suggested by \citet{10.1093/mnras/stu647}. 
Moreover, \citet{2016A&A...585A...6G} described a kinetic formation of the monomer by Al$_2$O$_2$+H$_2$O. 
However, the oxidation of Al$_2$O$_2$ to the alumina monomer, Al$_2$O$_3$, is hampered by substantial endothermicities. 
In the case of an oxidation by H$_2$O, the reaction is endothermic by $\Delta H_{r}$=+85 kJ mol$^{-1}$. 
Though the oxidation by OH is less endothermic (+28 kJ mol$^{-1}$), the reverse reaction (Al$_2$O$_3$+H) proceeds on much faster timescales owing to the large abundance of atomic hydrogen (H).
From an energetic perspective, the comparatively stable compounds Al$_2$O and Al$_2$O$_2$ represent energetic 
bottlenecks in the synthesis of the alumina monomer. Hence, an efficient alumina formation route does not involve Al$_2$O and Al$_2$O$_2$, or, does not proceed via the monomer.\
We consider also the possibility that the monomer forms via the species AlO$_2$. The reaction enthalpy of AlO$_2$+AlOH to Al$_2$O$_3$ and 
H is exothermic by 72 kJ mol$^{-1}$ at 0 K and stays exothermic up to $T=2700$ K.
However, the formation of AlO$_2$ itself is hampered. The oxidation of AlO by OH is endothermic by 36 kJ mol$^{-1}$ as discused above. AlOH is even harder to oxidise and requires for OH an energy of at least 80 kJ mol$^{-1}$. Oxidations by other species such as H$_2$O, CO$_2$ and O$_2$ have even larger endothermicities as indicated in Figure \ref{Reactscheme}.\  
Depending on the environment conditions, H$_2$O, CO$_2$, and O$_{2}$ might be more abundant than OH, and can thus 
increase their reaction fluxes and oxidation efficiencies. Furthermore, for temperatures $T>0$, vibrational and rotational contributions of the reacting species can change the order of the favoured oxydiser. The oxidation of Al$_2$O$_2$ to Al$_2$O$_3$
is endothermic at 0 K, but at 4000 K the free energies of reaction would allow for oxidations by OH and CO$_2$, but not by H$_2$O and O$_2$. Oxidations by atomic O are energetically most favourable. However, for molecules and small clusters, such reactions require a third body M as catalyst to absorb the excess energy and stabilise the reaction product. 
This is a consequence of the low pressures prevailing in AGB circumstellar envelopes making an autocatalysis unlikely. 
Larger-sized clusters ($n\ge$4) can be oxidized by atomic O more easily, since their density of states is high allowing for ro-vibrational relaxations without the presence of a third body M. 
Thus, although these oxidations by atomic O are the most exothermic ones, they proceed rather slowly 
for the smallest species, representing the starting point of a bottom-up approach, and require sufficiently high gas densities. For a typical termolecular association rate these gas densities need to be of the order of $>10^{14}$ cm$^{-3}$ in order to compete with other bimolecular reactions.

\subsection{Chemical-kinetic network}    
Starting with the atoms Al, O, and H and assuming a pure atomic gas phase mixture, the molecules AlO and AlOH can form via 
\begin{equation}
\centering
\mathrm{Al+O+M \rightarrow AlO+M}  
\label{al+o+}
\end{equation}
\begin{equation}
\centering
\mathrm{AlO+H+M \rightarrow AlOH+M},
\label{alo+h}
\end{equation}

\noindent where M denotes an inert molecule acting as a catalyst and removing the reaction excess energy.
The reaction {Al+OH+M~$\rightarrow$~AlOH+M} is in principle also possible, once OH has formed and is available as a reagent.
As an alternative to a purely atomic gas, thermodynamic (chemical) equilibrium abundances can also be used as a starting point. However, the choice of the initial conditions (atomic versus thermodynamic equilibirum (TE) abundances) has only minor effects on the final abundances, as will be shown in Section \ref{non-eq}.
From RRKM unimolecular decomposition calculations of AlO and AlOH, we fit and deduce kinetic rates for the reverse processes of reactions \ref{al+o+} and \ref{alo+h}. We find activation barriers of 477 kJ mol$^{-1}$  and 466 kJ mol$^{-1}$ for the dissociation of AlO and AlOH, respectively.
Atomic Al can also react with OH in a bimolecular reaction to form AlO
\begin{equation}
\centering
\mathrm{Al + OH \rightarrow AlO + H}.
\label{al+oh}
\end{equation}

A series of trajectories were run on reaction \ref{al+oh}, using the Atom Centered Density Matrix Propagation  (ADMP) molecular dynamics model \citep{doi:10.1063/1.1514582}. 
Because of the severe change in reduced mass of the system (H is one of the products), the reaction dynamics are constrained even though 
the reaction is exothermic by 72 kJ mol-1. 
The reactions takes place on both singlet and triplet surfaces. 
Although there is an energy barrier on the triplet surface, the barrier is `late' (i.e. between the intermediate AlOH and the products {AlO+H}) and, importantly, the barrier height is 50 kJ mol$^{-1}$ below the height of the reactants {Al + OH}.
There are two requirements for successful reaction: 
the OH must be vibrating with at least 2 quanta of vibrational energy (corresponding to a vibrational temperature of $\sim$2$\times$5300 K); and the collision energy needs to be modest (to allow angular momentum to be conserved), corresponding to a kinetic temperature of 500 K or less. These two requirements are counter to each other: high temperature is needed for a significant OH vibrational population, but low temperatures to favour the modest collision energy. 
The rate coefficient was then constructed by multiplying a typical collision frequency (5 $\times$ 10$^{-10}$ cm$^3$ s$^{-1}$) 
by the probability that the collision energy is less than 1.5 k$_{B}$T where T=500 K, and the probability that the OH has at least 2 vibrational quanta. This results in the Arrhenius expression of reaction 161 in Table \ref{network}. 
The alternative product channel AlH+O is very endothermic (+125 kJ mol$^{-1}$) and thus not included in our network.\\  
Once formed, the concentrations of AlO and AlOH are primarily regulated by the reaction

\begin{equation}
\centering
\mathrm{AlO + H_2 \rightarrow AlOH + H}.
\label{alo+h2}
\end{equation}

This reaction is exothermic by $-$45 kJ mol$^{-1}$. 
Despite its exothermicity, experiments lead by \citet{osti_7157939} showed a negigible reactivity of AlO with H$_2$ and hence, give an upper limit of the bimolecular rate constant of 5$\times$10$^{-14}$ cm$^3$s$^{-1}$ at room temperature. A more recent study \citep{Mangan2021} experimentally determined the reaction rate of \ref{alo+h2} including a detailed characterisation of the PES, showing good agreement with the theoretical study of \citet{doi:10.1021/jp111826y}.  
The products AlOH + H form predominantly through H atom abstraction via a linear Al$-$O$-$H$-$H transition state lying 45 kJ mol$^{-1}$ above the reagents {AlO+H$_2$}. 
Despite the reaction barrier of 45 kJ mol$^{-1}$, the best fit with respect to the experimental data points results in a lower barrier height of 31 kJ mol$^{-1}$. 
Another product of reaction \ref{alo+h2} is Al+H$_2$O. 
However, its branching ratio is negligibly small (0.008\%).
The resulting kinetic rate for reaction \ref{alo+h2} is temperature-dependent and has an activation energy corresponding to an equivalent temperature of 2030 K.\\
 
Another process impacting the AlO/AlOH balance is the reaction

\begin{equation}
\centering
\mathrm{AlO + H_{2}O \rightarrow AlOH + OH},
\label{alo+h2o}
\end{equation}

having a reaction enthalpy of +11.6 kJ mol$^{-1}$.
\citet{Mangan2021} found that this reaction proceeds predominantly via an AlO$-$H$_2$O adduct with a small submerged barrier re-arranging to Al(OH)$_2$, which finally dissociates to AlOH and OH without a barrier. 
A direct pathway via a quasi-linear transition state involves a barrier of 60 kJ mol$^{-1}$ and is not competitive. 
An alternative product channel of \ref{alo+h2o} is OAlOH + H showing also a slightly endothermic reaction enthalpy (+8.6 kJ mol$^{-1}$). Since a direct reaction route via a OAlOH$_2$ transition state involves a significant barrier of 87 kJ mol$^{-1}$,
the dominant OAlOH production channel takes course via the dissociation of Al(OH)$_2$, similar to in the AlOH production channel. The formation of AlO$_2$ and H$_2$ is very endothermic.\\

The reaction of AlO+OH can also form AlOH. However, although it is exothermic by 56 kJ mol$^{-1}$, it was found to proceed rather slowly. 
ACMP trajectories on the triplet surface (which connects with the products AlOH + O($^{3}$P) in their electronic ground states)
show that this reaction requires the AlO to have at least 1 quantum of vibrational excitation corresponding to a vibrational temperature of 1370 K.  
A low collision energy is also required, corresponding to a kinetic temperature of 600 K or less, and the reaction cross section 
has a relatively small impact parameter ($<$ 0.5 \AA{}). 
The resulting rate coefficient between 1000 and 2000 K is around 3 $\times$ 10$^{-12}$ cm$^3$ s$^{-1}$ with a small temperature dependence. 
The alternative reaction channel forming AlO$_2$ + H is endothermic by 35 kJ mol-1 and is characterised by a large decrease in reduced mass, 
and so is not competitive with the AlOH + O product channel.
Finally, we examine alumina dimer (i.e. Al$_4$O$_6$) formation pathways that do not involve the unfavourable alumina monomer (see Section \ref{viability}).
We find an enhanced stability of (AlO)$_x$, $x=1-4$, clusters with an Al:O stoichiometry of 1:1, compared
to small Al$_x$O$_y$, $x,y=1-4$, $x\ne y$ clusters with a different stoichiometry than 1:1 (see Table \ref{ALXoYs}). These findings are consistent 
with the results of \citet{2005EPJD...32..329P} and \citet{2015JPCA..119.8944L}.
Consequently, it is instructive to consider reactions within the stability valley of the (AlO)$_x$, $x=$1$-$4, clusters. 
The successive addition of AlO molecules  

\begin{equation}
\centering
\mathrm{Al_{x}O_{x} + AlO + M \rightarrow Al_{x+1}O_{x+1} + M}
\end{equation}

\noindent is energetically favourable as the number of strong Al$-$O bonds ($\sim$ 500 kJ mol$^{-1}$ per bond) increases naturally with size $x$. 
However, this process requires a third body M and is therefore only effective in the densest circumstellar regimes (typically $n_{gas} > 10^{14}$ cm$^{-3}$), or once x is large enough.  

\begin{figure}[H]
  \includegraphics[scale=0.4]{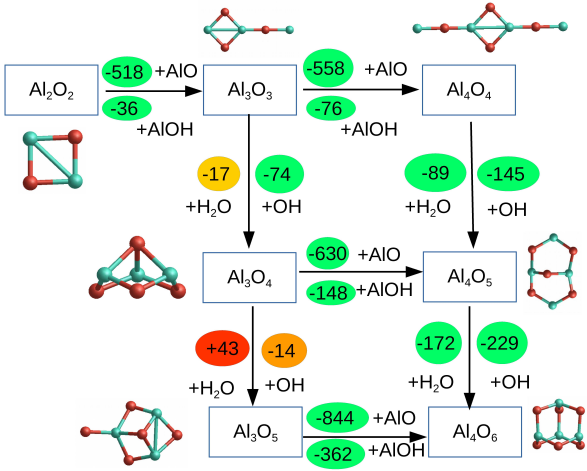}
\caption{Reaction scheme: Formation routes towards the alumina dimer. Reaction enthalpies 
are given in kJ mol$^{-1}$ and the color-coding is the same as in Figure \ref{Reactscheme}, where very exothermic reactions are colour$-$coded in green, slightly exothermic reactions in orange, and endothermic reactions in red \label{dimerdirect}.}
\end{figure}

The successive addition of AlOH molecules
\begin{equation}
\centering
\mathrm{Al_{x}O_{x} + AlOH \rightarrow Al_{x+1}O_{x+1} + H}
\end{equation}

\noindent proceed as bimolecular reactions, but have significantly lower heat of reactions (lower by 482 kJ mol$^{-1}$ at $T=0$ K), corresponding to the dissociation energy of {AlOH~$\rightarrow$~AlO+H} as compared to the AlO addition, since the ejection of an H atom is energetically expensive.    
Nevertheless, for $x=$1$-$3, the reaction with AlOH is energetically viable with enthalpies of $-$65.9 kJ mol$^{-1}$ ($x=$1), 
$-$36.0 kJ mol$^{-1}$ ($x=$2), and $-$76.0 kJ mol$^{-1}$ ($x=$3). 
For $x=$4 (Al$_4$O$_4$+AlOH $\rightarrow$ Al$_5$O$_5$+H) the enthalpy is still slightly exothermic by $-$3.9 kJ mol$^{-1}$, but at higher temperatures ($>$ 100 K) the reaction becomes endothermic/endergonic. It is thus probable that a subsequent cluster growth reaction departs from an Al:O stoichiometry of 1:1.    
Therefore, starting with Al$_2$O$_2$ (see Figure \ref{dimerdirect}) the formation of Al$_3$O$_3$ is energetically viable.  
The oxidations of Al$_3$O$_3$ and Al$_4$O$_4$ are both exothermic and can readily form Al$_3$O$_4$ and Al$_4$O$_5$. However, the oxidations of Al$_3$O$_4$ represent an energetic bottleneck, since only the oxidation by OH is energetically viable.\\

Once the alumina dimer (Al$_4$O$_6$) is formed, we assume that it coagulates to the alumina tetramer (Al$_8$O$_{12}$) in a very exothermic reaction.
The alumina dimer can also react with other Al-bearing species than itself to form products containing 5 to 7 Al atoms. Due to the exothermicity of many cluster growth reactions, these processes are likely to occur. A characterization of these intermediate Al-O-H cluster species and related kinetic pathways will be the subject of a future study.  However, owing to the exponentially increasing number of possible reaction pathways, we do not explicitly take these intermediate species into account in this study, but subsume them by an association reaction. This allows us to neatly link the kinetically controlled regime with the larger-sized (Al$_2$O$_3$)$_n$ clusters with n$\ge$4.


\subsubsection{Kinetic rate evaluation}
For large reactive systems, it is very expensive to calculate the PES including all possible and unknown transition states. Therefore, we assess kinetic rates involving more than 6 atoms with collision and capture theory.
Exothermic reactions without a barrier of the form A+B can proceed at the collision frequency, which is given as
\begin{equation}
\centering
k_{coll}(T) = \sigma_{AB}\sqrt{\frac{8k_{B}T}{\pi\mu_{AB}}}
\label{kcoll} 
\end{equation} 

where $\sigma_{AB}=\pi(r_{A}+r_{B})^2$ is the (geometrical) cross section, $\mu_{AB}=\frac{m_{A}m_{B}}{m_{A}+m_{B}}$ is the reduced mass 
and k$_{B}$ 
the Boltzmann constant. The radii r$_{A}$ and r$_{B}$ of species A and B are determined from the calculated structures. 
In Table \ref{cradii}, we provide two sets of cluster volumes and radii. The first set is derived from atomic 
core coordinates and calculated by Delauney triangulation. The second set exhibits larger values, as it includes also ``electron interaction'' volumes derived from atomic van-der-Waals radii. 
Owing to its T$^{0.5}$ temperature dependence, the collisional frequency is increased by a factor of $\sqrt{10}\simeq$3.16 for a characteristic (circum-)stellar temperature of 3000 K, as compared to room temperature. At these elevated temperatures the collision rate can take large values in many cases and is rather simplistic. 
In the kinetic network shown in Table \ref{network}, we included two collision rates proceeding via the formation of intermediates that decompose in either direction without a barrier (denoted as \textit{collision}). 
However, the majority of the reactions considered in this study have a more complex PES and are not well described by simple collision theory.\\
In addition to a pure geometric collision also long-range interactions 
between the reactants can contribute to the cross section. These interactions can be taken into account by using van der Waals radii, which are larger than the geometric radii and account for the spatial occupancy of the bond atoms. 
Generally, the long-range interaction effect can also be included directly in the rate expression.   
For example, in the reaction AlO+H$_2$ dipole-induced dipole forces or Debye forces, denoted as k$_{DiD}$, and dispersive or London forces, denoted as k$_{disp}$, act. If both reagents are polar (which is not the case for H$_2$) also dipole-dipole forces, denoted as k$_{DD}$, are present. 
To quantify and summarise the contribution of the long-range forces we use the form introduced by \citet{doi:10.1063/1.1899603}:  

\begin{equation}
\centering
k_{capt}(T)= 1.3 \times \max_{k}
     \left\{
     \begin{array}{l}
      k_{DD}\\
      k_{DiD} \\
      k_{disp}
      \end{array}
\right.  
\end{equation}

where $k_{capt}$ is the so called capture rate and corresponds to the upper limit for the rate of an exothermic reaction.
The detailed calculation of k$_{DD}$, k$_{DiD}$, and k$_{disp}$ including the factor of 1.3 can be found in \citet{doi:10.1063/1.1899603}.
In addition to the mass of the reagents, the calculation of capture rates requires the knowledge of the dipole moments, approximate polarisabilities, and vertical ionisation potentials.
In our network (in Table \ref{network}), the capture rate is used for 46 exothermic reactions (denoted as \textit{capture}), where the investigation of the PES is too expensive and TST- or RRKM-based rates are not obtainable.
The majority of the capture rates are dominated by dispersive forces and show a T$^{\frac{1}{6}}$ dependence. 
Dipole-dipole interactions, which follow a T$^{-\frac{1}{6}}$ dependence, are predominant for just three rates in our kinetic network.\\
In order to calculate the rates for the reverse (endothermic) reactions, we apply the principle of detailed balance (for more details see Eq. \ref{equil_constant}).
By fitting the reverse rate, we find Arrhenius-parametrized rate expressions (denoted as \textit{detailed balance} in Table \ref{network}).
In some cases the fitting results in unphysically high pre-exponential factors A of $>$2 $\times$ 10$^{-9}$ cm$^3$s$^{-1}$.
The large pre-exponential factors tend to occur in reactions where the reduced mass of the products is much smaller than the reactants (typically when one of the products is H or H$_2$). In these cases the conservation of angular momentum constrains the reaction cross section.
In such cases, we do not use the unphysically large detailed balance rates but apply a capture rate with energy barrier corresponding to 
the CBS-QB3 0K enthalpy (denoted as \textit{reverse capture} in Table \ref{network}). 

The basic chemical network consists of 54 atomic, molecular and cluster species and 163 individual reactions that can be found in Table \ref{network} of the Appendix.
We are aware that this network is not complete in terms of species (e.g. sulphur-containing species) and processes (e.g. ionisation) considered. 
However, we ran test calculations with an extensive kinetic network including 50 additional non-aluminum-bearing species and 286 additional reactions.  
Moreover, we carefully build up a chemical-kinetic network by using either rates from our extensive literature search, 
or, where these do not exist estimating rate coefficients using sound theoretical methods currently available.
Thereby, we always respect the balance between forward and backward reaction, instead of adopting rates from (astro-)chemical kinetic rate databases. 
Note that some rate expressions differ from the simple Arhenius parametrisation, as their complex temperature-dependence cannot be represented by an Arrhenius formulation.  
Furthermore, some rates are based on the Lindemann expression consisting of high- and low-pressure limiting rate terms. In circumstellar envelopes we can safely ignore the high pressure limit and use the low-pressure limit for the rate constant, as the prevailing pressures are orders of magnitude below 1 atm. 
Consequently, dissociations typically proceed via bimolecular channels as collisions with a body M.
However, a photolysis or photodissociation, i.e. a decay induced by high-energy photons, is also possible.
Finally, we include the photodissociations of AlO, AlOH, and Al$_2$O.
The photolysis rates of AlO, AlOH, and Al$_2$O were estimated using time-dependent density function theory (TD-DFT) \citep{1996CPL...256..454B}. 
The vertical excitation energies and transition dipole moments were calculated for transitions from the ground state of each molecule up to the first 30 electronically excited states. 
The resulting absorption cross section for each molecule was then convolved up to its dissociation threshold with a model stellar irradiance flux from the MARCS data-base for an evolved star with T$_\star$ = 2500 K \citep{2008A&A...486..951G}.
Oxygen-rich AGB stars typically show moderately lower effective temperatures (T = 2000$-$2400 K), but there are no MARCS models with T$<$2500 K available. 
The photolysis thresholds were set to correspond to the bond dissociation energies, giving 252, 238 and 213 nm for AlO, AlOH and Al$_2$O, respectively. 
Note that these are upper limits to the dissociation threshold wavelengths, because a photon with more than the bond energy may be required depending on the position of the upper dissociating electronic state of the molecule. 
The photolysis rate was then computed as a function of temperature by red-shifting the photolysis threshold to reflect the increasing internal energy of the molecule with temperature. The resulting photodissociation rates at 1 stellar radius are listed in Table \ref{network}.
For radial distances further out in the envelope, we assume a geometrical dilution ($\propto R^{-2}$) of the stellar radiation field, and no attenuation by dust.

\subsection{Equilibrium abundances}
\label{equ}
To compare our chemical-kinetic derived non-equilibrium abundances (see Section \ref{non-eq}) with thermodynamic equilibrium (TE) abundances, 
we perform calculations with the chemical equilibrium software GGchem \citep{2018A&A...614A...1W}. 
Moreover, we add the alumina-related cluster species presented in this study to the list of molecules. 

For convenience, we provide the fitting parameters a, b, c, d, e in the form presented in \citet{2018MNRAS.479..865S} for the term $\frac{-\Delta_f G(T)^0}{RT}$, i.e.   

\begin{equation}
\centering
\frac{-\Delta_f G(T)^0}{RT} = \frac{a}{T} + b\ln(T)+ cT + dT^{2}.
\end{equation}

The fitting parameters a, b, c, d, e for all aluminium-bearing species used in this study are provided in Table \ref{dgfits} of the Appendix.
Note that $\Delta_f G(T)^0 \neq \Delta_f G(T)$, since $\Delta_f G(T)$ includes, like the JANAF-NIST 
thermochemical tables, a scaling to the atomic (elemental) heats of formation with their corresponding stoichiometric factors. Thus, $-\Delta_f G(0)^0$ corresponds to the real, unscaled free energies of formation at 0 K.
We perform TE calculations for different C/O ratios of 0.4, 0.7, and 1.0, and for a typical atmospheric (photospheric) pressure of 10$^{-5}$ bar as a function of temperature pertaining to the dust formation zone in oxygen-rich AGB stars.

\begin{figure}
\includegraphics[scale=0.7]{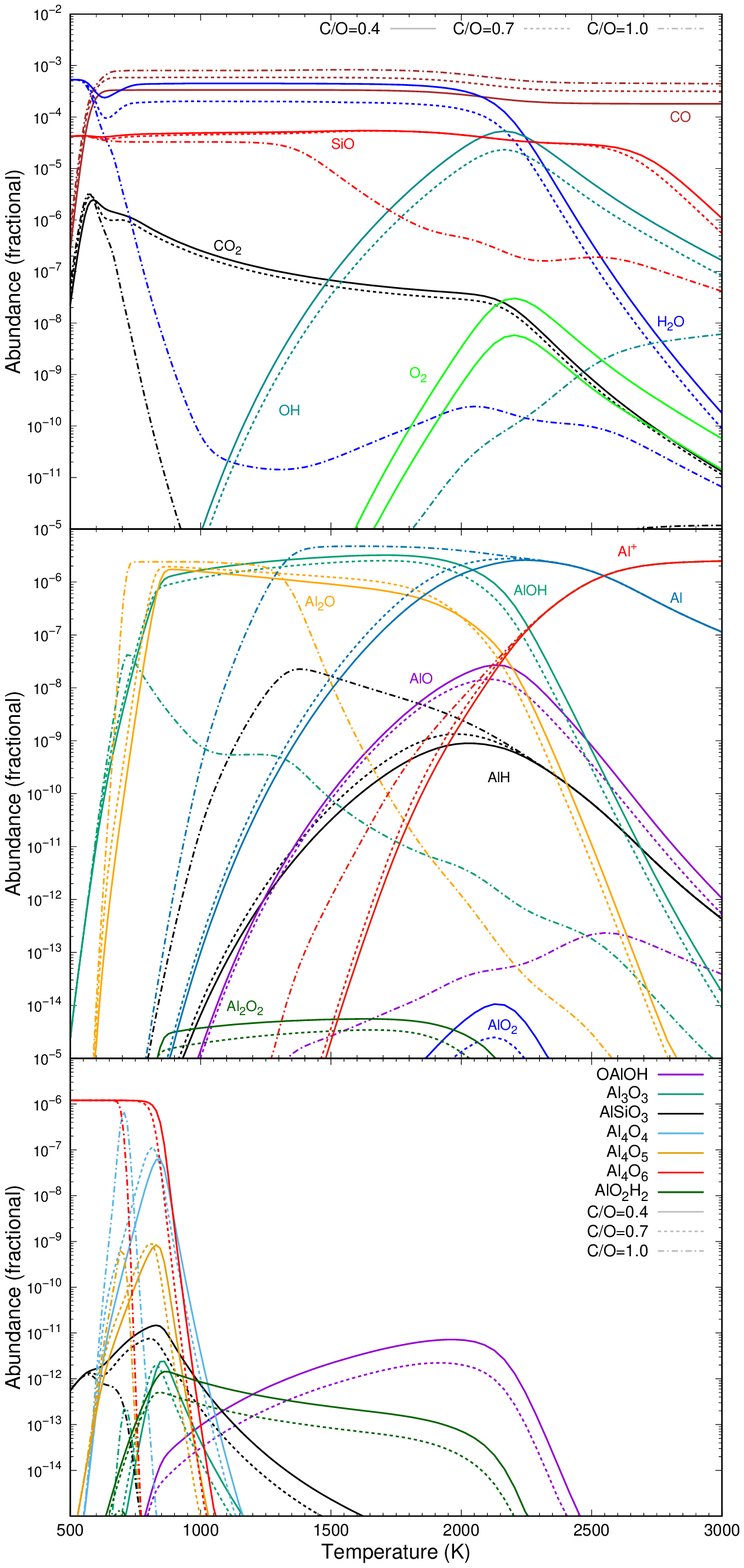}
\caption{\textit{Upper panel}: Thermodynamic equilibrium (TE) abundances of the abundant molecules CO, CO$_2$, H$_2$O (black), OH, and O$_2$ 
as a function of temperature for different C/O ratios. 
\textit{Middle panel}: TE abundances of the aluminium-bearing molecules AlO, AlOH, Al$_2$O, and  AlO$_2$ as well as Al and 
Al$^{+}$, as a function of temperature for different C/O ratios. 
\textit{Lower panel}: TE abundances of the aluminium-bearing clusters OAlOH, Al(OH)$_2$, Al$_3$O$_3$, Al$_4$O$_4$,
Al$_4$O$_5$, Al$_4$O$_6$, and AlSiO$_3$ participating in the formation of the alumina dimer (Al$_{4}$O$_{6}$) as a function of temperature for different C/O ratios. Straight lines correspond to C/O=0.4, dashed lines to C/O=0.7, and dashed-dotted 
line to C/O=1.0, respectively. The pressure is kept constant at 10$^{-5}$ bar. \label{maineq}.}
\end{figure}

First, we present the abundances of the prevalent molecular species CO and SiO including the oxidation agents as OH, H$_2$O, CO$_2$,
and O$_2$ (see upper panel of Figure \ref{maineq}).    
The most abundant molecules (apart from H$_2$) in oxygen-rich conditions are CO and H$_2$O 
showing fractional abundances above 10$^{-4}$. At temperatures T$\ge$2200 K the OH and H abundances increase at the cost of H$_2$O.
SiO is also fairly abundant with a value of $>$10$^{-5}$ up to T=2700 K. CO$_2$ and O$_2$ play a comparatively minor role hardly exceeding 10$^{-6}$ on the abundance scale. 
CO$_2$ is moderately abundant at lower temperatures (T$\le$2200), and O$_2$ only at higher temperatures. Note that CO$_2$ and O$_{2}$ not only show lower equilibrium abundances than H$_2$O and OH, but also the corresponding kinetic oxidations 
are comparatively less exothermic in the kinetic models (see Figure \ref{Reactscheme} and related discussion). Therefore, H$_2$O and OH should represent the primary oxidation species.   
The C/O=1.0 models (dot-dashed lines), characterising atmospheres of chemical transient S-type AGB stars, show generally lower abundances of the prevalent, oxygen-bearing molecules, except for CO, where the abundance is higher. This result is not unexpected in equilibirum conditions, as a C/O close to unity leaves little oxygen available to form other molecules.  
  
The TE abundances of the aluminium-bearing molecules are shown in the middle panel of Figure \ref{maineq}. 
They can be roughly grouped in three temperature zones: 
at high temperatures above T$\ge$2500 K, aluminium is predominantly in ionized form (Al$^{+}$), 
in the intermediate temperature range (T$=$2000$-$2500 K) atomic Al is the primary aluminium carrier, 
and for lower temperatures (T$\le$2000 K)) the molecules AlOH and Al$_2$O are the most abundant gas phase species.
The TE AlO abundance reaches its maximum ($\sim$10$^{-8}$) around T=2130 K whereas AlO$_2$ has negligible abundances. 
In the C/O=1.0 case, AlOH is shifted towards lower temperatures, whereas AlO peaks at higher temperatures with respect to the C/O=0.4 and C/O=0.7 cases. 

In the bottom panel of Figure \ref{maineq}, clusters acting as potential intermediates (OAlOH, Al$_3$O$_3$, AlSiO$_3$, Al$_4$O$_4$, Al$_4$O$_5$, Al$_4$O$_6$, AlO$_2$H$_2$) in the cluster nucleation route are included.    
At temperatures below T$<$800 K the equilibrium alumina dimer (Al$_4$O$_6$) abundance dominates the aluminium content. The steep decrease beyond T$=$800 K is accompanied with a sharp rise of the species AlOH and Al$_2$O (see middle panel). 
Moreover, except for OAlOH and AlO$_2$H$_2$, the abundances of the intermediates are negligible for temperatures above T$=$1000 K.\\
In summary, under the conditions of thermodynamic equilibrium, alumina dimers (and larger compounds with same stoichiometry) do not form in significant quantities at temperatures above T$=$800 K. This is inconsistent with observations of the higher temperatures found in the dust formation zone. Therefore, chemical equilibrium fails to predict the onset of alumina dust formation, assuming that the nucleation proceeds via the stable and promising dimer cluster.




\subsection{Non-equilibrium chemical-kinetic models}
\label{non-eq}
For the non-equilibrium computations we use a chemical-kinetic reaction network (instead of assuming chemical equilibrium), which is given in the Appendix \ref{network}. 
Furthermore, we expose the chemical species to different hydrodynamic trajectories characterized by varying thermodynamic conditions (i.e. density and temperature).
The trajectories include a non-pulsating outflow model represented by a $\beta$-velocity law, and a pulsation model that will be discused in more detail. Moreover, we distinguish between two model stars, SRV representing Semi-Regular Variable AGB stars and MIRA representing MIRA-type pulsating AGB stars. Apart from their pulsation properties (amplitude and period), 
which are generally shorter for SRVs than for MIRAs, other quantities also differ for these two types of AGB stars. 
They include temperature, density, and mass-loss rate. Following stellar evolution models, the differences in these quantities suggest that SRV and MIRA represent two successive evolutionary states of the same type of AGB star \citep{1992A&A...263...97K}.

\begin{table}
\caption{Parameters for the model stars SRV and MIRA used to calculate hydrodynamic trajectories. Number densities n$_0 \equiv$ n(1R$_\star$) are given in units of cm$^{-3}$, temperatures T$_0 \equiv$ T(1R$_\star$) in units of K, the terminal v$_{\infty}$ and shock velocity v$_s$ in units of km s$^{-1}$ and the pulsation period (pp) in units of days. Number densities and temperatures correspond to their values at the photosphere (1 R$_\star$)\label{models}.}
\begin{tabular}{ c | r r | r | r r}
Model & n$_0$ & T$_0$ & v$_{\infty}$   & v$_s$  & pp \\ 
\hline
SRV   & 1.e14 & 2400  &  5.7           &   10 & 332 \\
MIRA  & 4.e14 & 2000  & 17.7           &   20 & 450 \\
\end{tabular}
\end{table}

For reasons of comparability and consistency, we choose a fixed C/O ratio of 0.75. We are aware that the C/O ratio is subject to changes during the evolution on the AGB. Recurrent Third Dredge-Up (TDU) and mixing episodes gradually enhance the carbon content at the stellar surface leading to an evolution-induced increase of the C/O ratio \citep{2015ApJS..219...40C}. 
Therefore, SRV type stars are expected to have a (slightly) lower C/O ratio than MIRA type AGB stars. 
According to the FRUITY stellar evolution models, a star with an initial mass of 1.5 M$_{\odot}$ and solar metallicity (Z=0.014) experiences five TDU mixing episodes with an interpulse period (i.e. the time between two TDUs) lasting about 1$-$2$\times$10$^5$ years \citep{2015ApJS..219...40C}. During the time span of an interpulse period, the C/O ratio is constant and takes values of 0.41 (TDU 1), 0.56 (TDU 2), 0.75 (TDU 3), 0.96 (TDU 4) and 1.00 (TDU 5). 
Therefore, our initial elemental composition corresponds to third of five TDU episodes of the FRUITY model described above.
Moreover, our choice of C/O=0.75 lies in beween the solar value of 0.55 \citep{2009ARA&A..47..481A} and a C/O ratio of 0.95 characteristic for an MS type AGB star, and hence reflects a certain progression in stellar evolution.
We performed test calculations for the two cases with solar C/O ratios (0.55) and for a stellar model resembling an AGB star of MS type (C/O=0.95). 
The lower C/O slightly shifts the chemistry towards higher abundances of O-bearing species, but qualitatively preserved the difference between SRV and MIRA models. 
In the stellar model resembling an MS star, no alumina clusters form and the aluminum chemistry is predominated by the halides AlF and AlCl instead of AlOH and AlO, as expected.  
Therefore, the main conclusions from our kinetic modelling study are not altered by these tests. 

\subsection{Non-pulsating outflow model}
The non-pulsating outflow models in our study are represented by a $\beta$-velocity law.
The $\beta$-velocity law represents a stellar outflow model, where the velocity increases monotonically with the distance from the star and accounts for the net acceleration of the wind. It should be kept in mind that the beta velocity approximation does not take into account stellar pulsations.  
Its general form is given by

\begin{equation}
v(r) = v_0 +(v_\infty -v_0) \times \left(1-\frac{r_0}{r}\right)^{\beta}
\end{equation}

where v$_0$ and v$_\infty$ are the initial and terminal velocities and $r_0$ the initial radius, often referred to as dust condensation radius.
Here, we assume that $r_0$=1 R$_\star$, v$_0$=v($r_0$)=1.5 km s$^{-1}$,  and $\beta$=1.0 for both model stars, SRV and MIRA. 
Typically, $r_0$ is found to be often greater than 1 R$_\star$. 
However, we intentionally choose the stellar surface as the starting point, since we do not impose a dust condensation radius, but aim to derive the onset of dust formation by the kinetic cluster growth.
Therefore, we naturally account for a dust formation \textit{zone} rather than a dust formation \textit{radius}. 
Moreover, our choice of $r_0$  allows a more direct comparison with the pulsating models, which is one of the main purposes of this study.
We choose a value of $\beta=1.0$ for both model stars, which is in accordance with \citet{Decin2010}.
We note that the value for $\beta$ is often (slightly) larger for oxygen-rich AGB stars (see e.g. \citet{2014A&A...561A...5K,Maerker}). 
A larger value of $\beta$ implies a slower wind acceleration and consequently, a longer timescale for the chemistry to occur and for the alumina clusters / dust to form. 
Therefore, $\beta$=1 corresponds to the conservative limit of a comparatively quickly accelerating oxygen-rich AGB wind.
 
The terminal wind speed v$_{\infty}$ is dependent on the model star and has typical values of 5$-$20 km s$^{-1}$. For the SRV model we assume v$_{\infty}$=5.7 km s$^{-1}$ and for the MIRA model v$_{\infty}$=17.7 km s$^{-1}$ based on the results of \citet{Maerker} for R Dor and of \citet{refId0D} for IK Tau. 
The initial velocity v$_0$ is consistent with a microturbulent velocity of 1.1$-$1.5 km s$^{-1}$ derived from the line broadening close to the stellar photosphere \citep{2004A&A...422..651S}.
The gas temperature T is given as a function of radius and follows a power law with an exponent $\alpha$=0.6 \citep{1997A&A...324..237W}:
\begin{equation}
T(r) = T_{\star}\left(\frac{r}{R_\star}\right)^{-\alpha}
\end{equation}

The radial density profile is derived from the scale height at the stellar surface, H$_{\star}$=$R_{gas}T_{\star}R_{\star}^{2}/\mu M_{\star}G$ of an ideal gas with a parameter $\gamma$=0.89 (see e.g. \citet{2016A&A...585A...6G} and reference therein) and is given as

\begin{equation}
n(r) = n_{\star}\exp\left(\frac{R_\star(1-\gamma^2)}{H_\star(1-\alpha)}\left(1-\frac{r}{R_{\star}}\right)^{\alpha-1}\right)
\end{equation}

Since the kinetic rate network is formulated as a set of differential equations in time, we transform the monotonically decreasing temperature and density into functions of time by using a constant $dt=dr/v(r)$.

\begin{figure}
\includegraphics[scale=0.7]{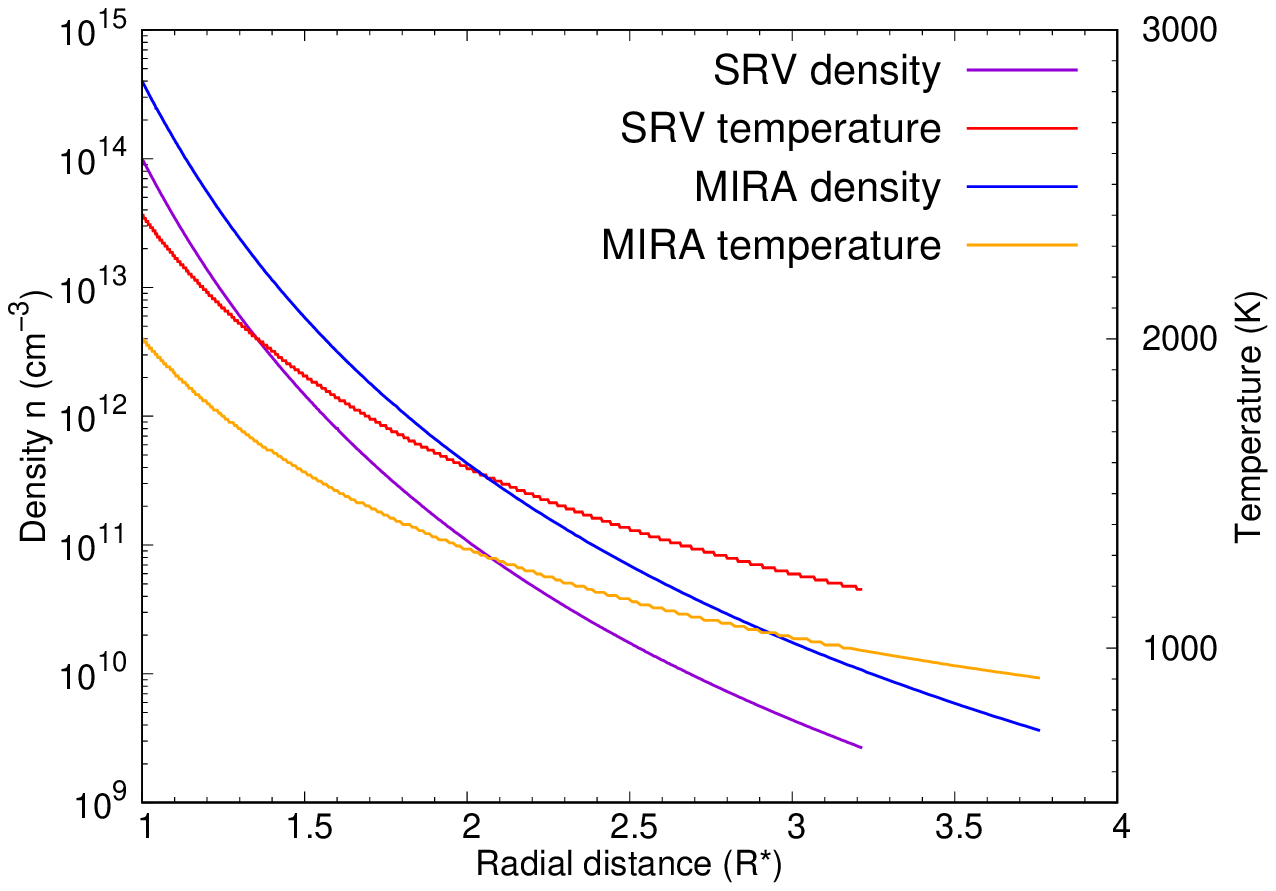}
\caption{Density and temperature profiles of the SRV and MIRA model star envelopes. These profiles apply to the non-pulsating models, as well as to the preshock conditions of the pulsating models \label{betahydro}}
\end{figure}

\begin{figure}
\includegraphics[scale=0.7]{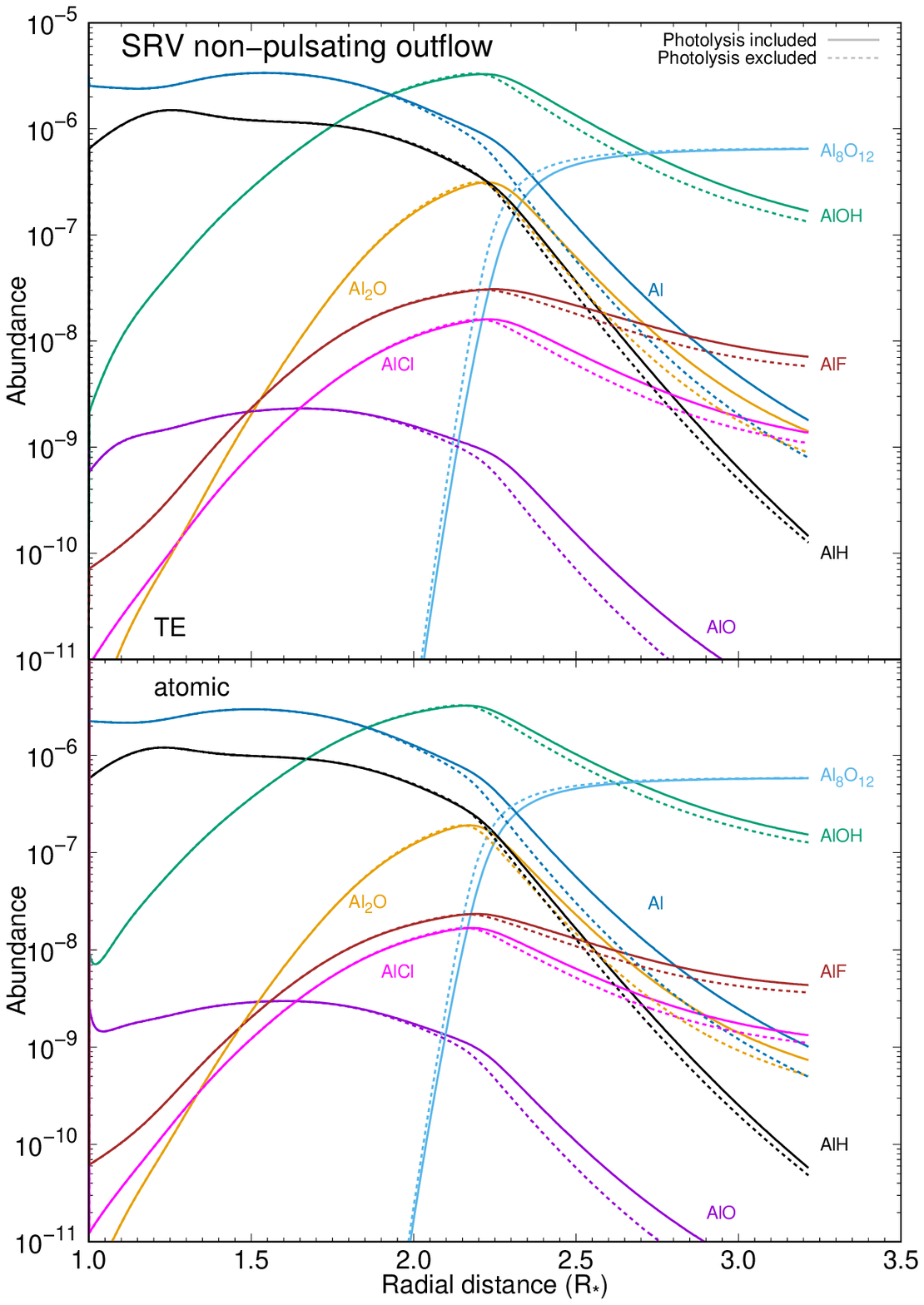}
\caption{Abundances of the prevalent aluminium-bearing species the SRV $\beta$-velocity model. Straight lines show abundances including the AlOH photolysis, dashed lines without it.\label{betaSRValu}}
\end{figure}

In the SRV non-pulsating outflow model (i.e. the SRV $\beta$-velocity law) the aluminum chemistry is dominated by atomic Al for $r<$1.8 R$_\star$, by AlOH for 1.8 R$_\star<r<$ 2.7 R$_\star$ and by alumina tetramers (Al$_8$O$_{12}$) for $r<$ 2.7 R$_\star$ (see Figure \ref{betaSRValu}). 
The exclusion of the photolysis reactions (shown in dashed lines) has a minor impact on the species abundances. Also the choice of the initial abundance distribution (TE versus atomic) does hardly affect the species abundances (top versus bottom panel).
Alumina tetramer clusters start to form around 2 R$_{\star}$ thereby impacting the other aluminum-bearing species. 
At 2.2 R$_{\star}$ a significant amount of clusters have formed, which is accompanied by the decrease of the Al-containing molecules.
AlO shows an abundance of a few times 10$^{-9}$ before the cluster formation becomes dominant. 
This is about an order of magnitude lower than predicted by recent ALMA observations \citep{refId0D}.
AlOH shows a strong increase in its abundance and reaches a maximum around 2 R$_\star$ and decreases again, when the the alumina tetramers form. In comparison to observations, AlOH is overpredcited in our SRV non-pulsating model, except for the innermost region inside 1.1 R$_\star$. AlCl reaches its peak fractional abundance in the cluster forming zone, where it agrees well with the observation for the semi-regular variable R Doradus.
In summary, the AlOH abundance is overpredicted and the AlO abundnace is underpredicted, both by about an order of magnitude, as compared with observations. Moreover, closest agreement between the SRV non-pulsating model and observations is found before and at the cluster formation zone, but not afterwards.


\begin{figure}
\includegraphics[scale=0.7]{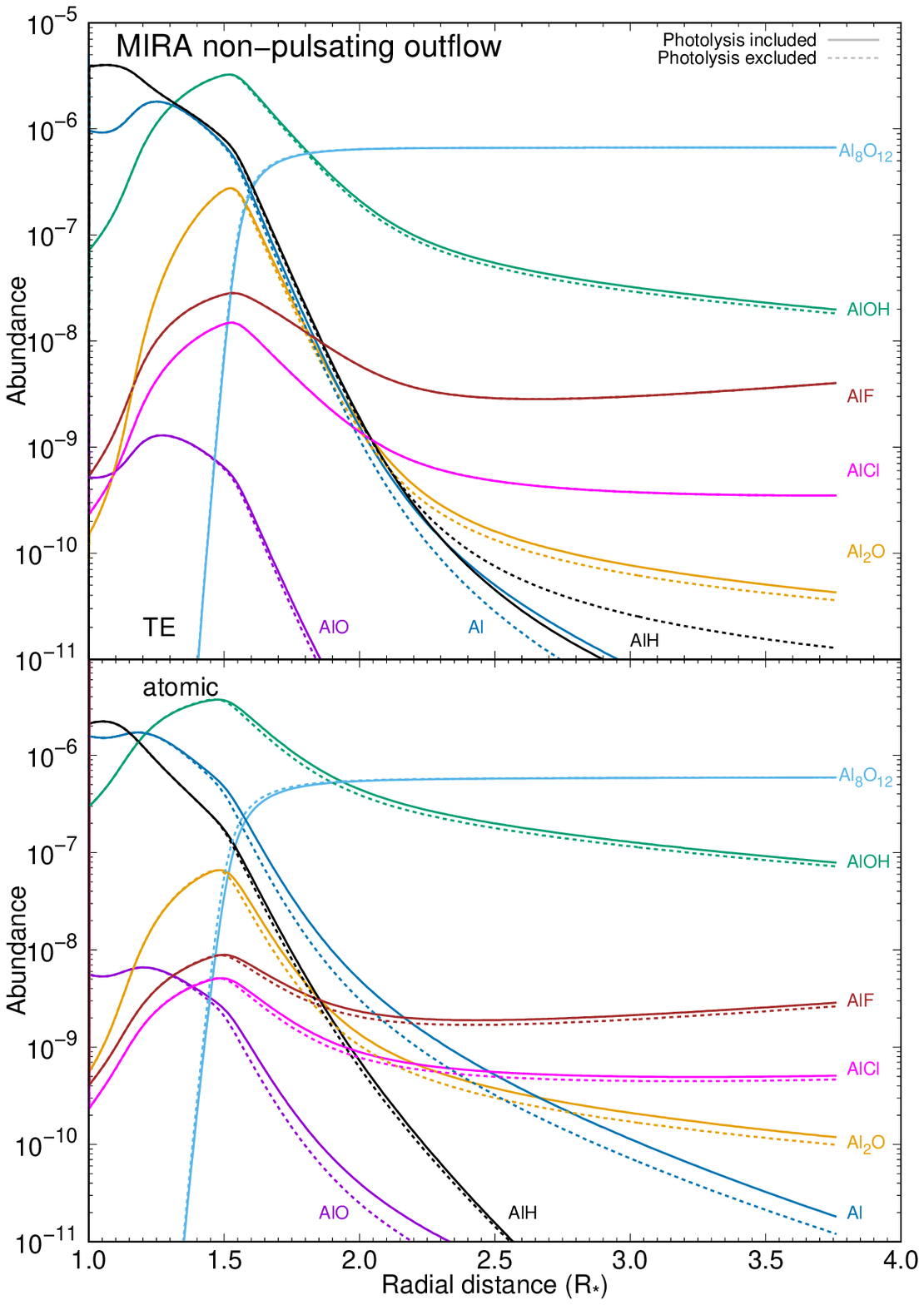}
\caption{Abundances of the prevalent aluminium-bearing species the MIRA non-pulsating outflow model. Straight lines show abundances including the photolysis, dashed lines without it.\label{betaMIRAalu}}
\end{figure}

The species abundances in the MIRA non-pulsating outflow model (i. e. $\beta$-velocity law) shown in Figure \ref{betaMIRAalu} have some common features with the SRV models, but also show some important variety. First, the molecular chemistry occurs closer to the star, which can be exemplified by the AlOH abundance peaking at 1.5 R$_\star$ and exceeding the atomic Al abundance for radii beyond 1.2 R$_\star$. This is closer to the star than in the SRV models and can be attributed to the lower temperatures and the higher gas densities in the MIRA models.
Between 1.4 and 1.7 R$_\star$ tetramer clusters form in significant amounts and dominate the Al content for $r>$2 R$_\star$, essentially independent of the photolysis. The cluster synthesis impacts the aluminium chemistry also in the MIRA non-pulsating outflow in decreasing their molecular abundances. The oxides AlO and Al$_2$O, as well as atomic Al and AlH are much stronger depleted than the aluminium halides AlF and AlCl, which even slightly increase for $r>$2.5 R$_\star$. 
Obviously, AlO, Al$_2$O, AlH, and atomic Al are molecular precursor to alumina clusters. 
AlO is only present in the innermost zones ($r<$1.8 R$_\star$) and decreses to negligible amounts, once the cluster have formed.
This is agreement with recent observation showing no AlO in the dust fromation zone / inner wind in Mira-type AGB stars like IK Tau \citep{refId0D,2020ApJ...904..110D}
We note overall larger abundances of AlO, AlOH, Al$_2$O and Al, when starting with a gas that is initially atomic (instead of assuming chemical equilibrium) for the MIRA non-pulsating models.
In comparison with \citet{refId0D}, the AlOH abundance in the MIRA non-pulsating model is a bit high showing better agreement in the TE case (upper panel in Figure \ref{betaMIRAalu}). 
 
\subsection{Pulsating model}
\begin{figure}
\includegraphics[scale=0.7]{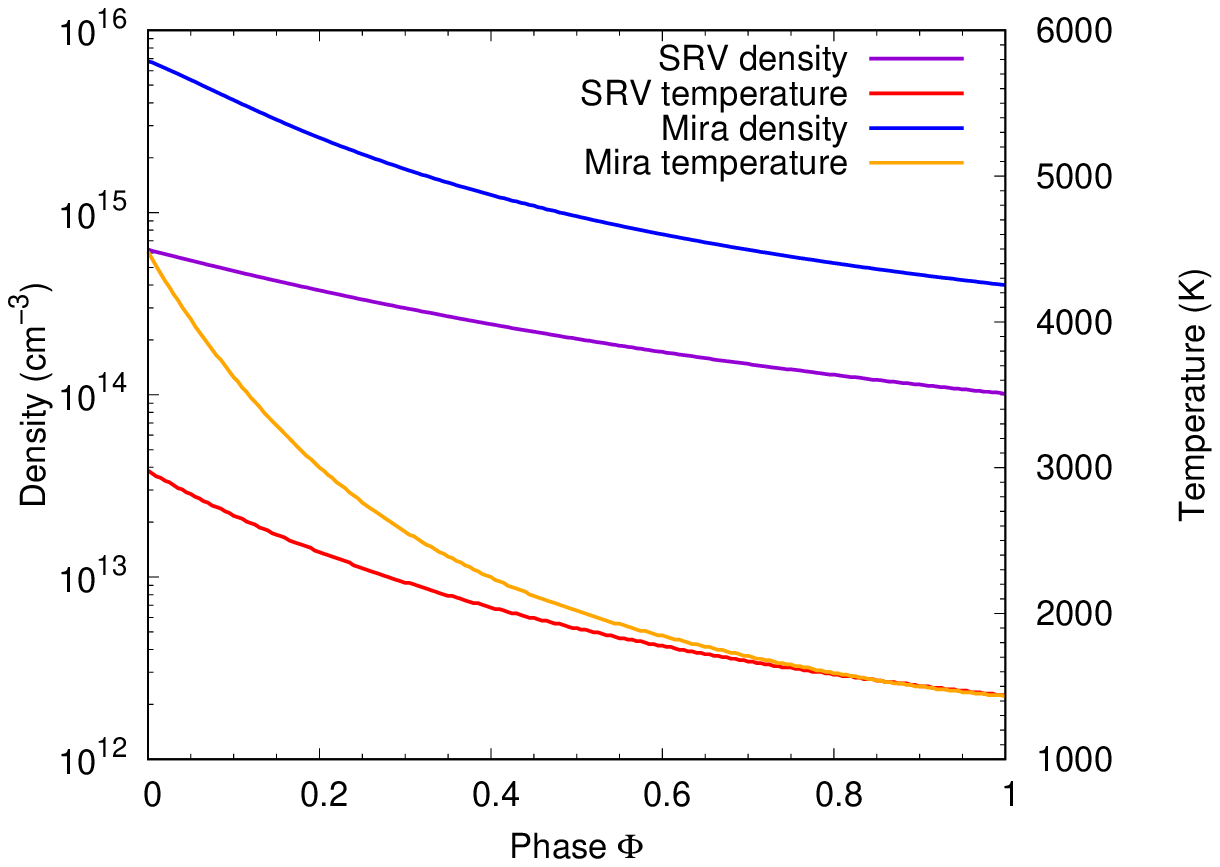}
\caption{The temperature and density profiles of the postshock gas at 1 R$_\star$ in the two models, SRV and MIRA, as a function of pulsation phase $\Phi$=t/P, where t is the time and P is the pulsation period\label{hydroL}.}
\end{figure}

In the pulsating model, we follow Lagrangian trajectories of an atmospheric gas parcel experiencing a pulsation-induced shock and a subsequent adiabatic expansion. The resulting trajectories correspond to ballistic and periodic excursions \citep{1985ApJ...299..167B}. 
This formalism has been widely used in combination with a chemical-kinetic newtork by \citet{1992ApJ...401..269C,1998A&A...330..676W,1999A&A...341L..47D,2006A&A...456.1001C,2016A&A...585A...6G} and applied to different circumstellar environments. We apply the pulsating model for both model stars, SRV and MIRA, starting at the stellar photosphere (1 R$_\star$).
In both model stars, gas density and temperatures decrease monotonically  with time (see Figure \ref{hydroL}) during one pulsation cycle. After a complete pulsation cycle the gas densities have relaxed to their pre-shock values.
Like in the non-pulsating $\beta$-velocity law, the SRV and MIRA model stars differ by their initial densities n$_0$ and temperatures T$_0$ (see Table \ref{models}). In addition to these quantities, shock velocities (defining the jumps in temperature and gas density) and pulsation periods differ for SRV and MIRA, and are required input quantities for the pulsating models. They are different for SRV and MIRA (see Table \ref{models}).
The physical timescale is given by the pulsation period, which typically lasts of the order of hundreds of days. In the case of the SRV model it is 338 days, and in the MIRA model 470 days.\\

\begin{figure}
\includegraphics[scale=0.7]{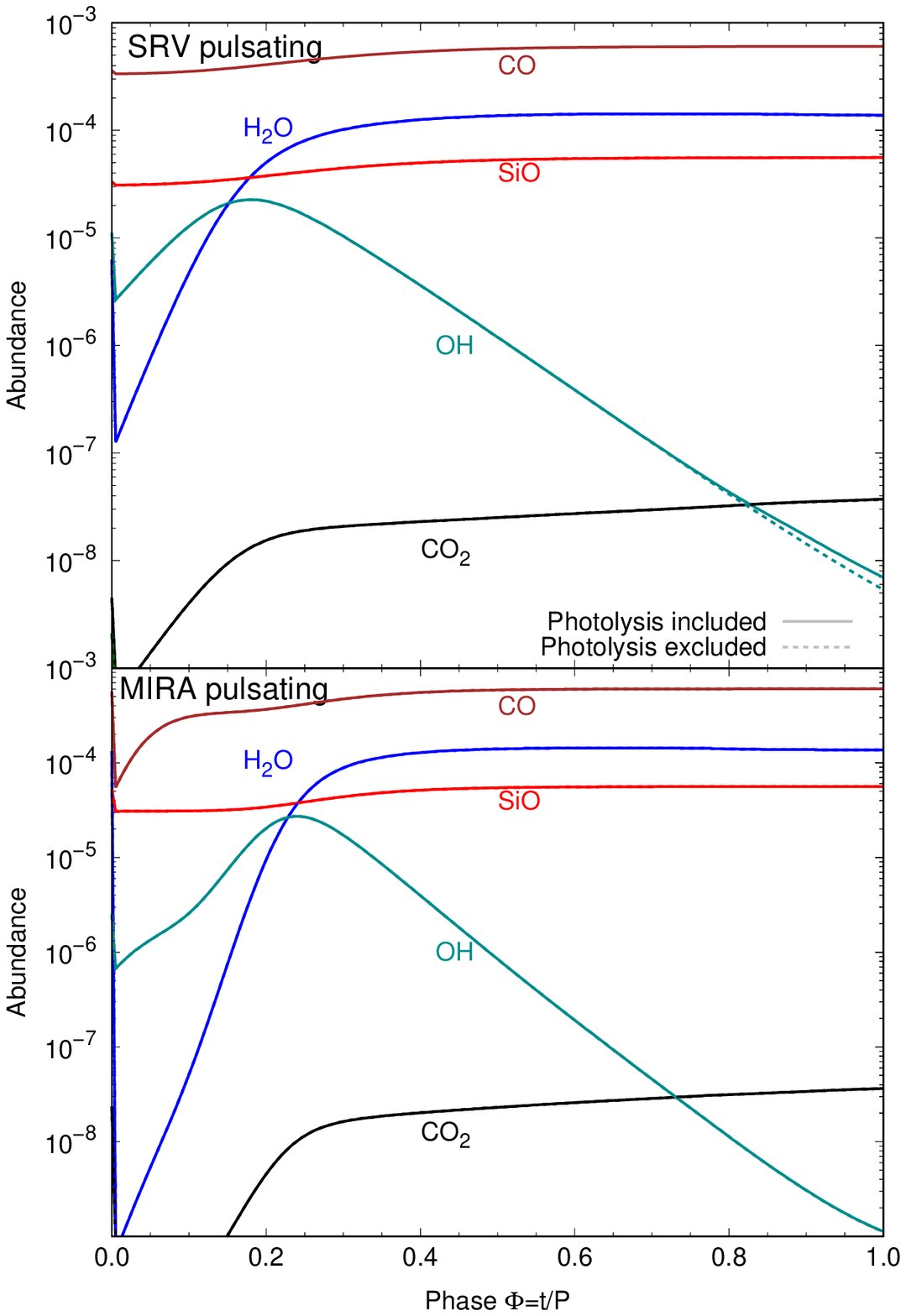}
\caption{The kinetic (non-equilibrium) abundances of the prevalent gas phase molecules in the postshock gas at 1 R$_\star$ of the pulsating models as a function of pulsation phase\label{SRVmain}. 
\textit{Upper panel}: SRV \textit{Lower panel}: MIRA}
\end{figure}

As a first step we follow the chemistry for one full pulsation cycle with a period P, starting at the immediate post-shock (phase $\Phi=$t/P=0.0) and ending just before the next shock arrives (phase $\Phi=$t/P=1.0).
The abundances of the prevalent oxygen-bearing species CO, CO$_2$, H$_2$O, OH, SiO for both models, SRV and MIRA, are displayed in Figures \ref{SRVmain}.

We find good agreement with observations and the temperature dependence of the main species also reflects the equilibrium calculations presented in Section \ref{equ}. 
Moreover, the inclusion of the three photolysis rates has only a minor effect (OH at late phase) on the prevalent species.
In both models (SRV and MIRA), similar trends with pulsation phase (time) are found. CO$_2$, H$_2$O, and CO reform in the postshock gas, whereas OH is destroyed.  
In both pulsation models, SRV and MIRA, molecular oxygen (O$_2$) 
plays a minor role with abundances not exceeding 1 $\times$ 10$^{-9}$.
On the one hand this result shows that the applied chemical-kinetic network produces good agreement with measured abundances. On the other hand, it shows that these main species are overall less affected by kinetic processes. However, this is not the case for the aluminium-bearing species including the (molecular) precursors of alumina dust, as will be shown below.


\subsubsection{SRV pulsating model}
\begin{figure}
\includegraphics[scale=0.7]{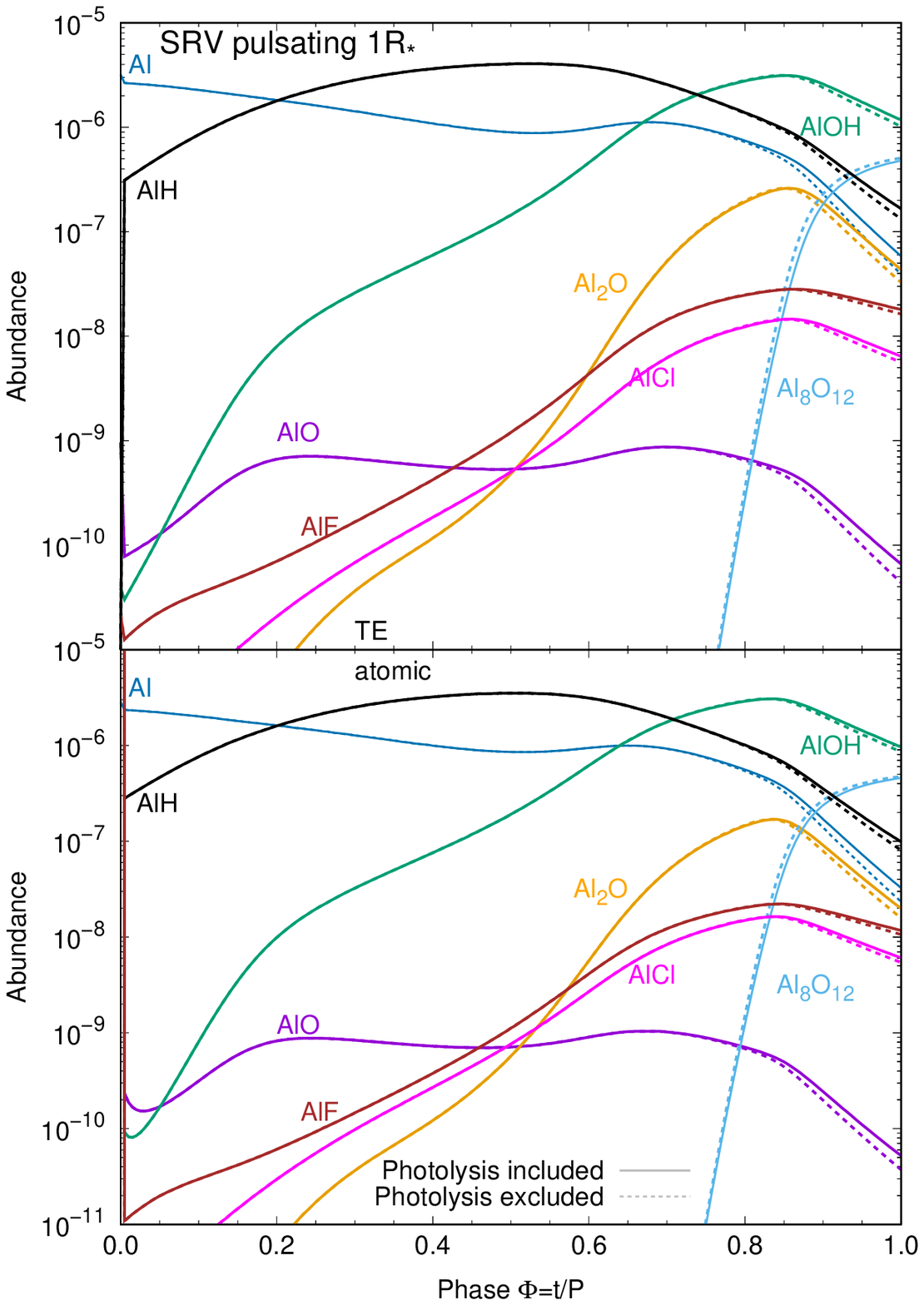}
\caption{The kinetic abundances of aluminum-bearing species in the postshock gas at 1 R$_\star$ of the pulsating SRV model as a function of pulsation phase $\Phi$. Dashed lines show the abundances, when the photodissociation is excluded.\textit{Upper panel}: initial abundances are given by chemical equilibirum (TE). \textit{Lower panel}: initial abundances correspond to a pure atomic gas composition.\label{SRV1R}}
\end{figure}

In the SRV pulsating model at 1 R$_\star$ (see Figure \ref{SRV1R}), aluminium hydride (AlH) and atomic Al controls the aluminium content in the early post-shock gas ($\Phi<0.6$) 
and AlOH is the most abundant aluminum-bearing species at later phases $\Phi>0.6$. 
A few dozen days later, at $\Phi>0.75$, alumina tetramers start forming and account for 71.8 \% of the aluminium content at $\Phi= 1.0$ (AlOH accounts for 22.1 \%).
Note, that an alumina tetramer contains 8 Al atoms, whereas the molecules AlO and AlOH bear just one Al atom. These stoichiometries need to be taken into account, when inspecting the species contributions to the total aluminum content. 
Before the clusters are forming, the AlO abundance at 1 R$_\star$ is of the order of few times 10$^{-10}$, and decreases afterwards. This is still about two orders of magnitude lower than observed, but also still very close to the star. In contrast, 
AlOH shows large abundances at late phase ($\Phi>0.7$) exceeding the observational abundance by 2 orders of magnitude. 


\begin{figure}
\includegraphics[scale=0.7]{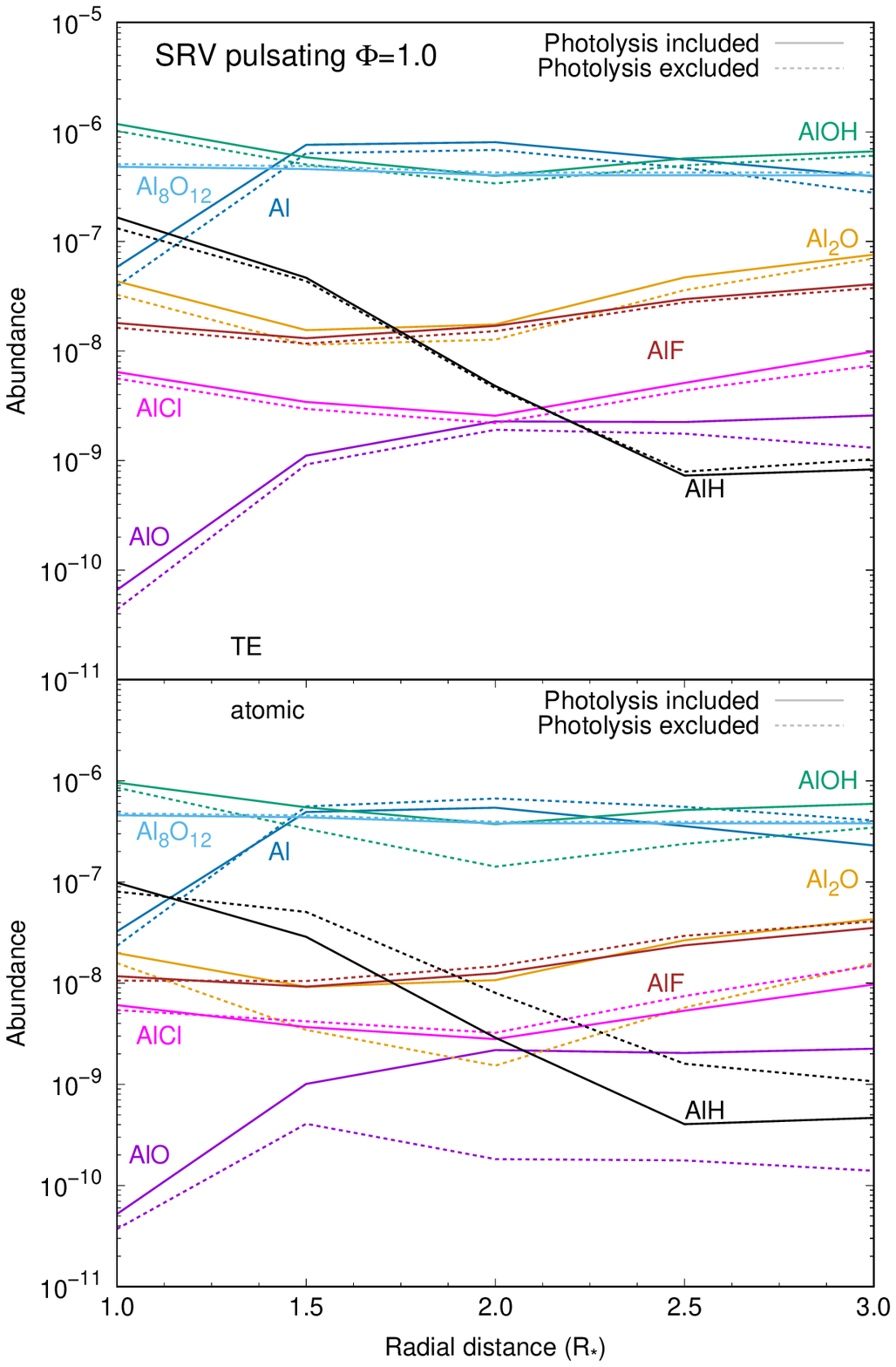}
\caption{The kinetic (non-equilibrium) abundances of the aluminum-bearing species in the postshock gas of the SRV model after a complete pulsation cycle (Phase $\Phi=$1.0) as a function of radial position. Dashed lines show the abundances, when the AlOH photodissociation is excluded. \textit{Upper panel}: initial abundances are given by chemical equilibirum. \textit{Lower panel}: initial abundances correspond to a pure atomic gas composition.\label{SRVpp}}
\end{figure}

One of the main drawbacks of the applied pulsating model is that its trajectory is purely periodic, as it returns to its initial radial position after a complete pulsation period. Therefore, a consecutive (continuous) trajectory describing an outflow (i.e. mass loss) is not possible with the assumptions made of this pulsating model and a rescaling to larger radii is required. 
We are aware that pulsating models with a gradually outward directed flow exist (see e.g. \citet{2018A&A...619A..47L,universe7040080}), but they presuppose (imply) a certain degree of condensation. This assumption in turn uses pre-existing solids, whose formation is the subject of our present study. Therefore, a self-consistent coupling of these complementary methods is desirable, but it is beyond the scope of our current investigation.
Nevertheless, bearing the approximative nature of our approach in mind, we follow the evolution of the pulsating model in steps of 0.5 R$_\star$ up to a distance of 3 R$_\star$ from the star (see Figure \ref{SRVpp}).  
Most of the species abundances (Al$_2$O, AlF, Al$_8$O$_{12}$) are shaped and largely detyermined at the first pulsation at 1 R$_\star$, where the gas densities are highest and the pulsation is strongest, and show little variation for larger radii. 
However, we find an increasing trend for AlO and a decreasing trend of AlH with distance from the star in the case.
The reaction AlO+H$_2$ plays an important role. 
In the consecutive pulsation model, the H$_2$ amount decreases gradually with the distance of the star, as a consequence of the decreasing gas density. In addition, the decreasing gas temperatures reduce also the efficiency of AlO+H$_2$ reaction with a barrier of 2092 K. 
In other words, with increasing distance from the star, less and less AlO is consumpt by the AlO+H$_2 \rightarrow$ AlOH+H reaction, owing to the lack of molecular H$_2$, and leads to a moderate increase of AlO up to a fractional abundance of 2 $\times$ 10$^{-9}$.\\

\subsubsection{MIRA pulsating model}
\begin{figure}
\includegraphics[scale=0.7]{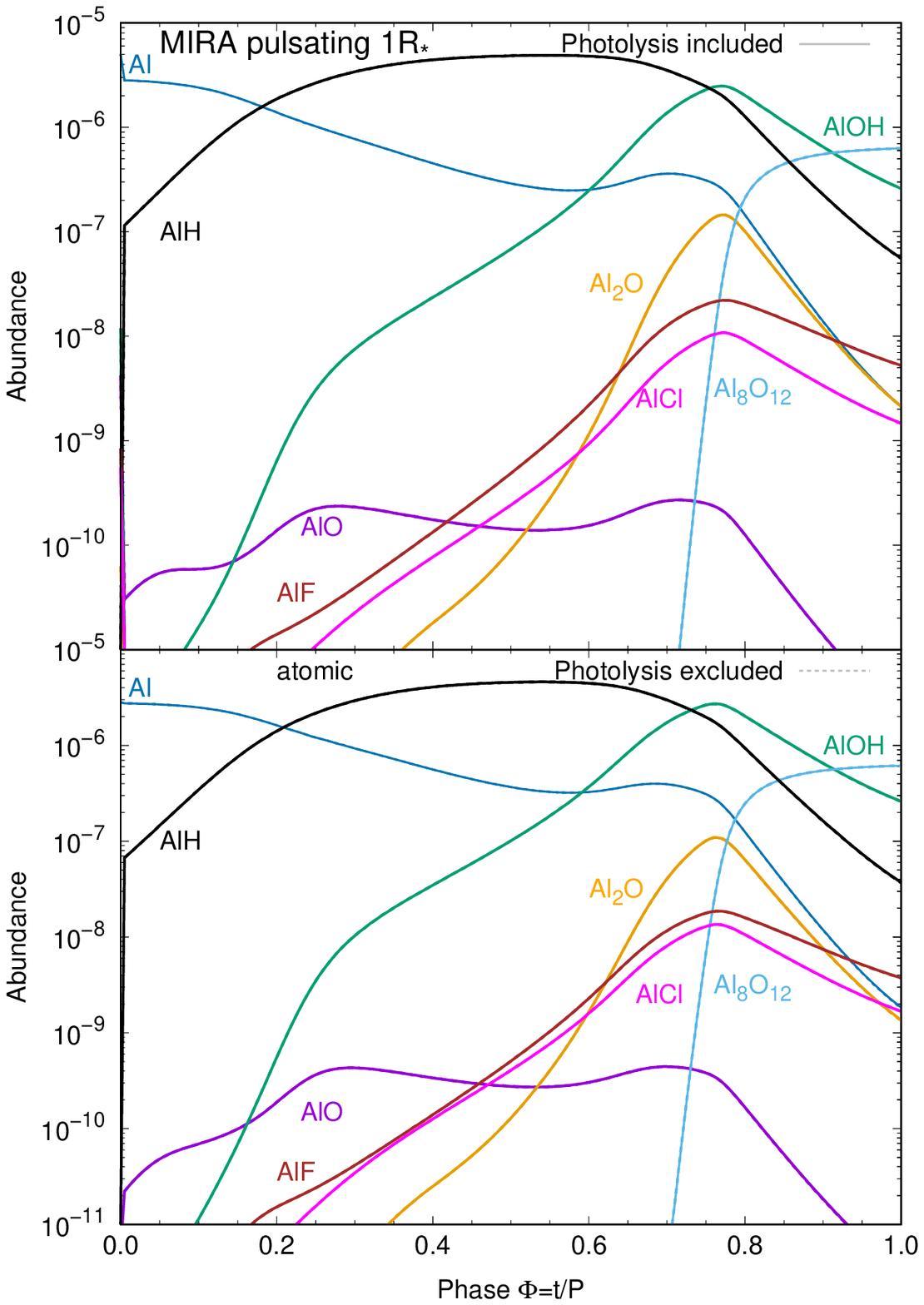}
\caption{The kinetic (non-equilibrium) abundances of the aluminum-bearing species in the postshock gas of the MIRA model as a function of pulsation phase $\Phi$.
Dashed lines show the abundances, when the AlOH photodissociation is excluded. \textit{Upper panel}: initial abundances are given by chemical equilibirum. \textit{Lower panel}: initial abundances correspond to a pure atomic gas composition\label{MIRA1R}.}
\end{figure}

In Figure \ref{MIRA1R}, the abundances of the main aluminium-bearing species in the MIRA pulsating model at 1 R$_\star$ are shown. 
In comparison with the SRV pulsating model, the effect of the photolysis reactions is entirely negligible for the MIRA pulsation at 1 R$_\star$. 
The alumina tetramer clusters start to form in significant amounts at phase $\Phi=$0.7 and account for 93.8 \% of the total aluminium content at phase $\Phi=$1.0.
The Al-containing molecules AlOH, Al$_2$O, AlCl, and AlF show similar abundance profiles as in the SRV case and peak, when the clusters start to form. These peaks appear sharper and the increase of Al$_8$O$_{12}$ steeper than in the SRV pulsating models.
On the other hand, the pulsation period in the MIRA model (470 days) is longer than in the SRV model (338 days).
Moreover, we note practially no difference, when starting with an pure atomic gas or molecular TE abundances.

\begin{figure}
\includegraphics[scale=0.7]{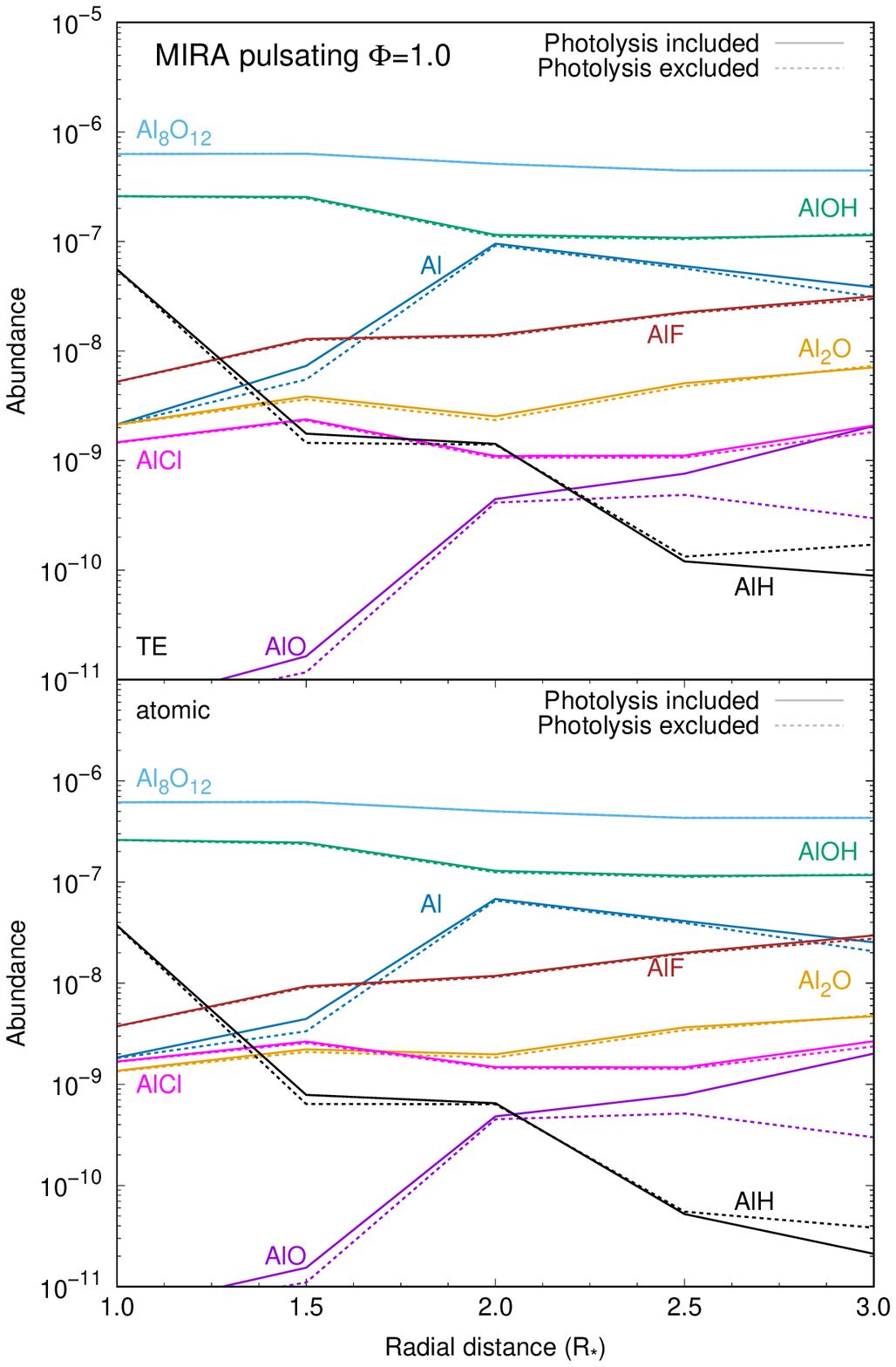}
\caption{The non-equilibrium abundances of the aluminum-bearing species in the postshock gas of the MIRA model after a complete pulsation cycle (Phase $\Phi=$1.0) as a function of radial position. Dashed-dotted lines show the abundances, when the AlOH photodissociation is excluded. \textit{Upper panel}: initial abundances are given by chemical equilibirum (TE). \textit{Lower panel}: initial abundances correspond to a pure atomic gas composition.\label{MIRApp}}
\end{figure}

In Figure \ref{MIRApp}, the MIRA model abundances of the consecutive pulsating model are shown. Clearly, the aluminium content is dominated by the alumina clusters and AlOH, independent of the inclusion of photolysis and of the initial gas mixture (TE or atomic).
We note that the photolysis processes start to play a role for $r>$1.0 R$_\star$, particularly in the case of an atomic initial gas. 
This can be explained, as the photolysis reactions decrease as $r^{-2}$ and therefore decrease less rapidly than the total gas density showing an exponential decline with radial distance.
At 3 R$_\star$ the AlO abundance is similar in both pulsating models, SRV and MIRA. 
However, for $r < 2R_\star$, we note that AlO is orders of magnitude less abundant in the MIRA model when compared to the SRV model, reflecting recent observations.
The halides AlF and AlCl show comparable amounts in the SRV and MIRA models.

\subsection{Homogeneous nucleation to the bulk limit (corundum)}
\label{Nucleation}
As indicated in the previous sections, the chemical-kinetic cluster growth is formulated up to the size of the alumina tetramer cluster, Al$_8$O$_{12}$. The subsequent nucleation is treated implicitly.
The choice of the alumina tetramer as the size linking the chemical-kinetic regime with the thermodynamically described larger (Al$_2$O$_3$)$_n$ clusters appears arbitrary at first glance. 
However, by inspecting the reaction enthalpies, we find values of $<$ $-$800 kJ/mole (corresponding to equivalent temperatures $\sim$10$^5$ K) for the association of two Al$_2$O$_3$ monomers and for the association of two Al$_4$O$_6$ clusters. As pointed out in Section \ref{viability}, the alumina monomer is comparatively unfavourable in terms of thermodynamics and kinetics. 
Therefore, the favourable association of two Al$_4$O$_6$ clusters with a negligible reverse dissociation rate represents an ideal transition from a chemical-kinetic controlled regime to a cluster treatment based on 
thermodynamics. In addition, the density of rovibrational states, which increases with the number of atoms contained in the cluster, is large enough for $n>$ 4 to relax the necessity of a chemical-kinetic treatment.
In the following, the cluster properties (energies, bond lengths, coordination, vibration modes) are 
analysed with increasing cluster size and compared
with the bulk limit.
In the case of alumina, the bulk limit corresponds to its most stable crystalline solid form, which is $\alpha$-alumina. 
The chemical-kinetic treatment beyond the tetramer is possible in principle and the choice of the tetramer as end point is somewhat arbitrary. 
However, in the chemically-rich gas mixture of circumstellar envelopes, the dust nucleation is likely to proceed via several species (i.e. heterogeneously). 
Therefore, we present the results for the homogeneous alumina nucleation in the form of size-dependent (up to $n=10$) cluster characteristics, their interpolation towards larger sizes ($n>10$), and the comparison to the crystalline bulk limit ($n\rightarrow\infty$).\\

 $\alpha$-alumina, or corundum, is the thermodynamically most stable crystalline form (i.e. polymorph) of solid Al$_2$O$_3$ and corresponds to our reference bulk limit.
Other alumina polymorphs (like $\theta$, $\delta$ and $\gamma$) are metastable or `transitional' and 
transform to corundum at temperatures higher than $\sim T=1000-1200$ K \citep{doi:10.1111/j.1151-2916.1998.tb02581.x}.
As alumina clusters grow in size, two phase transitions occur: from amorphous to the cubic $\gamma$-alumina form, and from $\gamma$-alumina to corundum ($\alpha$-alumina).   
Calorimetry experiments indicate that alumina becomes crystalline at a size of $\sim 40$\AA{} approximately \citep{doi:10.1021/jp405820g}.
\\

Many generic cluster properties (e.g. potential energies, melting temperatures), G($n$), can be 
approximated by a series expansion in the form of
\begin{equation}
\centering
G(n)= a_{0} + a_{1}n^{-\frac{1}{3}} + a_{2}n^{-\frac{2}{3}}+ a_{3}n^{-1} + ...,
\label{G_N}
\end{equation} 

\noindent where $n$ is the cluster size, or the number of Al$_2$O$_3$ formula units, $a_{0}=G_{bulk}$ is a characteristic constant for the corresponding bulk phase of the cluster, and $a_{1}$ is proportional to the fraction of surface atoms divided by `volume' atoms \citep{Johnston2002}.
In the case of energies, $a_{1}$ is often ascribed to the surface tension. In the case of simple, closely-packed clusters higher-order terms (second order and higher) can be neglected. Therefore, Eq. \ref{G_N} is classically approximated to describe the energy of a cluster with a single and simple size dependence as 

\begin{equation}
\centering
E(n)= a_{0} + a_{1}n^{-\frac{1}{3}}.
\label{surftens}
\end{equation} 

\begin{figure}
	\includegraphics[scale=0.7]{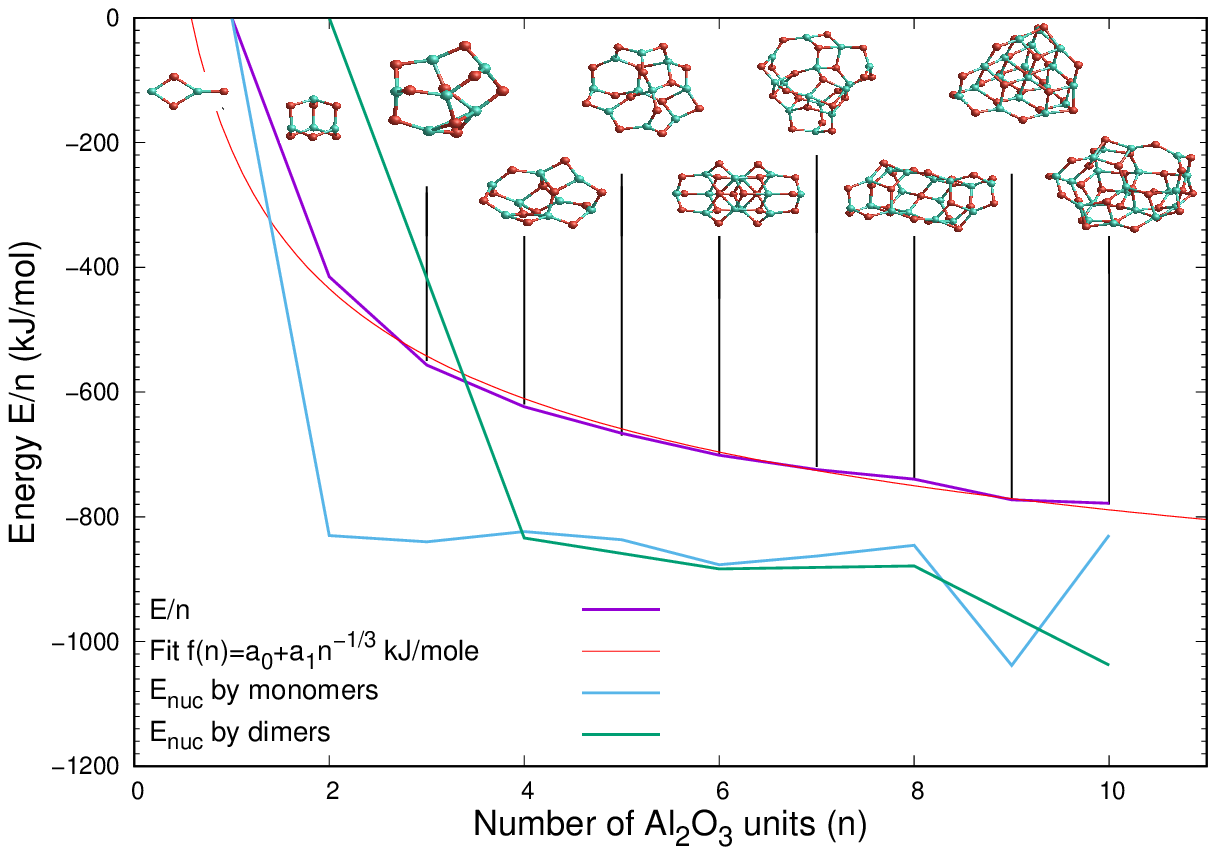}
	\caption{Cluster energies (in kJ mol$^{-1}$), normalised with respect to the Al$_{2}$O$_{3}$ monomer (which is set to 0), are shown with purple staight lines. Nucleation energies (in kJ mol$^{-1}$) are shown by monomer addition (in green) and by dimers (in blue).
		Fit with $a_{0} = -1287.9$ kJ mol$^{-1}$ and $a_{1} = 1075.42$ kJ mol$^{-1}$ of the normalised binding energies in red.
		\label{engs}}
\end{figure}

The electronic cluster energies (normalised to size $n$) are shown in Figure \ref{engs}.
Note that the \textit{cluster} energy, calculated as 
\begin{equation}
\centering
E=E(n)-nE(Al_{2}O_{3})
\end{equation}

\noindent with a subsequent scaling of the alumina monomer (Al$_2$O$_3$) energy to zero, 
is different from the electronic \textit{binding} energy calculated as
\begin{equation}
E_{b}=E(n)-n(2E(Al)-3E(O)), 
\end{equation}
taking also into account the bonds inside a stoichiometric Al$_2$O$_3$ unit.

As a consequence of an increasing number of bonds with size, \textit{cluster} and \textit{binding} energies decrease monotonically with size n. By definition, these two energies differ only by a constant value and show the same relative tendencies.
In Figure \ref{engs}, the cluster energies at $T=0$ K are shown as a function of cluster size n. 
We find that the calculated cluster energies can be roughly approximated by a first-order fit (Eq. \ref{surftens}), when excluding the monomer ($n=1$) from the fitting procedure. 
A best-fit constant $a_1$ of 1075.42 kJ/mol is found. 
Assuming a `monomeric radius' in the range of $r_{mon}=2.08-2.60$ \AA{} corresponding to the range given by geometric and van-der-Waals dimensions (see Table \ref{cradii}), surface tensions for different temperatures can be derived. 
Since the classical concept of surface tensions is not applicable to clusters on the (sub-)nano scale, we just provide the fit parameters $a_0 = -1287.9$ kJ mol$^{-1}$ and $a_1 = 1075.42$ kJ mol$^{-1}$. 

A closer inspection of the cluster energies reveals that some sizes ($n=3$,4,5) are more favourable, and some sizes ($n=8$,10) are less favourable than predicted by the fit to Eq. \ref{surftens}.  
For a temperature of $T=0$ K, the fit for the binding energies of the clusters (see Table \ref{summclusters}) with Eq. \ref{surftens}, we find a best fit with $a_{0} = -2906.48$ kJ mol$^{-1}$ and $a_{1} = 699.504$ kJ mol$^{-1}$. 
Moreover, we plot the nucleation energies of the clusters. The nucleation energy is defined as the difference in energy between the nucleating particles and the newly-formed cluster.
In the case of monomeric growth / nucleation, the nucleation energy is 

\begin{equation}
\centering
E_{nuc} = E(n)-E(1)-E(n-1).
\end{equation}

\noindent and in the case of dimer addition nucleation reaction it becomes 

\begin{equation}
\centering
E_{nuc} = E(n)-E(2)-E(n-2).
\end{equation}

The nucleation energies compare systems of the same size (number of atoms) and a normalisation is thus not required. Therefore, the monomer nucleation energy can be regarded as a finite-difference derivative of the cluster energy curve.
The monomeric nucleation energies show that the GM clusters with sizes $n=4$, 8 and 10, 
represent energetic bottlenecks, though the nucleation is still exothermic. 
In contrast, the GM clusters with sizes $n=2$, 3, 6, and 9 have comparatively large nucleation energies and are expected to form quickly. 
As the monomer ($n=1$) is not an outstandingly stable and abundant compound as has been pointed out in Section \ref{precursors}, we also plot the energy for the nucleation by alumina dimers.
The dimer nucleation scenario also indicates that the change in nucleation energy is largest for the formation of 
the tetramers ($n=4$) marking the end point of our chemical-kinetic description.
Generally, the nucleation is not constrained by the addition of monomers or dimers, but can proceed via polymers of any kind (see e.g. \citet{2019MNRAS.tmp.2040B}). Moreover, as previously mentioned, a heterogeneous nucleation, involving more than one chemical dust species, seems more realistic.\\

\begin{figure}
	\includegraphics[scale=0.7]{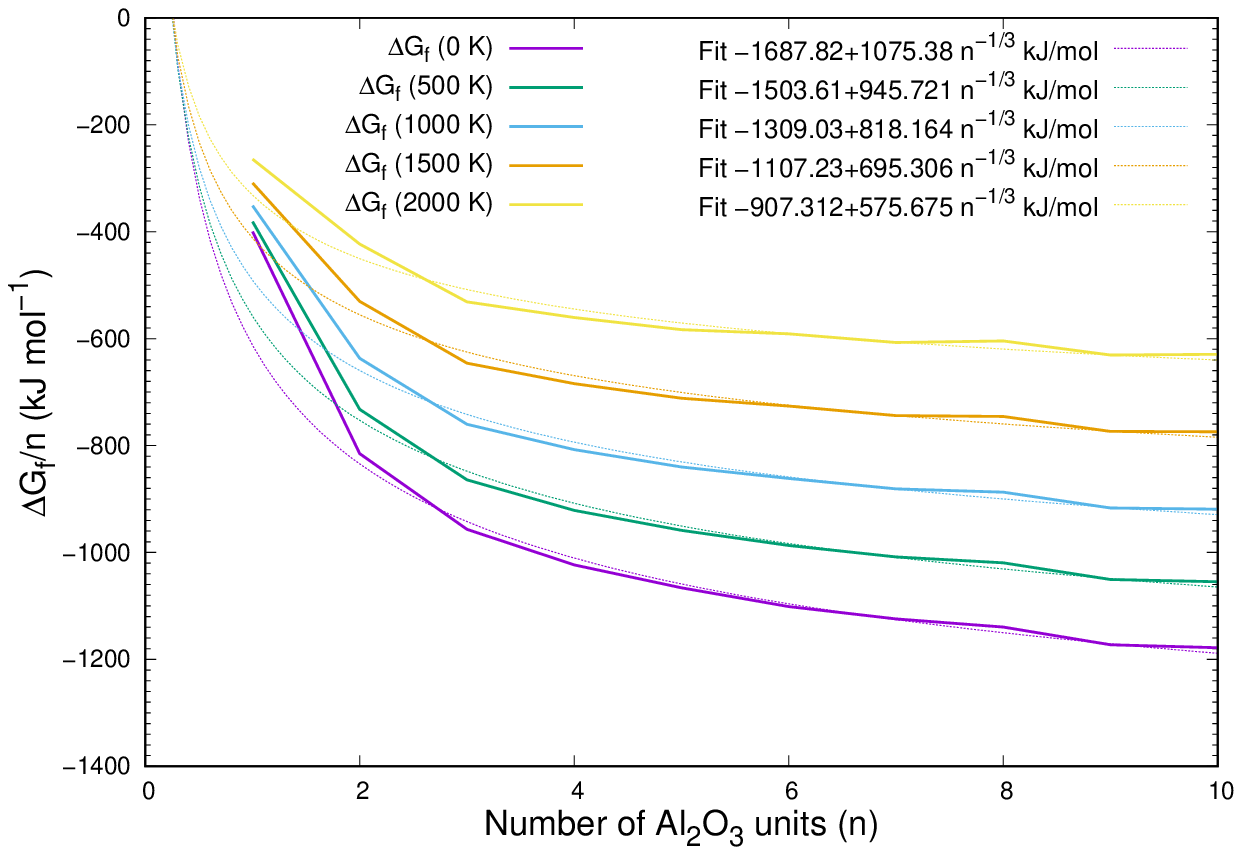}
	\caption{Gibbs free energies $\Delta$ G$_f$(T) of the clusters (in kJ mol$^{-1}$), normalised to the cluster size n, as a function of cluster size n, are shown for different temperatures.\label{DGfCluster}}
\end{figure}

In Figure \ref{DGfCluster}, we plot the Gibbs free energy of formation $\Delta$G$_f$(T) (scaled to the elemental heats of formation) as a function of the cluster size $n$ for temperatures $T=0$, 500, 1000, 1500 and 2000 K.
For $T=$0 K, we find the overall lowest Gibbs free energies, which are also fitted according 
to Eq. \ref{surftens}. The corresponding fit parameters are $a_0 = -1687.82$ and $a_{1} = 1075.38$ kJ mol$^{-1}$.
On the one hand, this result is consistent with the fits to the cluster energies in Figure 
\ref{engs}, showing an almost identical slope ($a_{1}$). On the other hand, 
the bulk parameter ($a_0$) is reasonably close to the standard free energy of formation of 1643.7 kJ mol$^{-1}$ as derived from electrochemical cell experiments \citep{Ghosh_1977} and to the energy of formation of 1663.6 kJ mol$^{-1}$ reported in the JANAF tables.
For larger temperatures, the bulk offsets $a_0$ becomes less negative, but also the slopes ($a_{1}$) become less steep. Furthermore, we note the same cluster sizes with comparatively enhanced ($n=3$,4) and reduced ($n=8$,10) favourability for temperatures $>0$ K.

\subsection{Geometric and electrostatic cluster properties}
\begin{figure}
\includegraphics[scale=0.7]{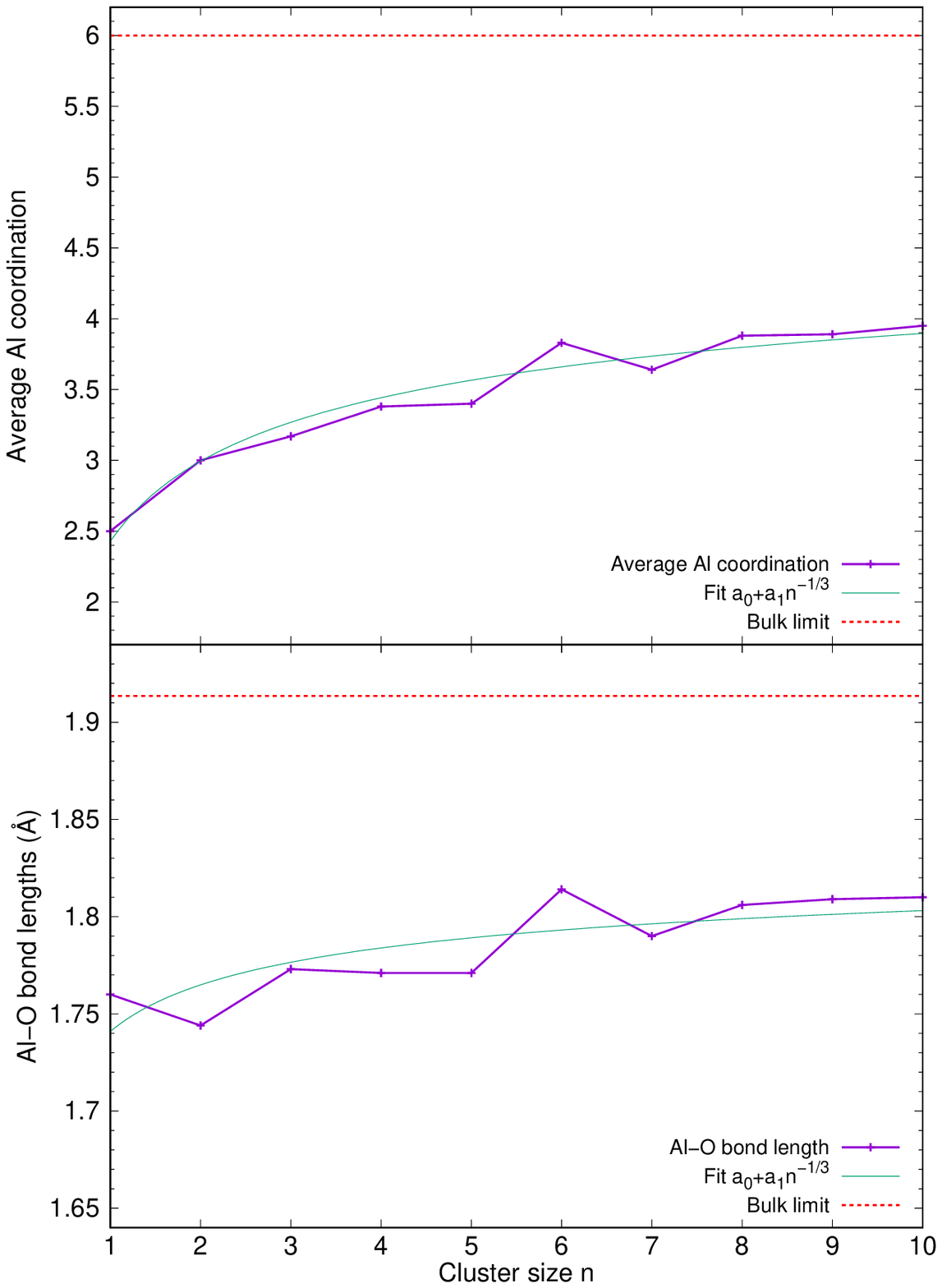}
\caption{Average Al coordination (top panel) and average AlO bond length in the GM clusters 
as a function of cluster size n. The values for the bulk limit ($\alpha$-alumina) are given 
in dashed red lines.\label{geomlimits}}
\end{figure}

In this section, we will discuss geometric (bond coordination and bond lengths) and 
electrostatic (atomic charges, dipole moments and vibration modes) properties of the GM 
clusters and compare them with each other and with corundum (i.e. the bulk limit).
The average atomic coordination is plotted in the upper panel of Figure \ref{geomlimits} 
and increases 
monotonically with the cluster size $n$, except for $n=6$ representing an outlier.
For the dimer ($n=2$) the coordination corresponds to the valence of the Al and O atoms. The 
atomic coordination inside the bulk (i.e. $\alpha$-alumina) is 6 for Al and 4 for O.
Hence, the atomic coordination of the clusters considered in this study are well below 
the bulk values. The average Al$-$O bond distances of the GM clusters are shown in the lower
panel of Figure \ref{geomlimits}. They show a less-regular pattern than the Al coordination, 
but also indicate an overall increasing trend. 
For sizes $n=3-5$ and for $n=8-10$ the average AlO distances are almost identical and sizes $n=1$ 
and $n=6$ represent outliers. 
Generally, it is not surprising that the average atomic coordination and AlO bond length 
increase with size $n$, as it reflects the decreasing surface-to-volume ratio with size 
(see Table \ref{summclusters}). 
If we apply a fit of the form of Eq. \ref{surftens} to the average Al coordination 
number $n_{Al}$, we find fitting parameters of $a_{0}=5.16747$ and $a_{1}=-2.73878$. 
For the average Al$-$O bond distance, we find fitting parameters of $a_{0}=1.8557605$ \AA{} 
and $a_{1}=-0.11595$ \AA{}.

Since the fit values of $a_{0}$ for both, average coordination and bond length, are below 
the value of $\alpha$-alumina ($\overline{n}_{Al}$=6, $\overline{d(AlO)}=1.91$ \AA{}), the bulk limits are not 
reached by these fits. We conclude that these fits are not suitable to find the approximate 
size, where the bulk limit is reached. However, they can serve as predictions for alumina
clusters with size $n>10$, which are computationally demanding to explore.

\begin{table}
	\caption{Point group (symmetry), average coordination numbers $\overline{n_{Al}}$ and $\overline{n_{O}}$, the average bond distance $\overline{d(AlO)}$ in \AA (\AA ngstr\"om), and 
		electronic binding energy E$_0$ per Al$_2$O$_3$ unit $n$ of the GM clusters in kJ mol$^{-1}$ for each cluster 
		size $n$ \label{summclusters}}
	\setcellgapes{2pt}\makegapedcells
	\begin{tabular}{r | c  c  c  c c}
		$n$ & Symmetry & $\overline{n_{Al}}$ & $\overline{n_{O}}$ & $\overline{d(AlO)}$ ($\AA$) & E$_0$/$n$ (kJ mol$^{-1}$) \\
		\hline
		1 & C$_{2\nu}$ & 2.5  & 1.67 & 1.760 & 1794.95 \\
		2 & T$_d$      & 3.0  & 2.00 & 1.744 & 2209.97 \\
		3 & C$_1$      & 3.17 & 2.11 & 1.773 & 2351.68 \\
		4 & C$_1$      & 3.38 & 2.25 & 1.771 & 2418.44 \\ 
		5 & C$_1$      & 3.4  & 2.26 & 1.771 & 2461.09 \\
		6 & C$_{2h}$   & 3.83 & 2.55 & 1.814 & 2496.35 \\
		7  &  C$_1$    & 3.64 & 2.43 & 1.790 & 2519.28 \\
		8 & C$_2$      & 3.88 & 2.58 & 1.806 & 2534.45 \\
		9 & C$_s$      & 3.89 & 2.59 & 1.809 & 2567.66 \\
		10 & C$_1$     & 3.95 & 2.63 & 1.810 & 2573.33\\
		\hline
		$\alpha$ & D$_{3d}$ & 6.0 & 4.0 & 1.91 & 3614.34 \\ 
		
	\end{tabular}
\end{table}

In Table \ref{summclusters}, we summarise our findings on the geometry of the GM clusters, 
including also the point group symmetries (given in the Sch\"onflies notation) and the binding energy
 per Al$_2$O$_3$ unit,  as a function of cluster size $n$.\\   

In Table \ref{mulliken}, we consider electrostatic properties of the GM clusters of 
by first examining the atomically partitioned Mulliken charges.
In general, we find significantly lower average magnitiudes of atomic charges for the PBE0-optimised structures than 
for the same clusters optimised with B3LYP. 
In both cases, the Mulliken average charges generally tend to decrease with cluster size $n$, 
with a slight increase between $n=6$ and $n=8$. The more accurate CBS-QB3 calculations (for $n=1-3$) exhibit larger atomic charges.
In general, the tendency for the (Mulliken) charges to decrease with size $n$ indicates a transition from more ionic clusters to a more covalent solid, which is not unexpected.


\begin{table}
	\caption{Average Mulliken charges of the Al atoms, $\overline{q}_{Al}$ in the GM clusters \label{mulliken}. 
		Owing to charge neutrality, atomic charges of O atoms can be computed as $\overline{q}_{O}=-\frac{3}{2}\overline{q}_{Al}
		$. Total dipole moments of the GM clusters are given in the units of Debye.}
	\setcellgapes{2pt}\makegapedcells
	\begin{tabular}{r | c c c | c c c}
		& \multicolumn{3}{c}{Mulliken charges} & \multicolumn{3}{c}{Dipole moments} \\
		n    & B3LYP & PBE0 & CBS & B3LYP & PBE0 & CBS \\
		\hline  
		1   & 0.60 & 0.58  & 1.21 & 2.59 & 2.46 & 2.42 \\
		2   & 1.01 & 0.89  & 1.21 & 0.00 & 0.00 & 0.00 \\
		3   & 0.67 & 0.59  & 1.61 & 1.23 & 1.01 & 1.49 \\
		4   & 0.61 & 0.51  &      & 1.46 & 1.56 &      \\
		5   & 0.58 & 0.48  &      & 3.93 & 3.91 &      \\
		6   & 0.37 & 0.24  &      & 0.00 & 0.00 & \\
		7   & 0.40 & 0.28  &      & 3.90 & 4.02 & \\
		8   & 0.42 & 0.30  &      & 3.57 & 3.64 & \\
		9   & 0.33 & 0.15  &      & 9.07 & 5.49 & \\
		10  & 0.30 & 0.20  &      & 5.67 & 5.60 & \\
	\end{tabular}
\end{table}

The dipole moments $\vec{D}$ of the GM clusters are crucial for astronomical observations, since only clusters with a $\vec{D} \ne 0$ can 
be detected by rotational lines. Therefore, the dimer ($n=$2) and the hexamer ($n=$6) GM isomers with a $\vec{D} = 0$ are not observable by pure rotational spectroscopy, even though they might be present.  Apart for $n=$9, the PBE0- and B3LYP-derived dipole moments agree with each other. For the symmetric nonamer GM cluster the B3LYP dipole moment is very large and does not agree with value for the PBE0 optimisation. By closer inspection, we find tiny differences between the B3LYP- and PBE0-optimised geometries, which could be the reason for for the discrpancy of these two functionals.

Finally, we aim to address the radiative properties of the GM clusters by investigating their vibrations
by using the rigid-rotor harmonic oscillator approximation and performing a harmonic vibrational analysis.
Anharmonic contributions might play a role for large oscillation amplitudes and low-lying excited electronic
states could contribute significantly, particularly at high temperatures. 
Recent investigations of the vibrational IR spectra of nano silicates revealed that anharmonic 
as well as thermal effects can have significant impact on the spectra
\citep{doi:10.1021/acsearthspacechem.9b00157,doi:10.1021/acsearthspacechem.0c00341}. 
However, an investigation of these anharmonic effects and temperature dependence are challenging and beyond the scope of this paper. 
In this study, we primarily aim to show the positions of the vibrational bands necessary to calculate partition functions, 
to exclude transition states, and
to explore whether the observed spectral alumina features at 11 $\mu$m and 13 $\mu$m  could arise from (small) alumina clusters.

\begin{figure}
 \includegraphics[scale=0.7]{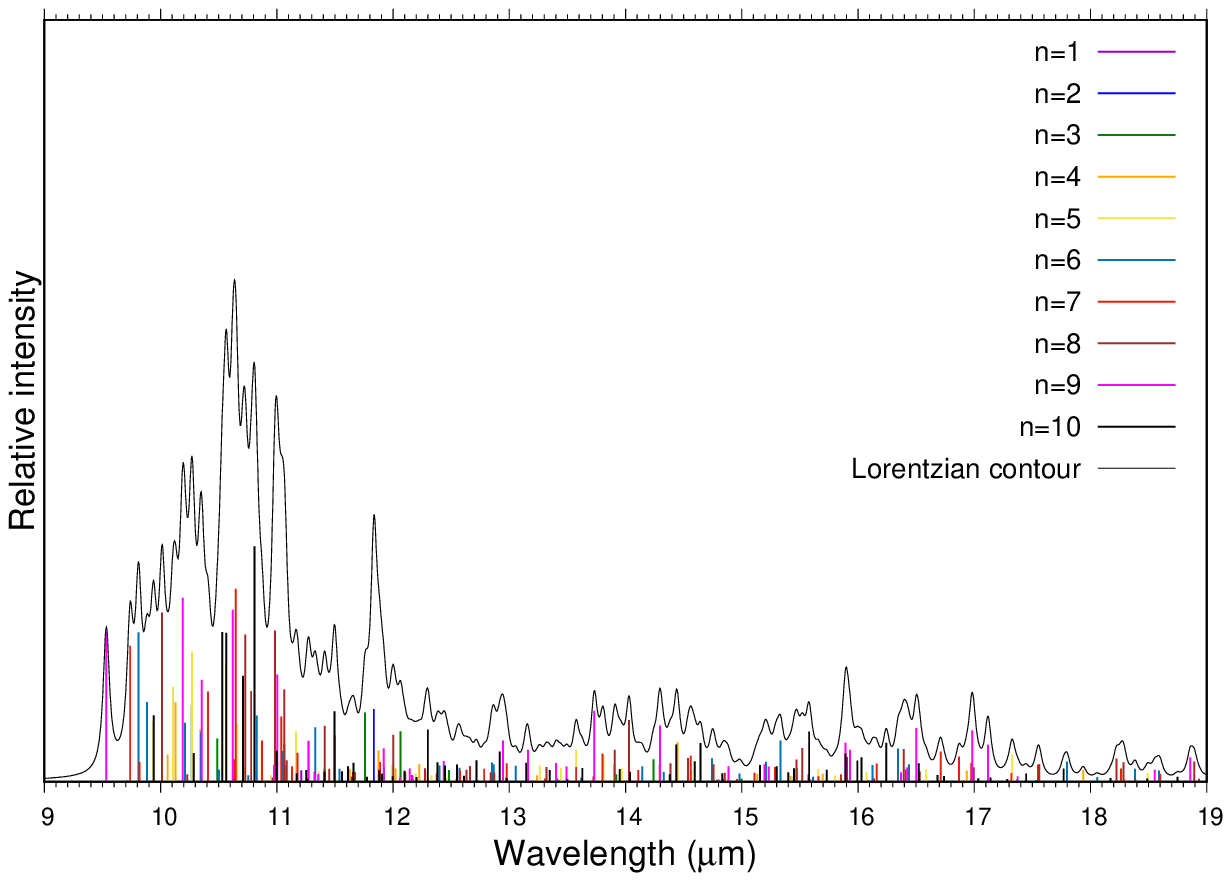}
\caption{Vibration modes of the (Al$_2$O$_3$)$_n$, $n=$1$-$10, GM clusters (color-coded) as a function of the wavelength (in $\mu$m).
The black contour corresponds to non-weighted (in size and metastable isomers) Lorentzian distribution . 
 \label{vibs}}
\end{figure}

In Figure \ref{vibs} the vibration modes of the (Al$_2$O$_3$)$_n$, $n=$1$-$10, GM clusters and their related intensities are shown as a function of wavelength. 
In order to transform the discrete vibration modes in a continuous spectrum and compare to an observed spectrum, line broadening in the form of a Lorentzian distribution with a half width $\gamma = 0.033 \mu$m, corresponding to a typical ALMA setup resolution in submillimeter range, is applied.
The contour (in black) comprises all unweighted (Al$_2$O$_3$)$_n$, $n=$1$-$10, GM clusters.
In principle, metastable clusters could also contribute to the species abundance and spectra. 
However, like the anharmonic and temperature effects, they are not included in this study.

The 13 $\mu$m feature seen in many M-type AGB stars principially could arise from small Al$_2$O$_3$ clusters.
However, our results indicate that the clusters under consideration here do not emit intensely in this wavelength region.
The 13 $\mu$m feature possibly originates from bulk-like alumina and heterogeneously mixed solids. 
This result is consistent with \citet{1997ApJ...476..199B}, who ruled out amorphous (non-crystalline) alumina 
as source of the 13 $\mu$m feature. Moreover, the authors derived laboratory optical constants for compact and porous 
amorphous solid alumina, resulting in a very broad emission feature around 11$-$12 $\mu$m, peaking at 11.5$-$11.8 $\mu$m. 
In this wavelength range our calculated IR intensity is comparatively low, though it is still stronger than at 13 $\mu$m.
\citet{B212654K} showed in an experimental study that large-sized aluminium oxide clusters (with sizes $n\simeq$15$-$40, but not
necessarily with a Al:O stoichiometry of 2:3) exhibit two main vibration modes at 11.8 $\mu$m and at 15 $\mu$m.
At 11.8 $\mu$m, our calculated IR spectra shows a local maximum (see Figure \ref{vibs}), which predominantly
arises from the dimers($n=2$). However, at 15 $\mu$m the overall IR intensity is rather low for the alumina clusters with $n=1-10$.
We also note that amorphous particles were considerably larger than our GM clusters.
Owing to the approximate nature that does not take into account anharmonicities or temperature effects, we 
cannot rule out small (impure) alumina clusters as a source of both spectral features, at 11 and 13 $\mu$m. 

\section{Discussion}
In Tables \ref{tableSRV} and \ref{tableMIRA}, we summarise our results for the TE calculations, the non-pulsating and the pulsating models for the two types of AGB stars, SRV and MIRA.
The abundances derived from recent observations are listed in Table \ref{tableOBS} for comparison.
In the SRV models, atomic Al is found to be important for all cases. In comparison, the MIRA models show lower, but still considerable amounts of atomic Al. 
Strong and broad doublet resonance lines (around 395 nm) of the neutral Al atom, likely originating from the photosphere, are found in Mira-variable stars \citep{1962ApJ...136...21M,2016AA...592A..42K}. 
Reliable abundance estimates of atomic Al are however lacking. Therefore, we confirm the presence of atomic Al close to the photosphere of M-type AGB stars.  
AlO is by far the most studied aluminum-bearing molecule in oxygen-rich AGB stars. 
AlO is a diatomic molecule with known transitions and has a large dipole moment.
It is thus not surprising that it  has been found in a number of late-type stars, since its discovery in red supergiants \citep{2009ApJ...694L..59T}. 
In Table \ref{tableOBS}, we list recent detections of AlO and give abundances, where available. 
\citet{2016AA...592A..42K} observed AlO in the archetypical star Mira instelf. The authors give a broad abundance range between 10$^{-9}$ and 10$^{-7}$,  owing to large uncertainties in the measurement of the column density of AlO. Their semi-analytical model results in an AlO abundance of {5~$\times$~10$^{-10}$} in Mira. 
We find generally lower AlO abundances in our kinetic MIRA models, except for in the consecutive pulsating model for $r>$2 R$_\star$ (see Table \ref{tableMIRA}). 
\citet{2017AA...598A..53D} conducted a search for AlO in 7 AGB stars, 
6 of which are MIRA-type (R Aqr, TX Cam, o Cet (Mira), R Cas, W Hya, IK Tau) and one of SRV type (R Dor). Definitively, they found AlO transitions in two stars (o cet and R Aqr) and tentatively,  in three further stellar sources.
The authors derive a source-averaged AlO fractional abundance of 4.5$\times$10$^{-8}$, based on a rotational diagram of {IK Tau}, also hinting at difficulties in deriving AlO abundances for individual stars. More recent ALMA observations \citep{refId0D,2020ApJ...904..110D} show the absence of AlO in the inner envelope of IK Tau, a MIRA-type variable, but its presence in R Dor, a typical SRV type star. In R Dor, they derive AlO fractional abundance of (3.7$-$7.8)$\times$ 10$^{-8}$. These values compare well with the SRV consecutive pulsating model. Moreover, we note higher AlO fractions in the SRV-type stars than in the MIRA-like AGB stars, which is consistently reproduced by our models for both types of trajectories, non-pulsating and pulsating models. 
AlOH represents the predominant Al-bearing molecule in both sets of models, SRV and MIRA. 
Its spectroscopic identification and a related abundance estimation by (radiative) transitions
is challenging, owing to the following reasons. 
At present, there are no spectroscopic constants avalable in the literature for the AlOH electronic bands in the optical regime.
Observations of rotational (and vibrational) transitions at longer wavelengths are feasible, but some rotational AlOH transitions are blended by rotational transitions of vibrationally excited 
TiO molecules and by rotational lines of a neutron-enriched isotopologue of SO$_2$ \citep{Decin1497}. 
In addition, it is debated that circumstellar AlOH also exists in a bent configuration \citep{0004-637X-863-2-139}, 
exhibiting different spectroscopic constants. 
At high temperatures, AlOH will be a quasi-linear molecule, because the difference in energy between the bent and linear forms is very small.
Therefore, an accurate observational abundance estimate is demanding and it tends to underestimate the AlOH content. 
This might explain the systematicallay larger AlOH model abundances in comparison with observations.
Overall, the AlOH/AlO ratio in our circumstellar models is larger than the observationally derived ratio indicating that the kinetic conversion of AlOH in AlO in our models is too inefficient.
Neither the photodissociations nor the bimolecular reactions AlOH+H and AlOH+OH (that produce AlO) are able to noticably transform the stable singlet AlOH molecule into the doublet AlO, although we used the most elaborate kinetic data available.  
Therfore, possible resorts include that
\begin{description}
\item{- the AlOH destroying (and AlO forming) reactions are described with too low (too large) rate expressions.} 
\item{- efficient, but unnoticed processes are missing in our present scheme.}   
\item{- the observationally derived AlO and AlOH abundances are artifacts of a rather simplistic treatment assuming a static peak abundances that follow a Gaussian distribution with an e-fold or 
dissociation radius.}
\end{description}
To test the first hypothesis, we varied the AlOH photodissociation rate. Best agreement with the observations of AlO and AlOH for all models (non-pulsating and pusalting, SRV and MIRA) is found, when we increased the photolysis rate by a factor of 10$^4$. 
However, this artificial upscaling of the photodissociation leads to a vanishing alumina tetramers/cluster production in the SRV pulsation models. In the non-pulsating models, alumina clusters are still forming, but show a later onset of their synthesis.\\

The ground states of Al$_2$O and AlO$_2$ are not observable by their rotational transitions. 
Our calculations indicate that Al$_2$O is likely to be an intermediate in the alumina cluster formation and shows noticeable abundances, 
whereas AlO$_2$ is negigible in abundance and cluster formation.
Therefore, there are no detections nor abundance estimates for these symmetric and linear molecules. 
For completeness, the aluminium-bearing halides AlF and AlCl are also listed in Tables \ref{tableSRV}, \ref{tableMIRA} and \ref{tableOBS}. 
These species become more important in S-type AGB stars and are not the main subject of this study examining oxygen-rich AGB stars. 
Nevertheless, the aluminium-bearing species AlF and AlCl might hamper the formation of aluminium oxide molecules and clusters 
by locking up aluminium. \citet{refId0D} found that AlCl is about 3 times more abundant in oxygen-rich SRV type stars. 
The $\beta$-velocity law at 3 R$_\star$ best fits these observations, but also the (single) pulsating models at 1 R$_\star$ reflect the magnitude and trends of these recent observations.

\begin{table}
\caption{Overview of the SRV model abundances for thermodynamic equlibrium (TE), the non-pulsating model ($\beta$-velocity), and the pulsating model. The fractional abundances (normalised to the total gas) are given as $a(b)=a\times10^{b}$ \label{tableSRV}}
\begin{tabular}{c | c | c c | c c }
   & TE & \multicolumn{2}{c}{non-pulsating} & \multicolumn{2}{c}{pulsating} \\
Species        &  1R$_\star$    & 2R$_\star$   & 3R$_\star$ & 1 R$_\star$ & 2 R$_\star$  \\
\hline
Al             &  3.2(-6)   & 1.7(-6)  & 4.7(-9)  & 5.8(-8) & 8.1(-7)  \\     
AlO            &  9.6(-10)  & 1.6(-9)  & 7.7(-12) & 6.6(-11)& 2.3(-9)  \\
AlOH           &  3.9(-11)  & 2.5(-6)  & 2.6(-7)  & 1.2(-6) & 3.4(-7)  \\
Al$_{2}$O      &  1.2(-14)  & 1.6(-7)  & 3.0(-9)  & 4.3(-8) & 1.8(-8)  \\
AlO$_{2}$      &  2.9(-16)  & 4.9(-16) & 4.5(-20) & 5.4(-18)& 7.4(-18) \\
Al$_8$O$_{12}$ &   0        & 2.0(-12) & 6.3(-7)  & 4.8(-7) & 4.0(-7)  \\
AlCl           &  2.8(-12)  & 1.1(-8)  & 1.9(-9)  & 6.4(-9) & 2.6(-9)  \\ 	
AlF            &  2.1(-11)  & 2.3(-8)  & 8.8(-9)  & 1.8(-8) & 1.7(-8)  \\
\end{tabular}
\end{table}

\begin{table}
\caption{Overview of the MIRA model abundances for thermodynamic equlibrium (TE), the non-pulsating model ($\beta$-velocity), and the pulsating model. The fractional abundances (normalised to the total gas) are given as $a(b)=a\times10^{b}$.\label{tableMIRA}}
\begin{tabular}{c | c | c c | c c }
   & TE & \multicolumn{2}{c}{non-pulsating} & \multicolumn{2}{c}{pulsating} \\
Species        & 1 R$_\star$    & 2R$_\star$   & 3R$_\star$ & 1 R$_\star$ & 2 R$_\star$  \\
\hline
Al             & 3.2(-6)    & 1.6(-9) & 8.6(-12) & 2.1(-9)  & 9.6(-8) \\
AlO            & 4.2(-9)    & 2.2(-12)& 2.6(-14) & 2.4(-12) & 4.5(-10) \\
AlOH           & 1.2(-8)    & 2.1(-7) & 3.2(-8)  & 2.6(-7)  & 1.2(-7) \\
Al$_{2}$O      & 6.3(-11)   & 1.7(-9) & 7.7(-11) & 2.1(-9)  & 2.5(-9) \\
AlO$_{2}$      & 1.2(-15)   & 2.3(-20)& 3.1(-24) & 7.3(-20) & 6.1(-20) \\
Al$_8$O$_{12}$ & 0          & 6.4(-7) & 6.7(-7)  & 6.3(-7)  & 5.1(-7) \\
AlCl           & 5.5(-10)   & 1.4(-9) & 3.8(-10) & 1.6(-9)  & 1.1(-9) \\      
AlF            & 8.4(-10)   & 5.9(-9) & 3.0(-9)  & 5.3(-9)  & 1.4(-8) \\
\end{tabular}
\end{table}

\begin{table} 
\caption{Overview of recent observations of the species AlO, AlOH and AlCl in the innner winds of 
oxygen-rich AGB stars of SRV and MIRA type. Note that a conversion factor of $\sim$0.46 is applied 
to some molecules to translate from relative (with respect to H$_2$) to fractional (with respect to the 
total gas including He) abdundances. The fractional abundances (normalised to the total gas) are given as a(b)=a$\times$10$^{b}$\label{tableOBS}}
\begin{tabular}{c | c c c}
Species  & SRV         & MIRA     & Reference \\
\hline
AlO      & 3.7(-8)     &  0       &  \citet{refId0D}   \\
         & 7.8(-8)     &  0       &  \citet{2020ApJ...904..110D} \\
         &             &  5.0(-10) &  \citet{2016AA...592A..42K} \\
         & 4.4(-8)     &  0       &  \citet{2017AA...598A..53D} \\ 
AlOH     & 6.9(-10)    & 2.1(-9)  &  \citet{refId0D}   \\
AlCl     & 9.2(-9)     & 4.1(-10) &  \citet{refId0D}   \\
\end{tabular}
\end{table}

About 50 years ago \citet{doi:10.1080/00102206908952206} suggested that the combustion of aluminium particles proceeds via the clustering of AlO molecules, followed by an evaporation of aluminum, to finally form alumina. Since then, Al combustion including related clustering and burning times have been extensively studied in the laboratory (see e.g. \citet{doi:10.1063/1.474085,Beckstead2005} and references therein). 
More recent studies have used aluminized fuels and modelled the (Al$_2$O$_3$)$_n$ cluster formation by numerical simulations \citep{Starik2015,SAVELEV2018223}.
These studies were predominantly performed under conditions that are radically different from circumstellar envelopes: 
in the combustion environment the pressure is at least 5 orders of magnitude higher, the bath gas is N$_2$, and there are high levels of oxidants (O$_2$ and CO$_2$).
Bearing in mind these intrinsic differences, a comparison of the data from aluminium combustion experiments with our study is not applicable. An adaptation of our model and tests against experimental data could be done in a future study.
Moreover, we note that the majority of the clustering rates used in this study (in particular reaction involving species with more than 4 atoms) is based on collision and capture theory. 
These theoretically derived rates could differ from currently lacking laboratory measurements. 
However, we aimed to minimize this effect by using the most recent and accurate experimental rate date available.\\ 
The stoichiometric (Al$_2$O$_3$)$_n$ show a gradual monotonic trend with cluster size $n$ in energy, atomic coordination, average bond length, and atomic charges. However, for all these quantities, we find outliers to these trends, and these outlier cannot be predicted by interpolations. 
Our derived fit parameter for the bulk limit ($a_0 = -1687.82$ kJ mol$^{-1}$) shows excellent agreement with the JANAF-NIST database and 
other experimental data \citep{Ghosh_1977}.
Calorimetric experiments on the specific surface areas show that two size-dependent phase transtion occur in alumina  
\citep{McHale788,doi:10.1021/jp405820g}.
The smallest clusters reside in the amorphous regime characterised by sizes $n<800$, or in terms of diameters, by $d<40$\AA{}.
For sizes $800<n<1900$ (or diameters of $d=40-116$\AA{}), alumina is predominantly in cubic $\gamma$-alumina crystals.
Only larger-sized alumina particles ($n>1900$, $d>116$\AA{}) are mainly present in the form of hexagonal corundum. 
Therefore, our cluster calculations up to a size of $n=10$ and their interpolation to larger sizes clearly pertain to the 
size-regime of the non-crystalline (amorphous) phase, and cannot be repesented by top-down derived (crystalline) energies and 
structures. 


\section{Conclusions}
\label{concl}
We modeled the nucleation of alumina dust particles from a bottom-up perspective, starting with atomic and molecular precursors and using a chemical-kinetic approach applied to different circumstellar gas trajectories representing two model stars, a Semi-Regular Variable (SRV) and a regular variable (MIRA).
In both sets of circumstellar models, SRV and MIRA, alumina cluster particles form efficiently, independent of their initial gas-phase compositions. 
The results of the non-pulsating monotonic outflow ($\beta$-velocity law) and the pulsating (ballistic trajectory) models show similar trends.
The non-pulsating models predict a later onset of cluster formation at larger radii (2$-$2.5 R$_\star$) in the SRV envelope, 
as compared with the MIRA model star, where the clusters form already at  1.5 R$_\star$. 
The pulsating models show that alumina clusters form yet at 1 R$_\star$ in the late post-shock gas.
In the SRV pulsating model, the amount of alumina clusters formed at 1 R$_\star$ is already substantial ($\sim$ 72 \% of the total aluminium content). However, in the MIRA pulsating envelope, the cluster synthesis is even more fruitful
showing that more than 90 \% of the total Al content is in condensable clusters after one pulsation period.
These results reflect the evolutionary sequence along the AGB showing an increase of the stellar mass-loss rate.
Furthermore, the models consistently show larger AlO abundances in the SRV case, as compared with the MIRA case,
reflecting the trends of the most recent ALMA observations. 
Still, the modelled AlOH abundances exceed those derived from observations by a about two orders of magnitude. 
A possible process, destroying the stable singlet AlOH molecule, is the photolysis of AlOH.
However, in  AGB stars with rather low temperatures ($<2500$ K) the AlOH photolysis is of secondary importance.
Therefore, it is predominantly the bimolecular reaction AlO+H$_2$ $\leftrightarrow$ AlOH+H determining the AlO/AlOH ratio.
The effectiveness of the latter process depends on the concentration of molecular hydrogen, which is 
generally larger in the denser MIRA models. Therefore, the AlO/AlOH ratio is primarly controlled by the H/H$_{2}$ in circumstellar envelopes.\\
Larger-sized (Al$_2$O$_3$)$_n$ clusters ($n=4-10$) and their interpolation to the bulk-limit ($n\rightarrow \infty$) are described by 
size-dependent properties, as a chemical-kinetic treatment becomes increasingly expensive and prohibitive.
We find that the cluster energy can be approximated by a fit of the form E(n)=$-$1687.82 +1075.38 n$^{-1/3}$ kJ mol$^{-1}$.
The offset of $-$1687.82 kJ mol$^{-1}$ is resonably close to experimental values for bulk alumina (corundum).
Nevertheless, it is very likely that the nucleation in oxygen-rich AGB stars proceeds heterogeneously involving additional species, that are different from pure aluminium-oxygen clusters. 
Moreover, we note that the small size regime under consideration here is still deeply in the non-crystalline (amorphous) 
size regime and that phase transitions could affect our bottom-up nucleation fit.\\





\section{Acknowledgements}
We acknowledge the referee H.-P. Gail for his valuable comments that helped improving the quality of this manuscript.
D.G., J.M.C.P, and L.D. acknowledge support from the ERC consolidator grant 646758 “AEROSOL” and STFC grant ST/T000287/1. S.T.B. acknowledges Spanish MICIUN/FEDER RTI2018-095460-B-I00 and Mar\' ia de Maeztu MDM-2017-0767 grants and a Generalitat de Catalunya 2017SGR13 grant.
We acknowledge the CINECA award under the ISCRA initiative, for the availability of high performance computing resources and support for the projects IsC65 ``SPINEL'' and IsC76 ``DUSTY''. 
We acknowledge the ``Accordo Quadro'' INAF-CINECA MoU 2017$-$2020 for the projects 
INA17$-$C5A28 and INA20$-$C7B35.

\bibliographystyle{aa}
\bibliography{biblio}

\begin{appendix}
\section{Equilibrium constant and detailled balance: }

\begin{equation}
\centering
K_{eq} =  \frac{k_f}{k_r} = \exp\left(-\frac{\Delta G_r(T)}{RT}\right)
\label{equil_constant} 
\end{equation}

where K$_{eq}$ is the dimensionless equilibrium constant, k$_f$ and k$_r$ are the forward and the reverse reaction rates, respectively. 

\section{Chemical-kinetic network}

\begin{table*}[!ht]
	
\caption{Kinetic rate network listing (1) the reaction number, (2) the reaction, (3) the CBS-QB3 heat of reaction (enthalpy) at T= 0 K, (4) the critical temperature T$_c$, where the free energy of reaction, $\Delta$G$_r$(T) changes its sign, (5) the reaction rate with the pre-exponential factor A given as $a(-b) \equiv a \times 10^{-b}$(for unimolecular reactions (photodissociations in units of s$^{-1}$, for bimolecular reactions in units of cm$^3$s$^{-1}$, for termolecular reactions in units of cm$^6$s$^{-1}$), and (6) the reference or method of calculation\label{network}.}
\begin{tabular}{r l r r l l}
Number & Reaction & $\Delta$ H$_r$(0K) & T$_{c}$ & Rate k & Reference / remark  \\
\hline
\rule{0pt}{4ex} 1 & {H + H + H$_2$ $\rightarrow$ H$_2$ + H$_2$} & -437  & 3800   &  8.85(-33)(T/300)$^{-0.6}$ & NIST \\
2 & {H$_2$ +  H$_2$ $\rightarrow$ H + H + H$_2$} & +437 & 3800   & 1.5(-9)(T/300)$^{0.34}\exp(-48346.4/T)$ & NIST \\
3 & {H + H + H $\rightarrow$ H$_2$ + H} & -437  & 3800 &  8.82(-33) & NIST \\
4 & {H$_2$ +  H $\rightarrow$ H + H + H} & +437 & 3800   & 2.54(-8)(T/300)$^{-0.10}\exp(-52555.6/T)$ & NIST \\
5 & {H + H + He $\rightarrow$ H$_2$ + He} & -437  & 3800 & 4.96(-33) & NIST \\
6 & {H$_2$ + He $\rightarrow$ H + H + He} & +437 & 3800  & 8.85(-10)$\exp(-48346.4/T)$ & NIST \\
7 & {OH + OH $\rightarrow$ H$_2$O + O} & -67.4 & 4400 & 1.65(-12)(T/300)$^{1.1}\exp(-50.5/T)$ & NIST \\
8 & {H$_2$O + O $\rightarrow$ OH + OH} & +67.4  & 4400 & 1.84(-11)(T/300)$^{0.95}\exp(-8573.7/T)$ & NIST \\
9 & {CO + OH $\rightarrow$ CO$_2$ + H} & -105  & 3000 & 3.52(-12)$\exp(-2630.2/T)$ & NIST \\  
10 & {CO$_2$ + H $\rightarrow$ CO + OH} & +105 & 3000 & 2.51(-10)$\exp(-13229.1/T)$ & NIST \\
11 & {OH + H + H$_2$O $\rightarrow$ H$_2$O + H$_2$O} & -494 & 4000 & 1.19(-30)(T/300)$^{-2.1}$ & NIST \\
12 & {H$_2$O + H$_2$O $\rightarrow$ OH + H + H$_2$O} &  +494 & 4000 & 2.66(-7)$\exp(-7500.0/T)$ & NIST \\
13 & {OH + H $\rightarrow$ H$_2$ + O} & -10.9 & 1100 & 6.86(-14)(T/300)$^{2.8}\exp(-1949.5/T)$ & NIST \\
14 & {H$_2$ + O $\rightarrow$ OH + H} & +10.9  & 1100 & 3.44(-13)(T/300)$^{2.67}\exp(-3159.3/T)$ & NIST \\
15 & {H$_2$ + OH $\rightarrow$ H$_2$O + H} & -56.5 & no & 1.55(-12)(T/300)$^{1.6}\exp(-1659.7/T)$ & NIST \\
16 & {H$_2$O + H $\rightarrow$ H$_2$ + OH} &  56.5 & no &  6.82(-12)(T/300)$^{1.6}\exp(-9719.8/T)$ & NIST \\
17 & {C + O $\rightarrow$ CO + h$\nu$} & -1073 & no & 1.58(-17)(T/300)$^{0.34}\exp(-1297.4/T)$ & Fit to \cite{1990ApJ...349..675D} \\ 
18 & {C + O + M $\rightarrow$ CO + M} & -1073 & no & 2.00(-34) & NIST \\
19 & {CO + M $\rightarrow$ C + O + M} &  1073 & no &  4.40(-10)$\exp(-98600.0/T)$ & \cite{doi:10.1063/1.1673286}, eq. 4.7 \\
20 & {H + O + M $\rightarrow$ OH + M} & -426  & 4000 & 4.36(-32)(T/300)$^{-1.00}$ & NIST \\
21 & {OH + M $\rightarrow$ H + O + M} &  426  & 4000 &  4.00(-9)$\exp(-50000/T)$ & NIST \\
22 & {CO + O + M $\rightarrow$ CO$_2$ + M} & -532 & 3800 & 1.20(-32)$\exp(-2160/T)$ & NIST \\ 
23 & {CO$_2$ + M $\rightarrow$ CO + O + M} &  532 & 3800 & 8.02(-11)$\exp(-26900/T)$ &  \cite{1998Willacy} \\ 
24 & {OH + H + M $\rightarrow$ H$_2$O + M} & -494 & 4000 & 2.59(-31)(T/300)$^{-2.00}$ & NIST \\
25 & {H$_2$O + M $\rightarrow$ OH + H + M} & 494 & 4000 & 5.80(-9)$\exp(-52920/T)$ & NIST \\ 
\hline
\rule{0pt}{4ex} 26 & {Al + O + M $\rightarrow$ AlO + M} & -498 & 4900 & 3.34(-32)(T/300)$^{-1.79}$ & RRKM fit \\
27 & {AlO + M $\rightarrow$ Al + O + M} &  498 & 4900 &  4.01(-10)$\exp(-57391/T)$ & \cite{GOMEZMARTIN201756} RRKM fit \\
28 & {AlO + H + M $\rightarrow$ AlOH + M} & -482 & 4900 & 1.32(-28)(T/300)$^{-3.08}$ & RRKM fit  \\
29 & {AlOH + M $\rightarrow$ AlO + H + M} &  482 & 4900 & 1.47(-7)$\exp(-55985/T)$ & RRKM fit \\


30 & {Al + CO $_2$ $\rightarrow$ Al + CO} & 33.6  & 800   &  6.32(-10)$\exp(-4522.0/T)$ & RRKM fit\\
31 & {AlO + CO $\rightarrow$ Al + CO$_2$} & -33.6 & 800   &  10$^{-16.6318+3.43135\log(T)-0.5786\log(T)^2}$ & \citet{doi:10.1021/acsearthspacechem.0c00197} RRKM fit\\ 
32 & {Al + O$_2$ $\rightarrow$ AlO + O} & -0.9 & no    & 1.68(-10)(T/300)$^{-0.26}$ & \cite{GOMEZMARTIN201756} \\
33 & {AlO + O $\rightarrow$ Al + O$_2$} &  0.9 & no    & 7.15(-12)$\exp(-945./T)$ & RRKM fit \\
34 & {Al + H$_2$O $\rightarrow$ AlOH + H} & -60.4 & no  & 1.7(-12)$\frac{-422.}{T}$+1.45(-10)$\exp(-2657./T)$ & \cite{Mangan2021} \\ 
35 & {AlOH + H $\rightarrow$ Al + H$_2$O} & 60.4  & no    & 4.31(-11)$\exp(-9457./T)$ & \cite{Mangan2021} \\
36 & {AlO + H$_2$ $\rightarrow$ AlOH + H} & -45.0  & no &  5.37(-13)(T/300)$^{2.77}\exp(-2190./T)$ & \cite{Mangan2021} \\ 
37 & {AlOH + H $\rightarrow$ AlO + H$_2$} & 45.0 & no & 8.89(-11)$\exp(-9092/T)$ & \cite{Mangan2021} \\
38 & {AlO  + O + M $\rightarrow$ AlO$_2$ + M} & -391 & 2900 &  6.74(-30)(T/300)$^{-2.53}$ & \citet{https://doi.org/10.1029/2020JA028792} RRKM fit \\
39 & {AlO$_2$ + M $\rightarrow$ AlO  + O + M} & 391 & 2900  & 4.51(-7)$\exp(-44166/T)$ & \citet{https://doi.org/10.1029/2020JA028792} RRKM fit \\
40 & {AlO + CO$_2$ $\rightarrow$ AlO$_2$ + CO} & 140 & no & 1.81(-10)(T/300)$^{-0.81}\exp(-18138.2/T)$ & detailed balance \\
41 & {AlO$_2$ + CO $\rightarrow$ AlO + CO$_2$} & -140 & no & 2.55(-12)(T/300)$^{0.17}$ & \citet{doi:10.1021/acsearthspacechem.0c00197} RRKM fit \\
42 & {AlO + O$_2$ $\rightarrow$ AlO$_2$ + O} &  106 & no & 3.54(-11)(T/300)$^{0.19}\exp(-13140/T)$ & detailed balance \\ 
43 & {AlO$_2$ + O $\rightarrow$ AlO + O$_2$} & -106 & no & 1.90(-10)(T/300)$^{0.17}$ & \citet{doi:10.1021/acsearthspacechem.0c00197} RRKM fit \\
44 & {AlO + AlO $\rightarrow$ Al$_2$O + O} & -70.2  & 1800 &  1.42(-9)(T/300)$^{0.17}$ & capture \\
45 & {Al$_2$O + O $\rightarrow$ AlO + AlO} &  70.2  & 1800 &  1.42(-9)(T/300)$^{0.17} \exp(-8447.4/T)$ & reverse capt \\
46 & {AlOH + AlOH $\rightarrow$ Al$_2$O + H$_2$O}  & -25.9 & 600 & 9.85(-10)(T/300)$^{0.17}$ & capture \\
47 & {Al$_2$O + H$_2$O $\rightarrow$ AlOH + AlOH}  &  25.9 & 600 & 9.85(-10)(T/300)$^{0.17}\exp(-3593.2/T)$ & reverse capt \\
48 & {AlO + AlOH $\rightarrow$ Al$_2$O + OH} & -14.3 & 500 &  1.00(-9)(T/300)$^{0.17}$ & capture \\
49 & {Al$_2$O + OH $\rightarrow$ AlO + AlOH} & 14.3 & 500 &   1.00(-9)(T/300)$^{0.17}\exp(-1727.3/T)$ & reverse capture \\
50 & {AlO + AlO + M $\rightarrow$ Al$_2$O$_2$ + M} & -547 & 3500 & 2.35(-33) & capture \\
51 & {Al$_2$O$_2$ + M $\rightarrow$ AlO + AlO + M} & 547  & 3500 & 1.45(-10)(T/300)$^{-0.63}\exp(-65887.9/T)$ & detailed balance \\
52 & {AlOH + AlOH $\rightarrow$ Al$_2$O$_2$ + H$_2$} & -21.0 & 400 & 9.8(-10)(T/300)$^{0.17}$ & capture \\
53 & {Al$_2$O$_2$ + H$_2$ $\rightarrow$ AlOH + AlOH} & 21.0  & 400 & 9.8(-10)(T/300)$^{0.17}\exp(-2524.6/T)$ & reverse capture \\
54 & {AlO + AlOH  $\rightarrow$ Al$_2$O$_2$ + H} & -65.9 & 1000 &  1.00(-9)(T/300)$^{0.17}$ & capture \\
55 & {Al$_2$O$_2$ + H $\rightarrow$ AlO + AlOH} & 65.9 & 1000 & 1.00(-9)(T/300)$^{0.17}\exp(-7926.3/T)$ & reverse capture \\

56 & {Al$_2$O + O$_2$ $\rightarrow$ Al$_2$O$_2$ + H} & 19.5 & no & 8.11(-12)(T/300)$^{1.69}\exp(-1965.8/T)$ & detailed balance \\
57 & {Al$_2$O$_2$ + O $\rightarrow$ Al$_2$O + O$_2$} & -19.5 & no   & 7.46(-10)(T/300)$^{0.17}$ & capture \\
58 & {Al$_2$O + H$_2$O $\rightarrow$ Al$_2$O$_2$ + H$_2$} & 4.9   & no   & 5.90(-11)(T/300)$^{1.37}\exp(-1391.5/T)$ & detailed balance (capture) 
\\
59 & {Al$_2$O$_2$ + H$_2$ $\rightarrow$ Al$_2$O + H$_2$O} & -4.9  & no   & 1.20(-11)$\exp(-17285/T)$ & TST calculation \\
\end{tabular}
\end{table*}
\begin{table*}[!ht]	
\begin{tabular}{r l r r l l}
60 & {Al$_2$O + OH $\rightarrow$ Al$_2$O$_2$ + H} & -51.5 & no   & 5.69(-10)(T/300)$^{0.17}$  & capture \\
61 & {Al$_2$O$_2$ + H $\rightarrow$ Al$_2$O + OH} & 51.5 & no    & 1.86(-07)(T/300)$^{-1.55}\exp(-6310.9/T)$ & detailed balance \\
62 & {Al$_2$O$_2$ + H$_2$O $\rightarrow$ Al$_2$O$_3$ + H$_2$} & 85.0  & no   & 1.78(-09)(T/300)$^{0.77}\exp(-11797.1/T)$ & detailed balance \\
63 & {Al$_2$O$_3$ + H$_2$ $\rightarrow$ Al$_2$O$_2$ + H$_2$O} & -85.0 & no   & 1.52(-9)(T/300)$^{0.17}$  & capture \\
64 & {Al$_2$O$_2$ + OH  $\rightarrow$ Al$_2$O$_3$ + H} & 28.2 & 4200  & 2.57(-11)(T/300)$^{1.78}\exp(-1977.4/T)$ & detailed balance \\
65 & {Al$_2$O$_3$ + H $\rightarrow$ Al$_2$O$_2$ + OH} & -28.2 & 4200  & 1.13(-9)(T/300)$^{0.17}$ & capture \\
66 & {AlO$_2$ + AlOH  $\rightarrow$ Al$_2$O$_3$ + H} & -72.5  & 2700  & 1.24(-9)(T/300)$^{0.17}$ & capture \\
67 & {Al$_2$O$_3$ + H $\rightarrow$ AlO$_2$ + AlOH} & 72.5 & 2700 & 1.24(-9)(T/300)$^{0.17}\exp(-8714.8/T)$ & reverse capture \\
68 & {AlO + H$_2$O $\rightarrow$ OAlOH + H} & 8.6 & 5400 & 2.03(-11)$\exp(-1360.0/T)$ & Mangan RRKM fit \\
69 & {OAlOH + H $\rightarrow$ AlO + H$_2$O}  & -8.6   & 5400 & 3.91(-09)(T/300)$^{-1.94}\exp(-725.6/T)$ & detailed balance \\
70 & {AlO + H$_2$O $\rightarrow$ AlOH + OH} & 11.5   & 1000  & 3.89(-10)$\exp(-1295.0/T)$ & \cite{Mangan2021} \\
71 & {AlOH + OH $\rightarrow$ AlO + H$_2$O} & -11.5  & 1000  & 6.05(-10)(T/300)$^{0.17}$ & capture  \\
72 & {AlO + H$_2$O + M $\rightarrow$ Al(OH)$_2$ + M} & -314 & 3100 & 2.40(-26)(T/300)$^{-3.87}$ & \cite{Mangan2021}  \\                
73 & {Al(OH)$_2$ + M $\rightarrow$ AlO + H$_2$O + M} & 314  & 3100 & 1.00(-10)$\exp(-17205/T)$ & Estimate \\
74 & {OAlOH + SiO $\rightarrow$ AlSiO$_3$ + H} & -6.0   &  100  & 1.22(-9)(T/300)$^{0.17}$ & capture \\
75 & {AlSiO$_3$ + H $\rightarrow$ OAlOH + SiO} & 6.0  &  100  & 1.22(-9)(T/300)$^{0.17}\exp(-724.1/T)$ & reverse capture \\
76 & {AlSiO$_3$ + AlO $\rightarrow$ Al$_2$O$_3$ + SiO} & -28.3  & no    & 1.64(-9)(T/300)$^{-0.17}$\tablefootnote[1]{dipole-dipole up to 1700K} & capture \\
77 & {Al$_2$O$_3$ + SiO $\rightarrow$ AlSiO$_3$ + AlO} & 28.3  & no &  6.28(-10)(T/300)$^{-0.18}\exp(-3110.2/T)$ & detailed balance \\
78 & {OAlOH + SiO + M $\rightarrow$ HAlSiO$_3$ + M} & -356 & 1500 &  2.03(-33) & capture \\
79 & {HAlSiO$_3$ + M $\rightarrow$ OAlOH + SiO + M} & 356 & 1500 & 2.72(-10)(T/300)$^{2.37}\exp(-35190.3/T)$ & detailed balance \\
80 & {HAlSiO$_3$ + H $\rightarrow$ AlSiO$_3$ + H$_2$} & -87.3  & no     &  1.13(-9)(T/300)$^{0.17}$ & capture \\ 
81 & {AlSiO$_3$ + H$_2$ $\rightarrow$ HAlSiO$_3$ + H} & 87.3  & no     & 5.07(-13)(T/300)$^{-1.04}\exp(-11390.8/T)$ & detailed balance \\
82 & {Al$_2$O$_3$ + SiO + M $\rightarrow$ Al$_2$O$_3$SiO + M} & -460  & 3000 & 1.38(-33) & capture \\
83 & {Al$_2$O$_3$SiO + M $\rightarrow$ Al$_2$O$_3$ + SiO + M}  & 460   & 3000 & 4.10(-11)(T/300)$^{-0.8}\exp(-53641.8/T)$ & detailed balance \\
84 & {Al$_2$O$_3$SiO + AlO $\rightarrow$ Al$_3$O$_4$ + SiO} & -200  & no & 1.13(-9)(T/300)$^{0.17}$ & capture  \\
85 & {Al$_3$O$_4$ + SiO $\rightarrow$ Al$_2$O$_3$SiO + AlO} &  200  & no & 1.67(-8)(T/300)$^{0.40}\exp(-19243/T)$ & detailed balance \\
86 & {Al$_3$O$_4$ + AlOH $\rightarrow$ Al$_4$O$_5$ + H} & -148  & 3200  & 1.22(-9)(T/300)$^{0.17}$ & capture  \\
87 & {Al$_4$O$_5$ + H $\rightarrow$ Al$_3$O$_4$ + AlOH} & 148 & 3200 & 1.22(-9)(T/300)$^{0.17}\exp(-17777.0/T)$& reverse capture \\
88 & {Al$_2$O$_2$ + AlO + M $\rightarrow$ Al$_3$O$_3$ + M} & -518 & 5000 & 2.38(-33)  & capture  \\       
89 & {Al$_3$O$_3$ + M $\rightarrow$ Al$_2$O$_2$ + AlO + M} & 518  & 5000 & 1.65(-12)(T/300)$^{-1.16}\exp(-62359.5/T)$ & detailed balance \\
90 & {Al$_2$O$_2$ + AlOH $\rightarrow$ Al$_3$O$_3$ + H} & -36.5 & no  & 1.40(-9)(T/300)$^{0.17}$ & capture \\
91 & {Al$_3$O$_3$ + H $\rightarrow$ Al$_2$O$_2$ + AlOH } & 36.5 & no  & 3.20(-11)$\exp(-4713/T)$ & TST calculation \\ 
92 & {Al$_2$O$_3$ + OH $\rightarrow$ Al$_2$O$_4$ + H} & -119 & 4200 & 6.63(-10)(T/300)$^{0.17}$\tablefootnote[2]{dipole-dipole up to 300K} & capture\\
93 & {Al$_2$O$_4$ + H $\rightarrow$ Al$_2$O$_3$ + OH} & 119 & 4200 & 3.18(-7)(T/300)$^{-0.96}\exp(-13422.8/T)$ & detailed balance \\
94 & {Al$_2$O$_3$ + H$_2$O  $\rightarrow$ Al$_2$O$_4$ + H$_2$} & -62.2  & 2000 & 7.84(-10)(T/300)$^{0.17}$ \tablefootnote[3]{dipole-dipole up to 400K} & capture \\
95 & {Al$_2$O$_4$ + H$_2$ $\rightarrow$ Al$_2$O$_3$ + H$_2$O} &  62.2  & 2000 & 7.84(-10)(T/300)$^{0.17}\exp(-7478.2/T)$ & reverse capture \\
96 & {Al$_2$O$_3$ + AlOH $\rightarrow$ Al$_3$O$_4$ + H} & -139   & 1600 & 8.96(-10))(T/300)$^{0.17}$ & capture \\
97 & {Al$_3$O$_4$ + H $\rightarrow$ Al$_2$O$_3$ + AlOH} & 139  & 1600 & 8.96(-10))(T/300)$^{0.17}\exp(-16655.4/T)$ & reverse capture \\
98 & {Al$_3$O$_3$ + H$_2$O $\rightarrow$ Al$_3$O$_4$ + H$_2$} & -17.1 & 250 & 7.83(-10)(T/300)$^{0.17}$ & capture\\
99 & {Al$_3$O$_4$ + H$_2$ $\rightarrow$ Al$_3$O$_3$ + H$_2$O} & 17.1 & 250 & 7.83(-10)(T/300)$^{0.17}\exp(-2077.8/T)$ & reverse capture \\
100 & {Al$_3$O$_3$ + OH  $\rightarrow$ Al$_3$O$_4$ + H} & -73.6 & 1000 & 7.01(-10)(T/300)$^{0.17}$ & capture\\
101 & {Al$_3$O$_4$ + H $\rightarrow$ Al$_3$O$_3$ + OH} & 73.6 & 1000 & 7.01(-10)(T/300)$^{0.17}\exp(-8869.2/T)$  & reverse capture \\
102 & {Al$_3$O$_3$ + AlO + M $\rightarrow$ Al$_4$O$_4$ + M} & -558 & 4400 & 1.76(-33) & capture \\
103 & {Al$_4$O$_4$ + M $\rightarrow$ Al$_3$O$_3$ + AlO + M} & 558 & 4400 & 1.88(-11)(T/300)$^{-1.18}\exp(-67158.6/T)$  & detailed balance \\
104 & {Al$_3$O$_3$ + AlOH $\rightarrow$ Al$_4$O$_4$ + H} & -76.4 & 2000 & 1.03(-9)(T/300)$^{0.17}$ & capture\\
105 & {Al$_4$O$_4$ + H $\rightarrow$ Al$_3$O$_3$ + AlOH} &  76.4 & 2000 &  1.03(-9)(T/300)$^{0.17}\exp(-9181.9/T)$  & reverse capture \\
106 & {Al$_4$O$_4$ + H$_2$O $\rightarrow$ Al$_4$O$_5$ + H$_2$} & -88.8 & 1100 &  8.46(-10)(T/300)$^{0.17}$ & capture\\
107 & {Al$_4$O$_5$ + H$_2$ $\rightarrow$ Al$_4$O$_4$ + H$_2$O} &  88.8 & 1100 &  8.46(-10)(T/300)$^{0.17}\exp(-10673.0/T)$  & reverse capture \\
108 & {Al$_4$O$_4$ + OH $\rightarrow$ Al$_4$O$_5$ + H} & -145 & 1700  & 7.57(-10)(T/300)$^{0.17}$ & capture\\
109 & {Al$_4$O$_5$ + H $\rightarrow$ Al$_4$O$_4$ + OH} &  145 & 1700  & 7.57(-10)(T/300)$^{0.17}\exp(-17464.4/T)$  & reverse capture \\
110 & {Al$_3$O$_4$ + AlO + M $\rightarrow$ Al$_4$O$_5$ + M} & -630 &  4500 & 2.07(-33) &  capture\\   
111 & {Al$_4$O$_5$ + M $\rightarrow$ Al$_3$O$_4$ + AlO + M} & 630 &  4500  & 1.04(-10)(T/300)$^{-1.15}\exp(-75773.8/T)$  & detailed balance \\
112 & {Al$_4$O$_5$ + H$_2$O $\rightarrow$ Al$_4$O$_6$ + H$_2$} & -172 & 2000 & 1.56(-9)(T/300)$^{0.17}$ & capture \\
113 & {Al$_4$O$_6$ + H$_2$ $\rightarrow$ Al$_4$O$_5$ + H$_2$O} & 172  & 2000 & 3.00(-9)$\exp(-21435.0/T)$ & detailed balance (truncated) \\ 
114 & {Al$_4$O$_5$ + OH $\rightarrow$ Al$_4$O$_6$ + H} & -229 & 2500 & 1.40(-9)(T/300)$^{0.17}$ & capture \\
115 & {Al$_4$O$_6$ + H $\rightarrow$ Al$_4$O$_5$ + OH} & 229  & 2500 & 1.40(-9)(T/300)$^{0.17}\exp(-27501.7/T)$ & reverse capture \\
116 & {AlO + AlH $\rightarrow$ Al$_2$O + H} &  -267 & no & 1.12(-9)(T/300)$^{0.17}$ & capture \\
117 & {Al$_2$O + H $\rightarrow$ AlO + AlH} & 267  & no & 1.59(-7)(T/300)$^{0.15}\exp(-31560.5/T)$ & detailed balance \\
118 & {AlO$_2$ + AlH $\rightarrow$ Al$_2$O$_2$ + H} & -353 & no & 1.40(-9)(T/300)$^{0.17}$ & capture \\
119 & {Al$_2$O$_2$ + H $\rightarrow$ AlO$_2$ + AlH} &  353 & no & 1.29(-6)(T/300)$^{-1.35}\exp(-42708.5/T)$ & detailed balance \\
\end{tabular}
\end{table*}
\begin{table*}[!ht]	
\begin{tabular}{r l r r l l}
120 & {Al$_2$O$_2$ + AlH $\rightarrow$ Al$_3$O$_2$ + H} & -73.0 & no & 1.62(-9)(T/300)$^{0.17}$ & capture \\
121 & {Al$_3$O$_2$ + H $\rightarrow$ Al$_2$O$_2$ + AlH} & 73.0 & no & 1.68(-10)(T/300)$^{-0.87}\exp(-7920.0/T)$ & detailed balance \\
122 & {Al$_2$O$_3$ + AlH $\rightarrow$ Al$_3$O$_3$ + H} & -317 & no & 1.04(-9)(T/300)$^{0.17}$ & capture \\
123 & {Al$_3$O$_3$ + H $\rightarrow$ Al$_2$O$_3$ + AlH} & 317  & no & 4.67(-8)(T/300)$^{-0.85}\exp(-37825.9/T)$ & detailed balance \\
124 & {Al$_2$O$_4$ + AlH $\rightarrow$ Al$_3$O$_4$ + H} & -272  & no & 8.27(-10)(T/300)$^{0.17}$ & capture \\
125 & {Al$_3$O$_4$ + H $\rightarrow$ Al$_2$O$_4$ + AlH} & 272 & no &  3.02(-6)(T/300)$^{-0.85}\exp(-33036.1/T)$ & detailed balance \\
126 & {Al$_3$O$_3$ + AlH $\rightarrow$ Al$_4$O$_3$ + H} & -117 & no & 1.22(-9)(T/300)$^{0.17}$ & capture \\
127 & {Al$_4$O$_3$ + H $\rightarrow$ Al$_3$O$_3$ + AlH} & 117  & no & 2.72(-9)(T/300)$^{-0.86}\exp(-13171.8/T)$ & detailed balance \\
128 & {Al$_3$O$_4$ + AlH $\rightarrow$ Al$_4$O$_4$ + H} & -256  & no & 1.44(-9)(T/300)$^{0.17}$ & capture \\
129 & {Al$_4$O$_4$ + H $\rightarrow$ Al$_3$O$_4$ + AlH} & 256 & no & 1.77(-10)(T/300)$^{-0.84}\exp(-29882.8/T)$ & detailed balance \\
130 & {Al$_3$O$_5$ + AlH $\rightarrow$ Al$_4$O$_5$ + H} & -387 & no & 1.15(-10)(T/300)$^{0.17}$ & capture \\
131 & {Al$_4$O$_5$ + H $\rightarrow$ Al$_3$O$_5$ + AlH} & 387 & no & 1.44(-07)(T/300)$^{-0.84}\exp(-46031.3/T)$ & detailed balance \\
132 & {Al$_3$O$_6$ + AlH $\rightarrow$ Al$_4$O$_6$ + H} & -682 & 5100 & 1.15(-9)(T/300)$^{0.17}$ & capture \\
133 & {Al$_4$O$_6$ + H $\rightarrow$ Al$_3$O$_6$ + AlH} & 682 & 5100 & 1.00(-10)$\exp(-81959.0/T)$ & capture reverse \\
134 & {Al + F + M $\rightarrow$ AlF + M} & -682 & no &  3.32(-32)(T/300)$^{-1.66}$ & RRKM fit \\
135 & {AlF + M $\rightarrow$ Al + F + M} & 682 & no & 5.76(-10)$\exp(-80937.0/T)$ & RRKM fit \\
136 & {Al + Cl + M $\rightarrow$ AlCl + M} & -515 & 5400 &  3.31(-31)(T/300)$^{0.2}$ & RRKM fit \\
137 & {AlCl + M $\rightarrow$ Al + Cl + M} & 515 & 5400 & 1.00(-9)$\exp(-619370/T)$ & RRKM fit \\
138 & {Al + H + M $\rightarrow$ AlH + M} & -301 & 3200 & 4.42(-31)(T/300)$^{0.34}$ & \cite{SWIHART200391} \\
139 & {AlH + M $\rightarrow$ Al + H + M} & 301 & 3200 & 1.00(-9)$\exp(-36206/T)$ & RRKM fit \\
140 & {AlCl + H $\rightarrow$ Al + HCl} & 82.8 & no &  1.00(-10)$\exp(-9959/T)$ & RRKM fit \\
141 & {Al + HCl $\rightarrow$ AlCl + H} & -82.8 & no & 1.52(-10)$\exp(-803/T)$ & \cite{doi:10.1021/j100340a019} \\
142 & {AlCl + H $\rightarrow$ AlH + Cl} & 214 & no &  8.20(-12)(T/300)$^{0.36}\exp(-25738.0/T)$ & detailed balance \\
143 & {AlH + Cl $\rightarrow$ AlCl + H} & -214 & no & 7.19(-12)(T/300)$^{0.5}$ & collision \\
144 & {F + H$_2$ $\rightarrow$ HF + H} & -133 & 3200 & 1.10(-10)$\exp(-450/T)$ & NIST \\
145 & {HF + H $\rightarrow$ F + H$_2$} & 133 & 3200 & 8.37(-11)(T/300)$^{0.6}\exp(-16357/T)$ & NIST \\
146 & {Cl + H$_2$ $\rightarrow$ HCl + H} &  4.3 & 400 & 3.87(-12)(T/300)$^{1.58}\exp(-1610/T)$ & NIST \\
147 & {HCl + H $\rightarrow$ Cl + H$_2$} & -4.3 & 400 & 2.41(-12)(T/300)$^{1.44}\exp(-1240/T)$ & NIST \\
148 & {Al + HF $\rightarrow$ AlH + F} &  269 & no & 6.34(-10)(T/300)$^{0.57}\exp(-31487.8/T)$ & detailed balance \\
149 & {AlH + F $\rightarrow$ Al + HF} & -269 & no & 3.92(-10)(T/300)$^{0.17}$ & capture \\
150 & {Al + HF $\rightarrow$ AlF + H} & -112 & no & 4.96(-10)(T/300)$^{0.17}$ & capture \\
151 & {AlF + H $\rightarrow$ Al + HF} &  112 & no & 1.79(-10)(T/300)$^{0.02}\exp(-12303.0/T)$ & detailed balance \\
152 & {AlO + Cl $\rightarrow$ AlCl + O} & -17.4 & no & 6.79(-12)(T/300)$^{0.5}$ & collision \\
153 & {AlCl + O $\rightarrow$ AlO + Cl} &  17.4 & no & 3.18(-12)(T/300)$^{0.48}\exp(-1833.5/T)$ & detailed balance \\
154 & {AlOH + Cl $\rightarrow$ AlCl + OH} & 38.5 & no & 6.90(-9)(T/300)$^{-0.79}\exp(-5043.7/T)$ & detailed balance \\  
155 & {AlCl + OH $\rightarrow$ AlOH + Cl} & -38.5 & no & 6.35(-9)(T/300)$^{0.17}$ & capture \\
156 & {AlOH + Cl $\rightarrow$ AlO + HCl} & 49.3 & no &  4.65(-9)(T/300)$^{-0.74}\exp(-6143.9/T)$ & detailed balance \\ 
157 & {AlO + HCl $\rightarrow$ AlOH + Cl} & -49.3 & no & 7.45(-10)(T/300)$^{0.17}$ & capture \\
158 & {Al$_4$O$_6$ +  Al$_4$O$_6$ $\rightarrow$ Al$_8$O$_{12}$} & -834 & no & 2.00(-9)& coagulation \\
159 & \textbf{Al$_8$O$_{12}$ $\rightarrow$ Al$_4$O$_6$ +  Al$_4$O$_6$} & 834 & no & 6.5(-3)$\exp(-100000./T)$  & detailed balance \\
160 & {AlO + OH $\rightarrow$ AlOH + O} & -55.9 & no & 3.0(-12) & Estimate from trajectory calculation \tablefootnote{Having a late barrier, a successful reaction on the triplet potential energy surface (PES) requires that AlO is vibrationallly excited, and a translational collision energy less than 7.5 kJ mol$^{-1}$. The resulting rate coefficient is estimated to be 3.0-3.7(-12) cm$^3$ s$^{-1}$ for T=1000-2000 K.} \\
161 & {AlOH + O $\rightarrow$ AlO + OH} &  55.9 & no & 1.50(-11)(T/300)$^{-0.97}\exp(-6861.8/T)$ & detailed balance \\
162 & {Al + OH $\rightarrow$ AlO + H} & -71.9 & no & 1.40(-10)$\exp(-9640./T)$ & Estimate from trajectory calculation \\
163 & {AlO + H $\rightarrow$ Al + OH} &  71.9 & no & 7.35(-11)(T/300)$^{-0.14}\exp(-17406.3/T)$ & RRKM fit \\
164 & {Al$_2$O + h$\nu_{\lambda<213 nm}$ $\rightarrow$  AlO + Al} & -568 & 4100 & 8.20(-7)$\exp(0.00273T)$ & \citep{Mangan2021} \\
165 & {AlO + h$\nu_{\lambda<252 nm}$ $\rightarrow$  Al + O} & -498 & 4900 & 9.68(-7)$\exp(0.00102T)$ & \citep{Mangan2021} \\
166 & {AlOH + h$\nu_{\lambda<238 nm}$ $\rightarrow$  AlO + H} & -482 & 4900 & 1.53(-4)($\exp(0.0014T)$)
&  \citep{Mangan2021} \\
\end{tabular}
\end{table*}

\begin{table*}
\caption{The cartesian coordinates x,y,z of (Al$_2$O$_3$)$_n$, $n=$1$-$10, GM clusters (in units of \AA{})\tablefootnote{Tables \ref{coordinates} are only available in electronic form at the CDS via anonymous ftp to cdsarc.u-strasbg.fr (130.79.128.5) or via http://cdsweb.u-strasbg.fr/cgi-bin/qcat?J/A+A/}.}
\begin{tabular}{l c c c}
1A  &   B3LYP/6-311+G(d) & -710.72608 a.u. \\
\hline
Al  &   0.00000 & 0.00000 & 1.69489 \\
..  &  ..       &  ..     & ...     \\  
\end{tabular}
\label{coordinates}
\end{table*}

\section{Thermo-chemical tables}
\begin{table}
\caption{The thermochemical tables of (Al$_2$O$_3$)$_n$, $n=$1$-$10, GM clusters as a function of temperature.\tablefootnote{Tables \ref{TDtables} are only available in electronic form at the CDS via anonymous ftp to cdsarc.u-strasbg.fr (130.79.128.5) or via http://cdsweb.u-strasbg.fr/cgi-bin/qcat?J/A+A/}}
\begin{tabular}{c c c c c c c}
1A & \\
 T(K)  &  S (J/molK)  & C$_p$ (J/molK)  & ddH (kJ/mol)   &    dH$_f$ (kJ/mol) &     dG$_f$ (kJ/mol)  & log K$_f$ \\
\hline
   0.00    &    0.000     &      0.000  &         0.000   &    -546.572   &   -546.572   &     $\infty$ \\
  ..     &    ..  & .. & .. &  ..& .. & ..  \\
\end{tabular}
\label{TDtables}
\end{table}

\section{Thermochemistry: Fit parameters}

\begin{table}
\caption{Fitting parameters a,b,c,d,e for the computation of $\Delta$G$_{f}^{0}$(T) used in equilibrium calculations\label{dgfits}.}
\begin{tabular}{r | c c c c c}
Species & a & b & c & d & e \\
\hline
AlO            & 5.99300E+04 & -1.37970E+00 & -1.73737E+00 & 3.91773E-04 & -2.51845E-08 \\
AlOH           & 1.18027E+05 & -2.05964E+00 & -8.69160E+00 & 5.90024E-04 & -3.04906E-08 \\
OAlOH          & 1.69811E+05 & -3.52430E+00 & -1.43673E+01 & 1.25241E-03 & -7.34479E-08 \\
HAlOH          & 1.34662E+05 & -4.57736E+00 & -4.35980E+00 & 1.66011E-03 & -9.78547E-08 \\
AlO$_2$H$_2$   & 2.08760E+05 & -5.10482E+00 & -1.50931E+01 & 1.92278E-03 & -1.11241E-07 \\        
AlO$_2$        & 1.07130E+05 & -2.61203E+00 & -8.89728E+00 & 7.50030E-04 & -4.93246E-08 \\
Al$_2$O        & 1.28411E+05 & -2.40921E+00 & -1.05621E+00 & 6.49789E-04 & -4.25611E-08 \\
Al$_2$O$_2$    & 1.86105E+05 & -3.31380E+00 & -1.99433E+01 & 1.58854E-03 & -1.06058E-07 \\
Al$_2$O$_3$    & 2.34139E+05 & -3.95729E+00 & -2.82239E+01 & 2.11009E-03 & -1.40019E-07 \\
Al$_2$O$_4$    & 2.99834E+05 & -4.30886E+00 & -4.27871E+01 & 2.51478E-03 & -1.66932E-07 \\
Al$_3$O$_3$    & 3.08682E+05 & -4.20634E+00 & -3.99769E+01 & 2.44693E-03 & -1.61816E-07 \\
Al$_3$O$_4$    & 3.69176E+05 & -5.56346E+00 & -5.20710E+01 & 3.40984E-03 & -2.28468E-07 \\
Al$_4$O$_4$    & 4.36049E+05 & -4.97450E+00 & -6.35660E+01 & 3.25225E-03 & -2.14255E-07 \\
Al$_4$O$_5$    & 5.05163E+05 & -6.53645E+00 & -7.58250E+01 & 4.29600E-03 & -2.85685E-07 \\
Al$_4$O$_6$    & 5.84165E+05 & -8.82800E+00 & -8.42817E+01 & 5.63090E-03 & -3.74702E-07 \\
AlSiO$_3$      & 2.67025E+05 & -4.60668E+00 & -2.59855E+01 & 2.35614E-03 & -1.54434E-07 \\
HAlSiO$_3$     & 2.75443E+05 & -1.00583E+01 & -7.73659E+00 & 6.85770E-03 & -8.55013E-07 \\
Al$_2$O$_3$SiO & 3.80803E+05 & -9.85869E+00 & -2.79375E+01 & 9.23519E-03 & -1.18218E-06 \\
AlF            & 8.19970E+04 & -1.31009E+00 & -2.43594E+00 & 4.12562E-04 & -3.03640E-08 \\
AlCl           & 6.20720E+04 & -9.43409E-01 & -4.14453E+00 & 2.44222E-04 & -1.87099E-08 \\
Al$_2$         & 1.09331E+04 & -7.39015E-01 & -5.44758E+00 & 1.35276E-04 & -1.05029E-08 \\
AlH            & 3.62088E+04 & -1.68480E+00 &  1.07115E+00 & 4.62791E-04 & -2.91925E-08 \\
Al$_3$O$_5$    & 4.22272E+05 & -6.47463E+00 & -6.17019E+01 & 4.04645E-03 & -2.69671E-07 \\
Al$_3$O$_6$    & 4.65554E+05 & -5.97223E+00 & -7.53001E+01 & 3.95721E-03 & -2.60889E-07 \\
Al$_4$O$_3$    & 3.59063E+05 & -3.47918E+00 & -5.55172E+01 & 2.29771E-03 & -1.50608E-07 \\
Al$_3$O$_2$    & 2.31189E+05 & -2.55393E+00 & -3.25776E+01 & 1.41095E-03 & -9.25233E-08 \\
\hline  
Al$_8$O$_{12}$  & 1.16740E+06 & -1.57762E+01 & -1.99568E+02 & 1.12760E-02 & -7.49050E-07 \\ 
Al$_{10}$O$_{15}$    & 1.48494e+06 & -1.94214E+01 & -2.57899E+02 & 1.41833E-02 & -9.41950E-07 \\
Al$_{12}$O$_{18}$(B) & 1.80753E+06 & -2.37453E+01 & -3.17029E+02 & 1.75208E-02 & -1.16698E-06 \\   
Al$_{14}$O$_{21}$    & 2.12820E+06 & -2.75694E+01 & -3.73757E+02 & 2.04905E-02 & -1.36329E-06 \\
Al$_{16}$O$_{24}$    & 2.44711E+06 & -3.23660E+01 & -4.30676E+02 & 2.40589E-02 & -1.60386E-06 \\ 
Al$_{18}$O$_{27}$    & 2.78902E+06 & -3.60898E+01 & -4.90629E+02 & 2.70106E-02 & -1.79999E-06 \\
Al$_{20}$O$_{30}$    & 3.10574e+06 & -4.06379E+01 & -5.45915E+02 & 3.04129E-02 & -2.02790E-06 \\

\end{tabular}
\end{table}

\section{Cluster geometries}

\begin{table}
\caption{Cluster volumes (in \AA{}$^3$) and radii (in \AA{}) as derived from atomic core coordinates
(Delauney triangulation) and from van-der-Waals interactions. The radii are calculated assuming spherical volumes.\label{cradii}}
\begin{tabular}{l | c c | c c }
$n$ & \multicolumn{2}{c}{Delauney} & \multicolumn{2}{c}{van der Waals} \\
     &  Volume & Radius &  Volume & Radius \\
\hline
1 & 0.0        & 0.0     &  73.73   &    2.60   \\ 
2 & 13.88      & 1.49    &  128.59  &    3.13   \\
3 & 29.63      & 1.92    &  185.49  &    3.54   \\
4 & 51.16      & 2.30    &  249.61  &    3.91   \\
5 & 80.00      & 2.67    &  308.37  &    4.19   \\
6 & 102.00     & 2.90    &  358.37  &    4.41   \\
7 & 142.74     & 3.24    &  347.3   &    4.36   \\ 
8 & 153.15     & 3.32    &  473.94  &    4.84   \\
9 & 154.17     & 3.33    &  520.73  &    4.99   \\
10 & 226.18    & 3.78    &  594.16  &    5.22   \\
\end{tabular}
\end{table}


\section{Cluster coordinates}


\end{appendix}
\end{document}